\documentclass[aps,11pt,article,amsmath,amssymb,nofootinbib]{revtex4-1}
\usepackage{braket}
\usepackage{color}
\usepackage{subfigure}
\usepackage{graphics} % standard graphics specifications
\usepackage{graphicx} % alternative graphics specifications
\usepackage{longtable} % helps with long table options
\usepackage{verbatim}
\usepackage{ulem}
\usepackage{bm} % special 'bold-math' package
\usepackage[colorlinks=true]{hyperref}
\usepackage{cleveref}

\newcommand{\nn}{\nonumber}
\newcommand{\drm}{\mathrm{d}}
\newcommand{\expect}[1]{{\langle #1 \rangle}}
\newcommand{\ie}{\textit{i.e., }}
\newcommand{\eg}{\textit{e.g., }}

\newcommand{\BQIC}{Berkeley Center for Quantum Information and Computation, Berkeley, California 94720 USA}
\newcommand{\deptPhys}{Department of Physics, University of California, Berkeley, California 94720 USA}
\newcommand{\deptChem}{Department of Chemistry, University of California, Berkeley, California 94720 USA}
\newcommand{\SNL}{Extreme-Scale Data Science and Analytics, Sandia National Laboratories, Livermore, California 94550 USA}

\begin{document}
\title{Quantum proportional-integral (PI) control}
\author{Hui Chen$^{1}$\footnote{Present address: Joint Institute of UMich-SJTU and Key Laboratory of
  System Control and Information Processing (MOE),
  Shanghai Jiao Tong University, Shanghai, 200240, China}}
\author{Hanhan Li$^{1,2}$\footnote{Present address: Google, 1600 Amphitheatre Parkway, Mountain View, CA 94043 USA}}
\author{Felix Motzoi$^{1,3}$\footnote{Present address: Forschungszentrum J\"ulich, Institute of Quantum Control (PGI-8), D-52425 J\"ulich, Germany}}
\author{Leigh Martin$^{1,2}$\footnote{Present address: Department of Physics, Harvard University, Cambridge, Massachusetts 02138, USA}}
\author{K. Birgitta Whaley$^{1,3}$}
\author{Mohan Sarovar$^{4}$}

\address{$^1$\BQIC}
\address{$^2$\deptPhys}
\address{$^3$\deptChem}
\address{$^4$\SNL}

\date{\today}
%%%%%%%%%%%%%%%%%%%%%%%%%%%%%%%%%%%%%%%%%%%%%%%%%%%%%%%%%%%%%%%%%%%%%%%%%%%%%%%%%%%%%%%
\begin{abstract}
Feedback control is an essential component of many modern technologies and 
provides a key capability for emergent quantum technologies. We extend existing approaches of direct feedback control in which the controller applies a function directly proportional to the output signal (P feedback),
to strategies in which feedback determined by an integrated output signal (I feedback), and to strategies in which feedback consists of a combination of P and I terms. The latter quantum PI feedback constitutes the analog of the widely used proportional-integral feedback of classical control. All of these strategies are experimentally feasible and require no complex state estimation. We apply the resulting formalism to two canonical quantum feedback control problems, namely, generation of an entangled state of two remote qubits, and stabilization of a harmonic oscillator under thermal noise under conditions of arbitrary measurement efficiency. These two problems allow analysis of the relative benefits of P, I, and PI feedback control. We find that for the two-qubit remote entanglement generation the best strategy can be a combined PI strategy 
when the measurement efficiency is less than one. In contrast, for harmonic state stabilization we find that P feedback shows the best performance when actuation of both position and momentum feedback is possible, while when only actuation of position is available, I feedback consistently shows the best performance, although feedback delay is shown to improve the performance of a P strategy here. 

\end{abstract}

\maketitle
%%%%%%%%%%%%%%%%%%%%%%%%%%%%%%%%%%%%%%%%%%%%%%%%%%%%%%%%%%%%%%%%%%%%
%%%%%%%%%%%%%%%%%%%%%%%%%%%%%%%%%%%%%%%%%%%%%%%%%%%%%%%%%%%%%%%%%%%%%%%%%%%%%%%%%%%%%%%
\section{Introduction}
\label{sec:Introduction}
%%%%%%%%%%%%%%%%%%%%%%%%%%%%%%%%%%%%%%%%%%%%%%%%%%%%%%%%%%%%%%%%%%%%%%%%%%%%%%%%%%%%%%%

The maturation of quantum technologies relies heavily on the development of advanced quantum measurement and control solutions. For this purpose, many concepts and solutions developed in classical control theory and practice can be carried over to the quantum domain. Recent examples of useful application of classical control concepts in the context of quantum systems are Lyapunov control \cite{DONG2012725,MIRRAHIMI20051987,Wang2010,KUANG20101257,SHARIFI2011522}, LQG control \cite{PhysRevA.60.2700,PhysRevA.62.012105,NURDIN20091837, Shaiju2007,PhysRevA.87.013815}, risk-sensitive control \cite{Helon2006,Yamamoto2009,PhysRevA.69.032108}, and filtering and smoothing for estimation and control \cite{BELAVKIN1999A405,edwards2005optimal,bouten2007introduction,Mathew2008Hinfinity,Nurdin2017,gough2012quantum, yanagisawa_quantum_2007, tsang_time-symmetric_2009, gammelmark_past_2013, guevara_quantum_2015, wheatley_improved_2015, huang_smoothing_2018}. 

Feedback control is particularly important for applications such as error correction, cooling, and stabilization of quantum systems. Feedback becomes most interesting when the control signals can be applied to a quantum system at timescales that are comparable to the timescale of the measurement. In this case, one must model the effects of intrinsic time evolution, measurement (including quantum backaction) and feedback control all at the same time, which results in interesting and complex dynamics. This typically leads to a description in terms of a continuous-in-time stochastic dynamical equation for the density matrix of the quantum system. 
The simplest type of feedback, in which the feedback operation is directly proportional to the measurement signal at the same time, leads to Markovian evolution of the system \cite{wiseman1993quantum, wiseman_quantum_1994}. This proportional feedback (often termed `direct' feedback) has been applied in theoretical analysis of many problems including state stabilization and cooling \cite{PhysRevA.64.063810, PhysRevB.68.235328, bushev_feedback_2006,PhysRevA.65.063803,Vitali:s}, quantum error correction \cite{PhysRevA.65.042301,PhysRevA.67.052310, chase_efficient_2008}, state purification \cite{combes_rapid_2006, combes_rapid_2010} and generation of entangled states \cite{mancini_optimal_2007, martin_deterministic_2015, PhysRevA.71.042309, martin2019single} and squeezed states \cite{VITALI2011848}, and has also been experimentally demonstrated \cite{PhysRevLett.89.133602,PhysRevA.70.023819,vijay2012stabilizing, martin_implementation_2019}. Recent work has extended quantum feedback control beyond proportional feedback to implementations based on estimation of the quantum state \cite{PhysRevA.74.012322, PhysRevLett.92.223004,thomsen2002continuous}, implementations using stochastic noise sources \cite{cardona2020exponential}, and to implementations using the most general form of feedback that does not include a time-delayed proportional  term
 \cite{zhang_locally_2018}. In the latter framework, referred to as Proportional and Quantum State Estimation (PaQS) feedback, the feedback operator can equivalently be expressed as a sum of independent deterministic and stochastic contributions. {This approach has also been extended to multiple measurement and feedback operators \cite{martin2019single}.}  In several instances, locally optimal feedback laws have been derived \cite{doherty2012quantum, PhysRevA.60.2700, PhysRevA.62.012105, PhysRevA.92.062321, martin2017optimal, zhang_locally_2018, jiang_optimality_2019}, with global optimality being shown in a smaller number of cases \cite{PhysRevA.92.062321,martin2017optimal,jiang_optimality_2019}.

As is the case for complex classical systems, the implementation of advanced, and particularly of optimal, feedback control solutions can be challenging, due to instrumentation and computation demands. Therefore, it is important to also develop heuristic control solutions in the quantum domain. In this paper we adapt one of the most widely used classical control heuristics, proportional-integral, referred to as PI feedback control \cite{aastrom1995pid}, to the quantum domain. 
In the classical domain both P and PI feedback are subsets of proportional, integral, derivative (PID) control, which includes options for modulating the feedback signal with both integrals and derivatives of the measurement signal, in addition to simple multiples of this. In classical PID control, the feedback signal is proportional to the function
\begin{align}
\label{eq:classPID}
f(t) = \alpha_p e(t) + \alpha_i  \int_0^\tau dt' \chi(t') e(t-t') + \alpha_d \frac{{\rm d}}{{\rm d}t}e(t),
\end{align}
where $e(t)$ is an \emph{error signal} that is usually derived from the measurement at time $t$, and $\alpha_{p}, \alpha_i, \alpha_d$ are real coefficients that dictate the relative weights of the proportional, integral and derivative information, respectively, in forming the control law at any time. These weights are usually tuned empirically to achieve good control performance, since their optimal values cannot be computed \emph{a priori} except for very simple systems. Intuitively, the integral portion is used to compensate for unused parts of the measurement signal at earlier times -- integration can increase the signal to noise ratio, can decrease the amount of time it takes to reach the steady-state, and can decrease overshoot of the desired set point. The third component of Eq. \eqref{eq:classPID}, derivative control, can increase the stability of a result by suppressing
slow deviations away from the desired target -- here the derivative attempts to anticipate the direction of change in the error. While PID control is not known to be optimal in any general setting, it has proven to be a very useful framework for formulating heuristic control laws in practice \cite{aastrom1995pid}.

In this paper we address the extension of the first two components of PID control to the quantum domain, formulating a quantum PI feedback law and analyzing the relative benefits of quantum PI, I, and P feedback in two canonical problems for quantum control, namely generation of  entanglement between remote qubits using local Hamiltonians and non-local measurements, and state stabilization of the harmonic oscillator in an external environment. In contrast to some earlier studies of these systems~\cite{PhysRevA.60.2700,martin_deterministic_2015,martin2017optimal}, our feedback implementations for these problems do not require any state estimation and only rely on simple integrals of the measured signal.
We allow for a time delay in the implementation of P feedback, as originally proposed by Wiseman~\cite{wiseman_quantum_1994}. A time delay between obtaining the measurement signal and implementing a feedback operation reflects common experimental constraints and is often regarded as being detrimental to proportional feedback \cite{ruskov2004maintaining,patti2017linear}. However we shall see that in the case of state stabilization of the harmonic oscillator, a time delay introduces additional flexibility of feedback that can be beneficial when the feedback control operations are restricted. We also examine the robustness of P feedback with respect to uncertainties in the time delay, in particular, to increases in the time delay beyond the ideal values for each protocol.

In general, our findings for these two classes of implementations show that adding an integral component to quantum feedback control can be useful in some but not all settings. This is different from the classical setting where adding an integral component to feedback control is almost universally beneficial \cite{aastrom1995pid}.   The different behavior of quantum systems can be rationalized by recalling a key difference between quantum and classical settings, which is the unavoidable presence of stochastic measurement noise in quantum systems.  In classical systems measurement noise can be minimized and even sometimes eliminated. However for quantum systems, any information gain from a measurement necessarily comes at the cost of added noise on the system.  The proportional component of feedback can be very effective at minimizing the impact of this added noise. In special cases, including entanglement generation for two qubits with unit efficiency measurements \cite{martin2017optimal} and the harmonic oscillator state stabilization with both position and momentum controls \cite{PhysRevA.60.2700}, P feedback can be used to cancel the measurement noise.  However, when this is not possible or when there are additional noise sources, we find that I feedback, or a combination of I and P feedback, can be more effective than P feedback.

We note that rigorous analysis of a quantum version of full PID control within the input-output analysis of controlled quantum stochastic evolutions has been recently presented by Gough \cite{gough2017quantum,gough2017non}.  In the current study of practical implementations, we do not investigate the full PID control in the quantum setting because the singular nature of the quantum measurement record makes the derivative terms ill-behaved and thus not useful for practical control implementations without further modifications. See Refs. \cite{PhysRevA.65.063803} and \cite{Vitali:s} for interesting applications of derivative-based feedback.
 
The remainder of the paper is organized as follows.  Sec.~\ref{sec:formalism} introduces notation and presents the general equation for PI feedback. Sec.~\ref{sec:TwoQubitEntanglementGeneration} discusses the control of entanglement of two remote qubits via a half-parity measurement and local feedback operations.  
Here we find that I feedback and PI feedback both show improved performance over P feedback alone. Sec.~\ref{sec:HO} investigates the control of state stabilization of a harmonic oscillator in a thermal environment, using feedback control on either both oscillator quadratures or a single quadrature. When control over both quadratures is possible, P feedback is found to perform better than pure I feedback control. 
When control over only the position quadrature is available, we find that time delay in P feedback can be beneficial, by allowing  an approximation of the average momentum of the state that can be used to generate a good control law.  However, despite this improvement of delayed P feedback over the direct, i.e., instantaneous, setting, a pure I feedback control strategy is nevertheless found to give better performance under the conditions of thermal damping.
In both Sections~\ref{sec:TwoQubitEntanglementGeneration} and ~\ref{sec:HO} we compare the results with prior work employing state estimation based feedback, 
and also analyze the robustness of the control law with respect to non-ideal time delay values. We close with a discussion and outlook for further work in Sec.~\ref{sec:discussion}.

%%%%%%%%%%%%%%%%%%%%%%%%%%%%%%%%%%%%%%%%%%%%%%%%%%%%%%%%%%%%%%%%
\section{Formalism}
\label{sec:formalism}
\begin{figure}[t!]
\includegraphics[width=0.5\columnwidth]{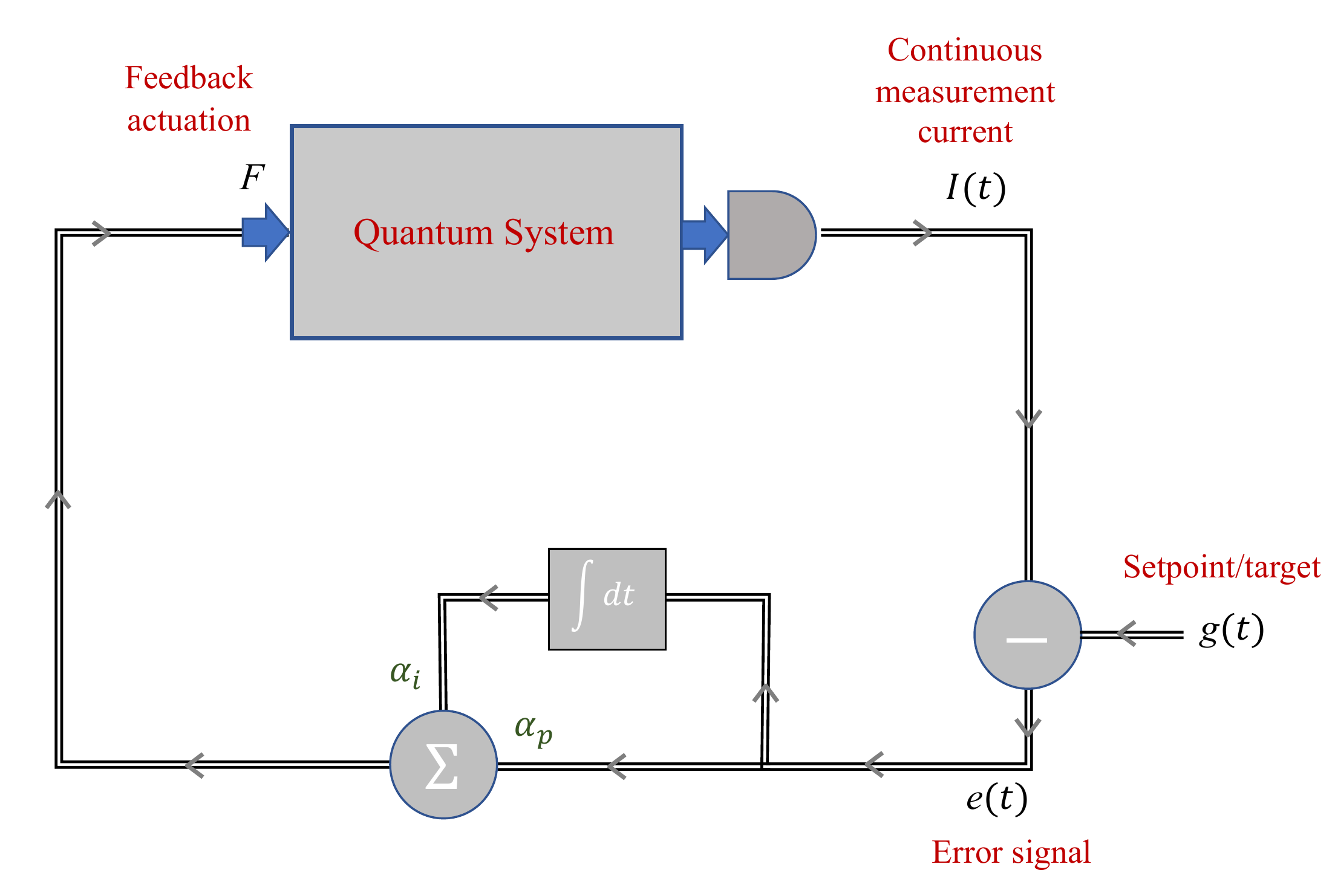}
\caption{Schematic block diagram of PI feedback control of a quantum system. First, the result of a continuous measurement on a quantum system is compared with a target value to form an error signal. This error signal is used to form two signals: (i) a scaled version obtained by multiplication by a real coefficient $\alpha_p\geq 0$, and (ii) a smoothed version obtained by integrating over a time interval and multiplication by a real coefficient $\alpha_i\geq 0$. These two signals are then additively combined and then used to condition actuation of the quantum system by an operator $F$.
\label{fig:block}
}
\end{figure}

%%%%%%%%%%%%%%%%%%%%%%%%%%%%%%%%%%%%%%%%%%%%%%%%%%%%%%%%%%%%%%%%
In this section, we will develop the formalism for a quantum system under continuous-in-time measurement (\eg  homodyne detection) and PI feedback control. Fig. \ref{fig:block} shows a block diagram of the feedback system that we aim to model.
We define $\rho$ to be the state of the system, $H$ the intrinsic Hamiltonian, $c$ the variable-strength measurement operator, and $\eta$ the measurement efficiency. We will set $\hbar=1$ throughout the paper.

The dynamics of the system conditioned on the measurement record, but without feedback control, is described by the following It\^o  stochastic master equation (SME) \cite{wiseman2010quantum}:
\begin{equation}
	\label{eq:sme_nf}
	[\drm \rho(t)]_{m}=-i[H,\rho(t)]+\mathcal{D}[c]\rho(t)\drm t
										+\sqrt{\eta}\mathcal{H}[c]\rho(t)\drm W(t),
\end{equation}
where $dW(t)$ are Wiener increments (Gaussian-distributed random variables with mean zero and autocorrelation $\mathbb{E}\{dW(s)dW(t)\} = \delta(t-s)dt$). The superoperators $\mathcal{D}$ and $\mathcal{H}$ in this equation are defined as $\mathcal{D}[A]\rho\equiv A\rho A^\dag -\frac{1}{2}(A^\dag A\rho+\rho A^\dag A)$ and $\mathcal{H}[A]\rho \equiv A\rho+\rho A^\dag-\mathrm{Tr}[(A+A^\dag)\rho]\rho$.
The corresponding measurement current can be written as \cite{wiseman2010quantum}
\begin{equation}
	j(t)=\langle c+c^\dagger \rangle(t)+\xi(t)/\sqrt{\eta},
	\label{eq:current}
\end{equation}
where $\xi(t) \equiv \drm W/\drm t$ is a white noise process. To emphasize the link between the measurement current and the conditional state evolution, the last term in Eq. \eqref{eq:sme_nf} is sometimes written as $\eta (j(t) - \expect{c+c^\dagger}(t))\mathcal{H}[c]\rho(t)\drm t$.

Before adding the feedback, we first define the error signal by analogy with classical PID control, as 
\begin{align}
    e(t) = j(t) - g(t),
\end{align}
where $g(t)$ is the setpoint or goal. This is often the desired  value of the observable $\langle c+c^\dagger \rangle(t)$ but could also be another target function. $g(t)$ is assumed to be a smoothly varying or constant function. Then the PI feedback operator in the quantum setting takes the form
\begin{align}
\label{eq:feedback_operator}
	\Big[ \alpha_p(t) e(t-\tau_P) + \alpha_i(t) \mathcal{J}(t) \Big] F, 
\end{align}
with some Hermitian operator $F$. 
Here $\alpha_p(t)$ and $\alpha_i(t)$ are time-dependent proportional and integral coefficients, respectively. This differs from classical PID control where the control coefficients are time-independent. Here, we will allow for time-dependence that is deterministic and independent of the measurement current, although in the following we will drop the time index on these coefficients for conciseness unless we wish to emphasize the time-dependence.  We have also included the freedom of having a time delay $\tau_P>0$ in the proportional component. While this is often viewed as an  experimental constraint on implementation of quantum feedback control protocols that is detrimental to performance~\cite{ruskov2004maintaining,patti2017linear}, we shall see below that for the harmonic oscillator state stabilization problem this can be used constructively to improve performance (subsection \ref{sec:oscillator_true_pro}). $\mathcal{J}(t)$ is the integrated error signal, 
\begin{align}
\label{eq:j_int}
	\mathcal{J}(t) = \int^{t}_{t-\tau_I} ds ~ w(t,s) \left[ j(s) - g(s)\right],
\end{align}
where $w$ is a smooth integration kernel that can be used to vary the contribution of the measurement current at past times, and $\tau_I$ is the integration time. We shall assume the kernels are $L^2$ integrable and normalize them such that $\int_0^{\tau_I} ds w(t,s) = 1$. Time-homogeneous kernels just depend on the time separation, $w(t,s) \rightarrow w(t-s)$. Typically, $w(t,s)$ decays with $t-s$ and puts decreasing weight on measurement results from further in the past. 

The action of this PI feedback only is captured by the following dynamics of the system density matrix $\rho(t)$:
\begin{equation}
	\label{eq:fb}
	[\dot{\rho}(t)]_{fb}=\mathcal{K} \rho \equiv -i \Big[ \alpha_p e(t-\tau_P) + \alpha_i\mathcal{J}(t) \Big] [F, \rho(t)],
\end{equation}
We now combine Eqs. \eqref{eq:sme_nf} and \eqref{eq:fb} to derive the SME for evolution under measurements and the PI feedback,   
using the general formalism developed
in Ref. \cite{wiseman_quantum_1994} and its extension to smoothed feedback signals in Refs. \cite{wiseman_quantum_1994,sarovar_2007}. For convenience we define the commutator superoperator $F^\times$ as $F^\times\rho\equiv [F,\rho]$. The time-evolved state after an infinitesimal time $dt$ is given by
\begin{equation}
	\rho(t+\drm t)=e^{\mathcal{K}\drm t} \{1-iH^\times\drm t+\mathcal{D}[c]\drm t+\sqrt{\eta}\mathcal{H}[c]\drm W(t) \} \rho(t). 
\end{equation} 
Note that this form ensures causality, since the feedback acts after the evolution due to measurement. The infinitesimal evolution equation is then obtained by expanding the exponential $e^{\mathcal{K}dt}$ in a Taylor series up to order $\drm t$. The first and second order terms in this expansion are:
\begin{align}
	\label{eq:Kdt}
	\mathcal{K}\drm t &= -i\Big[ \alpha_p e(t-\tau_P) + \alpha_i\mathcal{J}(t) \Big]\drm t F^\times \nn \\
	&= -i\alpha_p[\langle c+c^\dagger\rangle(t-\tau_P)\drm t -\drm W(t-\tau_P)/\sqrt{\eta} 
	- g(t-\tau_P)\drm t]F^\times -i\alpha_i  \mathcal{J}(t)\drm t F^{\times} \\
	\mathcal{K}^2 \drm t^2 &= -\Big[ \alpha_p e(t-\tau_P) + \alpha_i\mathcal{J}(t) \Big]^2 \drm t^2 F^\times F^\times \nn \\
	&= -\frac{\alpha_p^2}{\eta}\drm t	F^\times F^\times + O(\drm W \drm t),
\end{align} 
where to write the second line in each equality we have expanded $e(t)=j(t)-g(t)$, used the definitions $j(t)$ (Eq. \eqref{eq:current}), $\mathcal{J}(t)$ (Eq. \eqref{eq:j_int}), and the Ito rule $\drm W(s)\drm W(t) = \delta(t-s) \drm t$. 

Therefore, discarding all terms less than order $\drm t$, the evolution for the system conditioned on the measurement and subsequently acted upon by the PI feedback control is
\begin{align}
	\label{eq:deri_sme2}
	\rho(t+\drm t)&=\{1 -i\alpha_p[\langle c+c^\dagger\rangle(t-\tau_P)\drm t + \frac{\drm W(t-\tau_P)}{\sqrt{\eta}} - g(t-\tau_P)\drm t]F^\times -i \alpha_i \mathcal{J}(t) F^{\times} \drm t + \frac{\alpha_p^2}{\eta}\mathcal{D}[F]\drm t \} \nonumber\\
				 & \times \{1-iH^\times\drm t+\mathcal{D}[c]\drm t+\sqrt{\eta}\mathcal{H}[c]\drm W(t) \} \rho(t).
\end{align} 
Multiplying this expression out and again discarding all terms smaller than $O(\drm t)$, we find the following evolution for feedback with delay in the P component, $\tau_P>0$, is given by
\begin{align}
	\label{eq:sme_1}
	\drm \rho(t)=& \left\{-i[H,\rho(t)] +\mathcal{D}[c]\rho(t)
										+\frac{\alpha_p^2}{\eta}\mathcal{D}[F]\rho(t)
										-i \Big(\alpha_i \mathcal{J}(t) + \alpha_p e(t-\tau_P)\Big)[F,\rho] \right\}\drm t \nn \\
										&+\sqrt{\eta}\mathcal{H}[c]\rho(t)\drm W(t) 
\end{align}
For the zero time delay case, we go back to Eq. \eqref{eq:deri_sme2}, set $\tau_P=0$ and again multiply the expression out and discard terms smaller than $O(\drm t)$ to get \cite{wiseman_quantum_thesis_1994,wiseman1993quantum}
\begin{align}
	\label{eq:sme_2}
	\drm \rho(t)=& \left\{-i[H,\rho(t)] +\mathcal{D}[c]\rho(t)
									 +\frac{\alpha_p^2}{\eta}\mathcal{D}[F]\rho(t) 
									 -i \Big(\alpha_i \mathcal{J}(t) - \alpha_p g(t)\Big)[F,\rho]
									 -i\alpha_p[F,c\rho(t)+\rho(t)c^\dagger] \right\}\drm t \nn \\
									 & + \mathcal{H}[\sqrt{\eta}c- \frac{i\alpha_p}{\sqrt{\eta}} F]\rho(t) \drm W(t).
\end{align}
Note that in general it is not possible to obtain Eq. \eqref{eq:sme_2} by setting $\tau_P= 0$ in Eq. \eqref{eq:sme_1}. With zero time delay, the correlation between the feedback noise and measurement noise creates an order $\drm t$ term (proportional to $[F, c\rho(t) + \rho(t)c^\dagger]$) that is not present in the presence of time delay.

The two SMEs in Eqs.~\eqref{eq:sme_1} and \eqref{eq:sme_2} represent the evolution of the quantum system conditioned on a continuous measurement record, together with PI feedback based on that record.  Examining the terms proportional to $\alpha_i$, it is evident that the integral feedback component just adds a generator of time-dependent unitary evolution  to the system dynamics. This is in contrast to proportional feedback, which in addition to adding coherent evolution terms, also adds a dissipative evolution term and for $\tau_P =0$ also modifies the stochastic evolution term (the term proportional to $dW$ in Eq. \eqref{eq:deri_sme2}). This reflects the difference that in proportional feedback, the delta-correlated noise is directly fed back at each time instant, whereas in integral feedback, the feedback action is conditioned on a smoothed, tempered signal and thus is able to generate a conventional (time-dependent) Hamiltonian term. Note that while the latter is not necessarily smoothly varying in time, its increments are $O(dt)$.
We emphasize that these SMEs model feedback that requires no state estimation (usually a computationally expensive task), and thus are more suitable for application to experimental implementations. However, P feedback with $\tau_P=0$ will always be an approximation since any measurement and feedback loop will have finite delay. The $\tau_P=0$ limit is a good approximation if the delay is small compared to the intrinsic system evolution time scales. 

In this work, we 
simulate the above stochastic differential equations (SDE) describing evolution under PI feedback with a generalized Euler-Maruyama method. In the usual Euler-Maruyama method \cite{Klo.Pla-1992}, one generates a Wiener noise increment $\drm W(t)$ for each time step $[t, t+\drm t]$ and then updates the state according to the stochastic differential equation. In our generalized Euler-Maruyama method, for each time $t$ we keep a record of the noise up to time $\tau=\mathrm{max}(\tau_I,\tau_P)$ in the past, \ie $\drm W(t)$, $\drm W(t-\drm t)$, ... $\drm W(t-\tau)$). Then $\drm W(t-\tau_P)$ is accessible and $\mathcal{J}(t)$ can be calculated at each time $t$. The state is then updated according to the SME Eq. \eqref{eq:deri_sme2} as usual.
We normalize the density matrix at each time step to compensate for numerical round-off errors.

%%%%%%%%%%%%%%%%%%%%%%%%%%%%%%%%%%%%%%%%%%%%%%%%%%%%%%%%%%%%%%%%%%%%%%%%%%%%%%%%%%%%%%%%

%%%%%%%%%%%%%%%%%%%%%%%%%%%%%%%%%%%%%%%%%%%%%%%%%%%%%%%%%%%%%%%%%%%%%%%%%%%%%%%%%%%%%%

\section{Two-qubit Entanglement Generation}
\label{sec:TwoQubitEntanglementGeneration}
%%%%%%%%%%%%%%%%%%%%%%%%%%%%%%%%%%%%%%%%%%%%%%%%%%%%%%%%%%%%%%%%%%%%%%%%%%%%%%%%%%%%%%%%
In this section, we compare the performance of P feedback, I feedback, and PI feedback for the task of generating an entangled two-qubit state with a local Hamiltonian and non-local measurement.
This non-trivial state generation task was first addressed by measurement-based control with post-selection \cite{PhysRevLett.112.170501, PhysRevA.92.032308, dickel_chip--chip_2018}, then by P feedback and discrete feedback
\cite{martin_deterministic_2015,martin2017optimal} and most recently by PAQS control \cite{zhang_locally_2018}.  For perfect measurement efficiency $\eta =1$, the proportional feedback strategy with time-dependent $\alpha_p(t)$, was shown in Ref. \cite{martin2017optimal} to be globally optimal amongst all protocols that have constant measurement rate.
In this case, the measurement noise can be exactly canceled and the evolution converges deterministically to the target state.  
In the following, we consider the case where the measurement efficiency is not unity and the simplified setting where the feedback coefficients $\alpha_p$ and $\alpha_i$ are assumed to be time-independent. 
In this experimentally relevant setting, P feedback is not known to be globally optimal. Furthermore, the two-qubit system under measurement and feedback is not linear and therefore is representative of a more general class of quantum systems, in contrast to the linear setting of harmonic oscillator stabilization treated in the next section. This non-linearity makes analytical arguments for optimal feedback laws difficult and therefore we must resort to a numerical study. However, we ask the question: whether its advantageous to combine P and I feedback?

Consider two qubits subject to an intrinsic Hamiltonian $H=h_1\sigma_{z1}+h_2\sigma_{z2}$ and subject to negligible decoherence. In the following we will assume $h_1=h_2=h$. We measure the half-parity of the qubits \cite{PhysRevA.92.032308}, which allows a non-local implementation between remote qubits \cite{PhysRevLett.112.170501}. The relevant measurement operator $c$ is
\begin{equation}
	c=\sqrt{k}L_z = \frac{\sqrt{k}}{2}(\sigma_{z1}+\sigma_{z2}),
\end{equation}
where $k$ is the measurement strength and the associated measurement current is 
\begin{equation}
j(t)=2\langle L_z\rangle + \xi(t)/\sqrt{ k \eta}.
\label{eq:current_tq}
\end{equation}

The control goal is to stabilize the system in an entangled state, when starting from a simple product state, $\ket{\uparrow}\otimes \ket{\uparrow} \equiv \ket{\uparrow\uparrow}$. Given the exchange symmetry of the intrinsic Hamiltonian and the measurement operator (we will be careful to also maintain this symmetry with the feedback operator below), and since the initial state is exchange symmetric, we will remain in the symmetric triplet subspace of two qubits throughout the evolution. This subspace is spanned by the states $\ket{T_{-1}}=\Ket{\downarrow\downarrow}$, $\ket{T_0}=\frac{1}{2}(\Ket{\downarrow\uparrow}+\Ket{\uparrow\downarrow})$, and  $\ket{T_1}=\Ket{\uparrow\uparrow}$. Our goal is to evolve to, and stabilize the system in, the entangled state $\ket{T_0}$. 
As in Ref. \cite{martin_deterministic_2015} we use the intuition of rotating the system in the symmetric subspace and  choose a local feedback operator $F=L_{x}=\frac{1}{2}(\sigma_{x1}+\sigma_{x2})$. Applying a $L_x$ rotation can bring $\ket{T_{\pm 1}}$ closer to $\ket{T_0}$.

Since the control goal in this case is to prepare the state $\ket{T_0}$, and the deterministic part of the measurement under this state, $\bra{T_0}L_z\ket{T_0}$ is zero, we may set the goal to be $g(t)=\bra{T_0}L_z\ket{T_0}=0 ~~ \forall t$. Hence our error signal is $e(t)=j(t)$. Thus, we obtain the stochastic master equation that describes the evolution of the two-qubit system for both $\tau_{P}=0$ (Eq. \ref{eq:two_qubits}) and $\tau_{P}>0$ (Eq. \ref{eq:two_qubits_tau}) as shown in Appendix.~\ref{sec:app_TwoQubit}. We employ an exponential filter for the integral feedback: 
\begin{equation}
\mathcal{J}(t)=\frac{1}{\tau_I}\int_{t-\tau_I}^t j(s)\exp\left(-\frac{(t-s)}{\tau_I}\right)\text{d}s.
\end{equation}

\begin{figure}[t!]
\label{fig:twoqubits}
    \centering
    \subfigure[~P feedback, $\tau_P=0$]{
    \includegraphics[width=0.22\columnwidth]{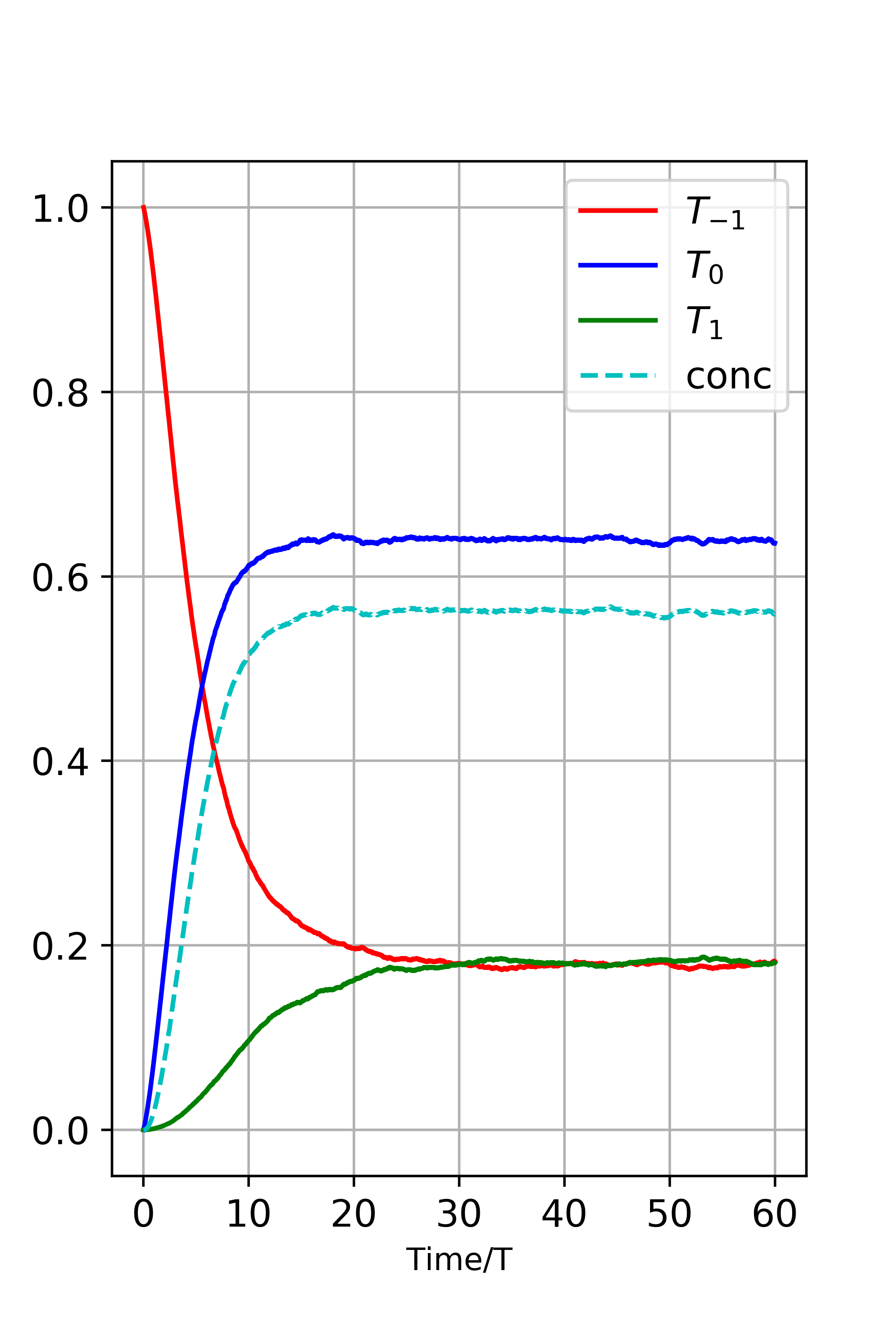}
    \label{fig:pro_nodelay}
    }
    \subfigure[~P feedback, $\tau_P>0$]{
    \includegraphics[width=0.22\columnwidth]{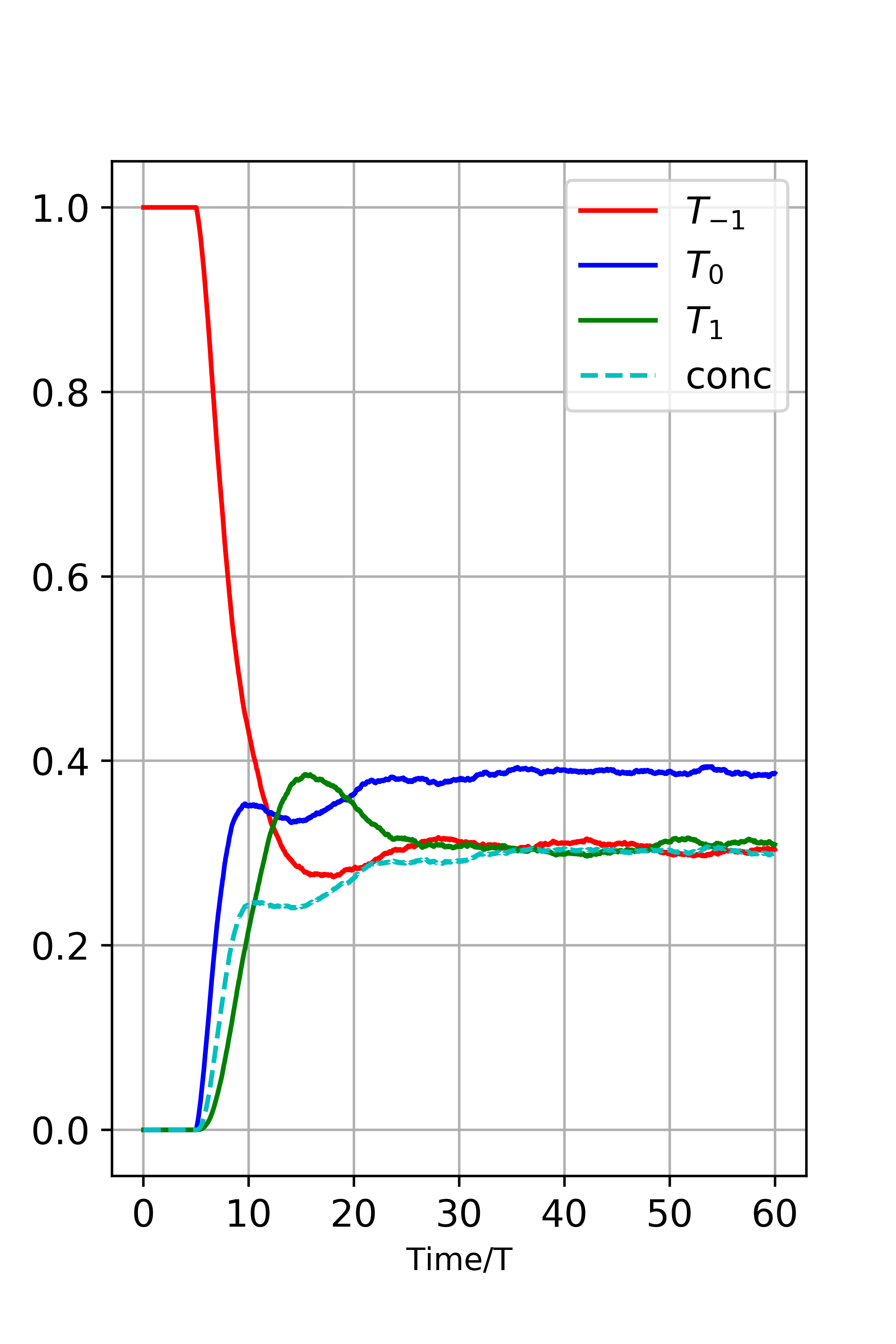}
    \label{fig:pro_delay}
    }
    \subfigure[~I feedback]{
    \includegraphics[width=0.22\columnwidth]{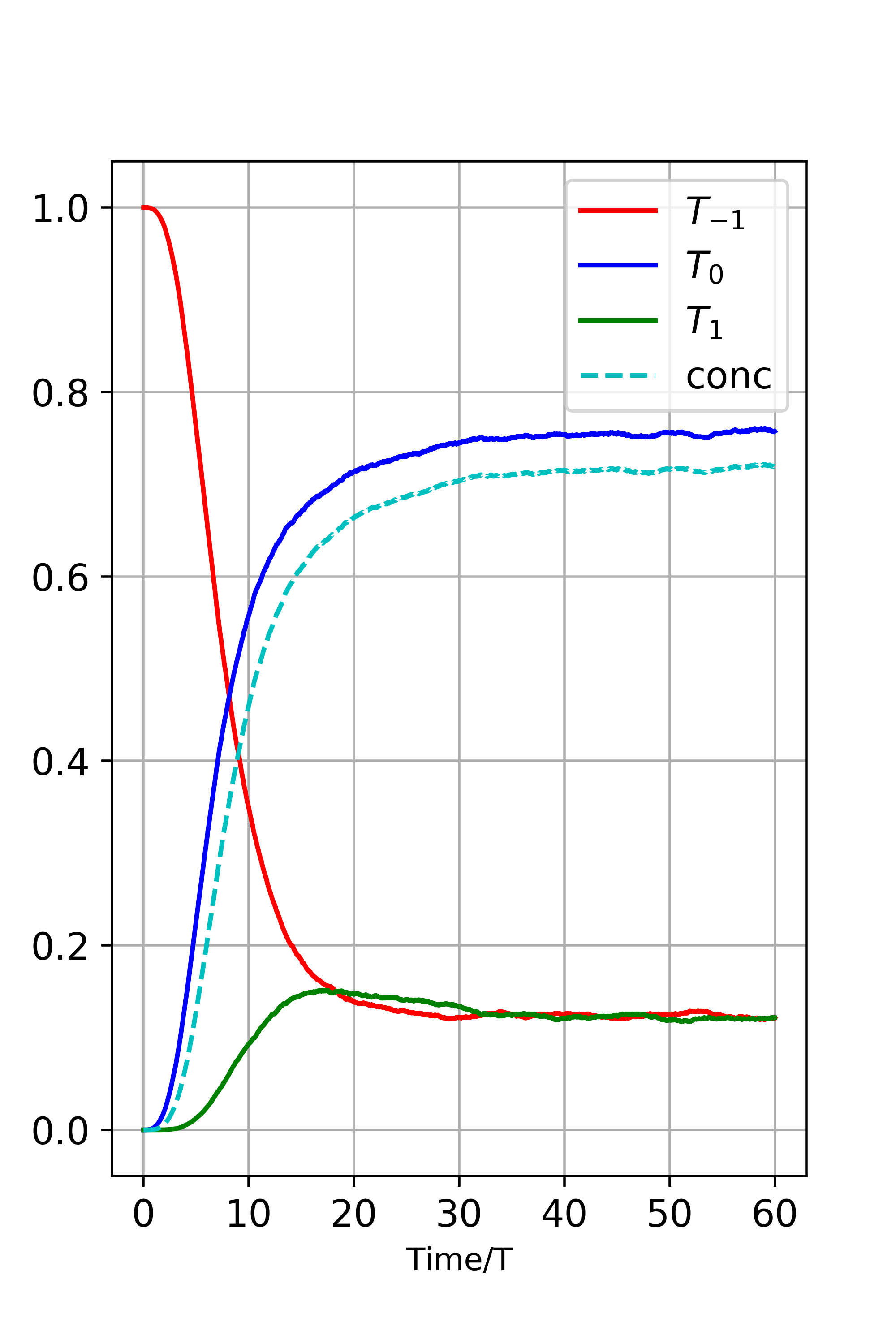}
    \label{fig:int}
    }
    \subfigure[~PI feedback]{
    \includegraphics[width=0.22\columnwidth]{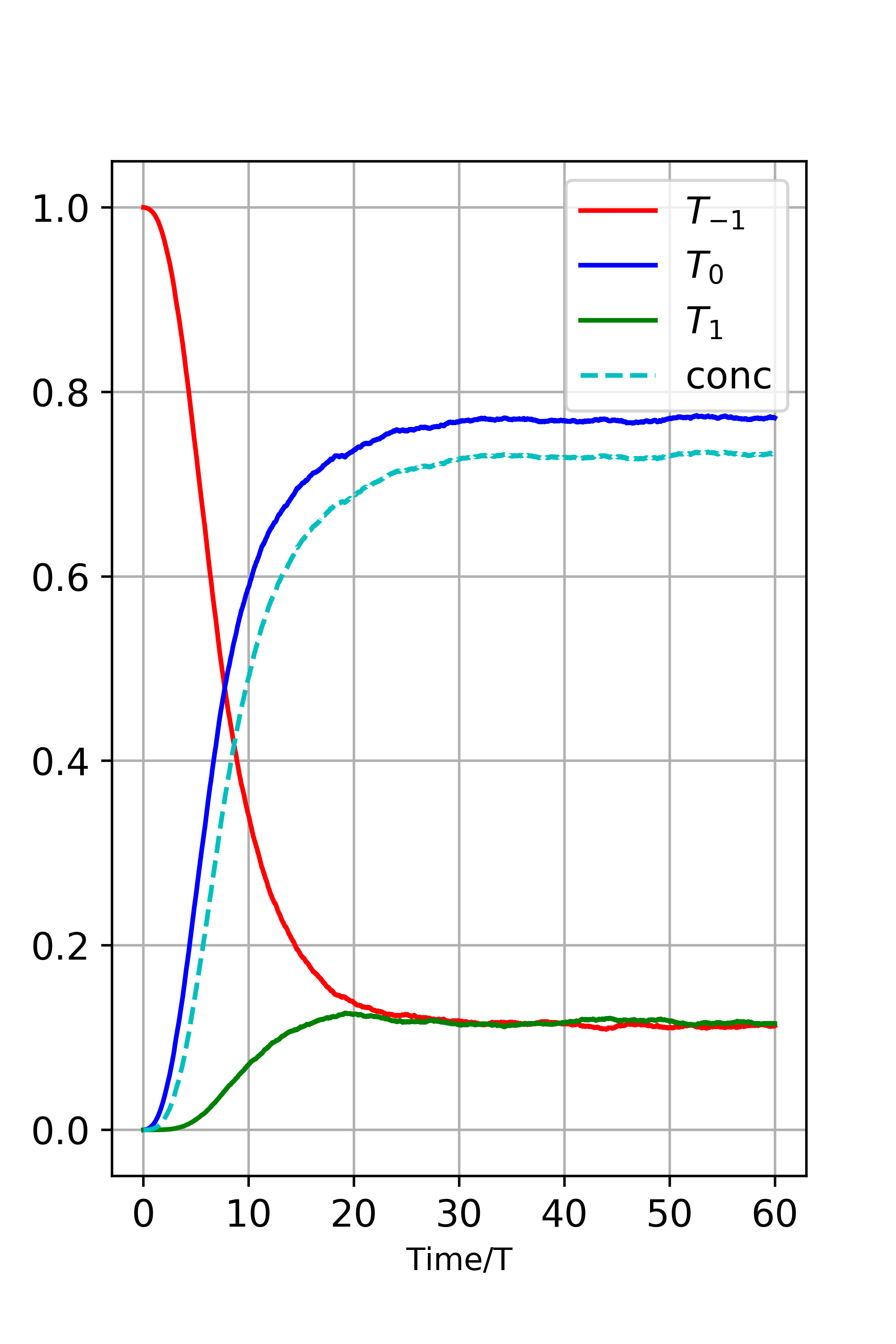}
    \label{fig:intPro}
    }
\caption{Average evolution of a two-qubit system under feedback control with Hamiltonian $h=0.1$. (a) shows the proportional feedback with $\alpha_p=0.2$ and $\tau_P=0$; (b) proportional feedback with $\alpha_p=0.2$ and $\tau_P=5$; (c) integral feedback with $\alpha_i=0.2$ and $\tau_I=3$ ;  (d)  combination of proportional feedback with $\alpha_p=0.03$ and integral feedback with $\alpha_i=0.17$, $\tau_I=3$. (a) is calculated by dropping the stochastic terms in  Eq. \eqref{eq:two_qubits}. (b) is calculated by averaging over 8000 trajectories simulated with Eq. \eqref{eq:two_qubits_tau}. (c) and  (d) are calculated by averaging over 8000 trajectories simulated with Eq. \eqref{eq:two_qubits}. For all plots the measurement efficiency is $\eta=0.4$ and the initial state is taken to be the unentangled state $T_1$.  The long time value of concurrence in (c) is $\sim 0.7196\pm0.0028$ and in (d) is $\sim 0.7289\pm0.0028$.}
\label{fig:tq_evo_ave}
\end{figure}

To assess the relative performance of the feedback strategies, we will look at the steady state average populations of the three triplet states as well as the average concurrence measure of entanglement. Given a two-qubit density operator $\rho$, the populations of the triplet states are given by $T_i=\Braket{T_i|\rho|T_i},\ i=-1,\ 0,\ 1$, and the concurrence is defined as \cite{PhysRevLett.78.5022,PhysRevLett.80.2245}
\begin{equation}
\mathcal{C}(\rho)\equiv\max(0,\lambda_1-\lambda_2-\lambda_3-\lambda_4),
\end{equation}
where $\lambda_1$, $\cdots$, $\lambda_4$ are the (non-negative) eigenvalues, in decreasing order, of the Hermitian matrix
$	R = \sqrt{\sqrt{\rho}\tilde{\rho}\sqrt{\rho}}
$ with
$	\tilde{\rho} = (\sigma_{y}\otimes\sigma_{y})\rho^{*}(\sigma_{y}\otimes\sigma_{y}), 
$ the spin flipped state of $\rho$.

Fig. \ref{fig:tq_evo_ave} shows these measures of the average evolution of the two-qubit system 
for the initial state $\ket{T_{1}}$, under the strategies of P feedback ($\alpha_p=1,\alpha_i=0$, panels (a) and (b)), I feedback ($\alpha_p=0,\alpha_i=1$, panel (c)), and PI feedback with a specific combination of $\alpha_p$ and $\alpha_i$ (panel (d)). The parameters of the system are $h=0.1, k=1, \eta=0.4$.  The choice of $k$ sets the units for the other rates in the model, $\eta$ was chosen to be consistent with current experimental capabilities \cite{PhysRevLett.112.170501}, and we vary $h$ later to see its effect on the conclusions drawn. The results in this section are for the $\ket{T_{1}}$ initial state. We have also simulated the protocols and their steady states starting from any mixture of product states in the triplet manifold (the initial states simplest to prepare in experiments) and the results are similar to those shown here for the $\ket{T_{1}}$ initial state.

Figs. \ref{fig:pro_nodelay} and \ref{fig:pro_delay} show the evolution under P feedback, with and without a time delay. We expect that there is little benefit in introducing a time delay in proportional feedback in this example, since there is no information in prior measurement currents that is germane to the control goal. Indeed this expectation is borne out by these figures; the performance of the time-delayed feedback is worse than without a time delay, $\tau_P=0$. Fig. \ref{fig:int} shows the performance under I feedback. The value of the integration time $\tau_I$ can be numerically optimized to yield maximum concurrence.  The plot in Fig. \ref{fig:int} uses $\tau_I = 3$, which is a near-optimal value for concurrence. 

Comparing Fig. \ref{fig:int} with Figs. \ref{fig:pro_nodelay} and \ref{fig:pro_delay}, it is evident that in the case of inefficient measurements, $\eta < 1$, an I feedback strategy is able to produce a significantly higher steady state average concurrence and target $T_0$ population than a P feedback strategy. Finally, in Fig. \ref{fig:intPro} we show the average behavior for a specific combination of P and I feedback, \ie of PI feedback, with $\alpha_p=0.03$ and $\alpha_i=0.17$. This combined PI feedback strategy performs slightly better than the pure I feedback strategy, thus outperforming both P and I strategies (the long time value of concurrence in Fig. \ref{fig:int} is $\sim 0.7196\pm0.0028$ and in Fig. \ref{fig:intPro} is $\sim 0.7289\pm0.0028$). We have plotted here the results of just one choice of $\alpha_p$ and $\alpha_i$ that combines P and I feedback. This particular choice was made to show that PI feedback can outperform P and I feedback based on a more general analysis of mixing the two types of feedback that we will detail below.
Note that the total feedback strength has been kept constant across all the settings shown in Fig. \ref{fig:tq_evo_ave}, specifically at $\alpha_i + \alpha_p=0.2$, in order to have a fair comparison. We also emphasize that these plots show average values of the state populations and concurrence, where the averages are computed over 8000 evolution trajectories.  For efficiency $\eta=0.4$, since none of these protocols achieves cancellation of the measurement noise, the individual trajectories of triplet state populations and concurrence show fluctuations for all four feedback strategies.

\begin{figure}[t!]
\centering
\subfigure[~Proportional feedback ]{
\includegraphics[width=0.25\columnwidth]{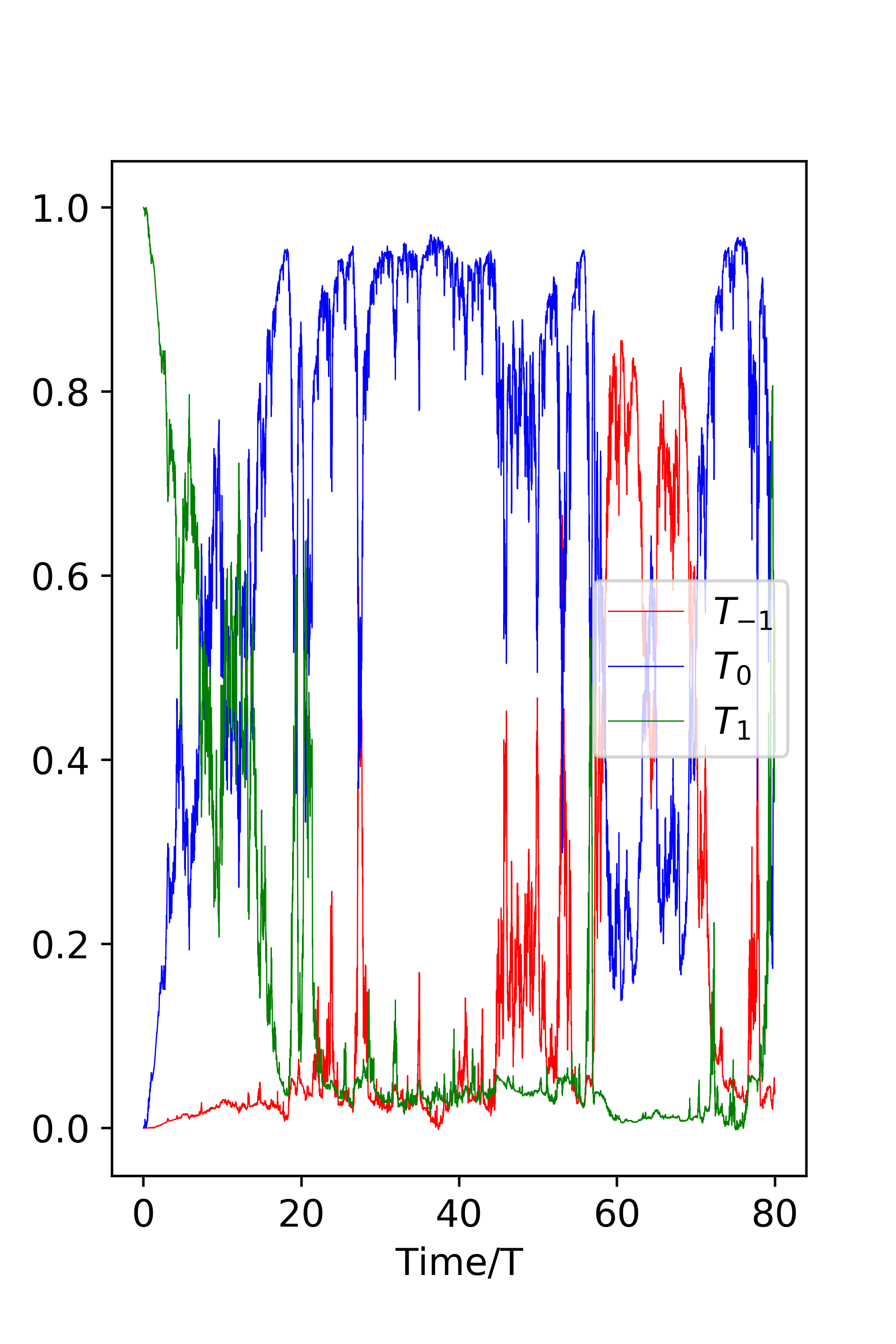}
\label{fig:single_proT0}}
\subfigure[~Integral feedback]{
\includegraphics[width=0.25\columnwidth]{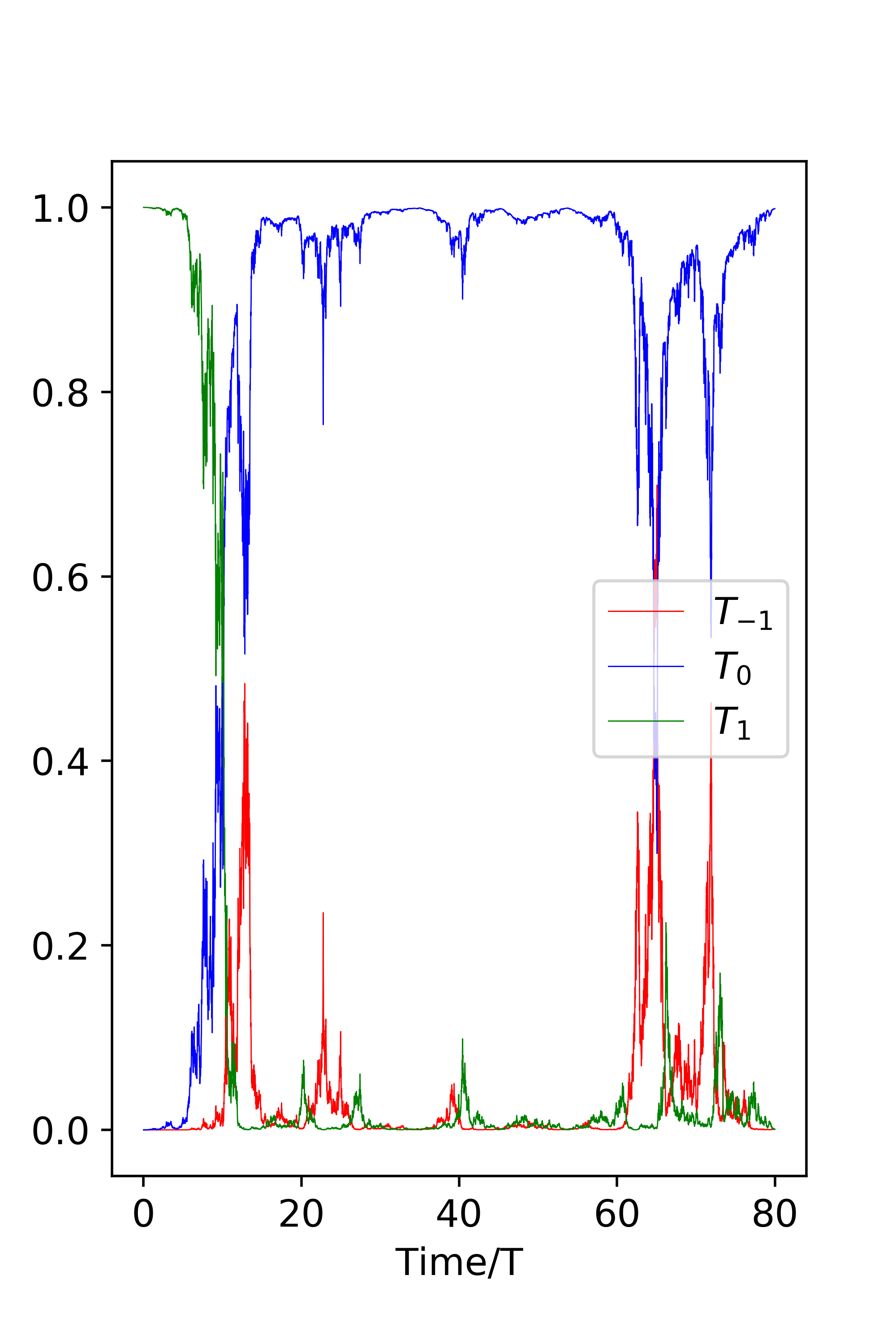}
\label{fig:single_intT0}}
\caption{Single trajectories of triplet state populations for proportional feedback ($\tau_P=0$) and integral feedback $(\tau_I=3)$. The measurement efficiency $\eta=0.4$ and initial state is 
an unentangled state $T_1$.}
\label{fig:Single_trajectory}
\end{figure}

Analysis of single trajectories reveals insight into the better performance of the I feedback strategy relative to the P feedback strategies.  Representative trajectories of the triplet state populations under I feedback and P feedback with zero time delay are shown in Fig. \ref{fig:Single_trajectory}. In general, the evolution under both feedback strategies drives the system towards the $\ket{T_0}$ state. The population $T_0$ can reach the value $1$ and remain there for some time period until a measurement noise fluctuation entering through the feedback term 
is large enough to drive it down. Under P feedback, we are conditioning feedback on the raw measurement and thus the $T_0$ population fluctuations can be large, which results in more frequent transitions out of the target state $\ket{T_0}$. In contrast, the integral component in the I feedback strategy smooths out the measurement current fluctuations, which reduces the probability of the feedback term kicking the system out of the target $\ket{T_0}$ state.
Consequently, as we analyze in detail below the ensemble average of the triplet population, $\mathbb{E}[T_0]$, will be larger for the integral control strategy than for the proportional control strategy. 

To understand this in more quantitative terms, we have given the evolution of these triplet populations and the associated off-diagonal elements of the density matrix in the triplet subspace under general PI feedback in Appendix \ref{sec:app_TwoQubit}. For the case of I feedback, \ie $\alpha_p=0$, $\alpha_i>0$, the evolution is:
\begin{equation}
\begin{aligned}
\drm T_{-1}&=\sqrt{2}\alpha_i \mathcal{J}(t) \text{Im}T_{0,-1}\drm t
-2\sqrt{\eta k}(1+\langle{L_z\rangle(t)})T_{-1}\drm W(t),\\
\drm T_0&=-\sqrt{2}\alpha_i \mathcal{J}(t)(\text{Im}T_{0,1}+\text{Im}T_{0,-1})]\text{d}t
-2\sqrt{\eta k}\langle L_z\rangle(t) T_0 \text{d}W(t),\\
\drm T_1&=\sqrt{2}\alpha_i \mathcal{J}(t)\text{Im}T_{0,1}\text{d}t
+2\sqrt{\eta k}(1-\langle{L_z\rangle}(t))T_{1} \text{d}W(t),\\
\text{d}T_{1,-1}&=\Big\{2[i2h-k]T_{1,-1}+\frac{i\alpha_i }{\sqrt{2}}\mathcal{J}(t)(T_{1,0}-T_{0,-1})\Big\}\text{d}t-2\sqrt{\eta k}\langle L_z\rangle(t) T_{1,-1}\text{d}W(t),\\
\text{d}T_{0,1}&=\Big\{-[i2h+\frac{k}{2}]T_{0,1} - \frac{i\alpha_i }{\sqrt{2}}\mathcal{J}(t)(T_{-1}-T_0+T_{-1,1})\Big\}\text{d}t
+\sqrt{\eta k}(1-2\langle L_z\rangle(t) )T_{0,1}\text{d}W(t),\\
\text{d}T_{0,-1}&=\Big\{[i2h-\frac{k}{2}]T_{0,-1}-\frac{i\alpha_i}{\sqrt{2}} \mathcal{J}(t)(T_0-T_{-1}-T_{1,-1})\Big\}\text{d}t-\sqrt{\eta k}(1+2\langle L_z\rangle(t) )T_{0,-1}\text{d}W(t).
\end{aligned}
\label{eq:intTwoQubit_trip}
\end{equation}
We suppress the time index of $T_i$ and $T_{i,j}$ here for notational conciseness. We cannot take the ensemble average (to obtain the average evolution) by simply dropping the stochastic terms in this case, because $\mathcal{J}(t)$ and $T_i$ and $T_{i,j}$ are correlated by virtue of the dependence of both on past Wiener increments. Moreover, due to their nonlinearity we cannot solve these equations directly.  However, we can use the following argument to show that Eq. \eqref{eq:intTwoQubit_trip} has a (unstable) steady state when $T_0=1$. Suppose at some time, $T_0$ reaches $1$ and we have $T_1 = T_{-1}=0$ (and thus $\expect{L_z}=0$). Then the coherences $T_{1,-1}, T_{0,1}, T_{0,-1}$ will be approximately zero also (since all populations other than $T_0$ are zero). As a result, in the above equations $\drm T_{-1}=\drm T_{0}=\drm T_{1}=\drm T_{1,-1} \approx 0$, and the only coherences that evolve are given by $\drm T_{0,1}=-\drm T_{0,-1}=i\frac{\alpha_i}{\sqrt{2}}\mathcal{J}(t)\text{d}t$. These coherences are generated by a non-zero $\mathcal{J}(t)$, and then go on to generate non-zero populations in the undesired states $T_{-1}$ and $T_1$. This perturbation away from the desired state is weak because of two factors: (i) $\mathcal{J}(t)$ can be made small when $T_0=1$, since the deterministic position of $j(t)$ is zero, and the averaging integral will dampen the fluctuations $\drm W(t)$ over the period $\tau_I$, and (ii) the coherences are dampened at a rate $k/2$, and therefore even when coherences are generated by non-zero $\mathcal{J}(t)$, they can be quickly dampened by the measurement induced dephasing before they generate non-zero populations in the undesired states.

\begin{figure}[t!]
\subfigure[]{
	\includegraphics[width=0.3\columnwidth]{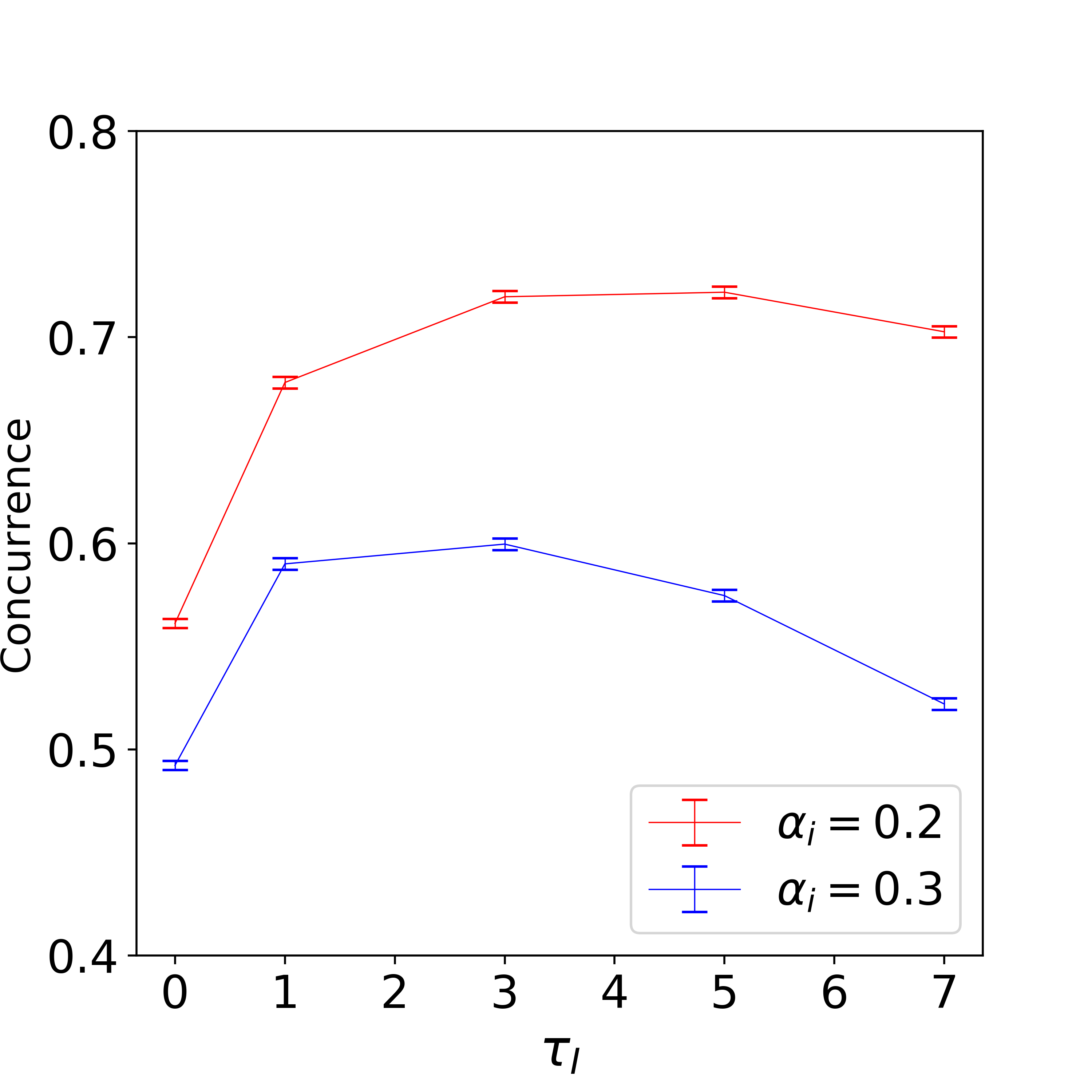}
	\label{fig:tq_conc_tau}}
\subfigure[]{
	\includegraphics[width=0.3\columnwidth]{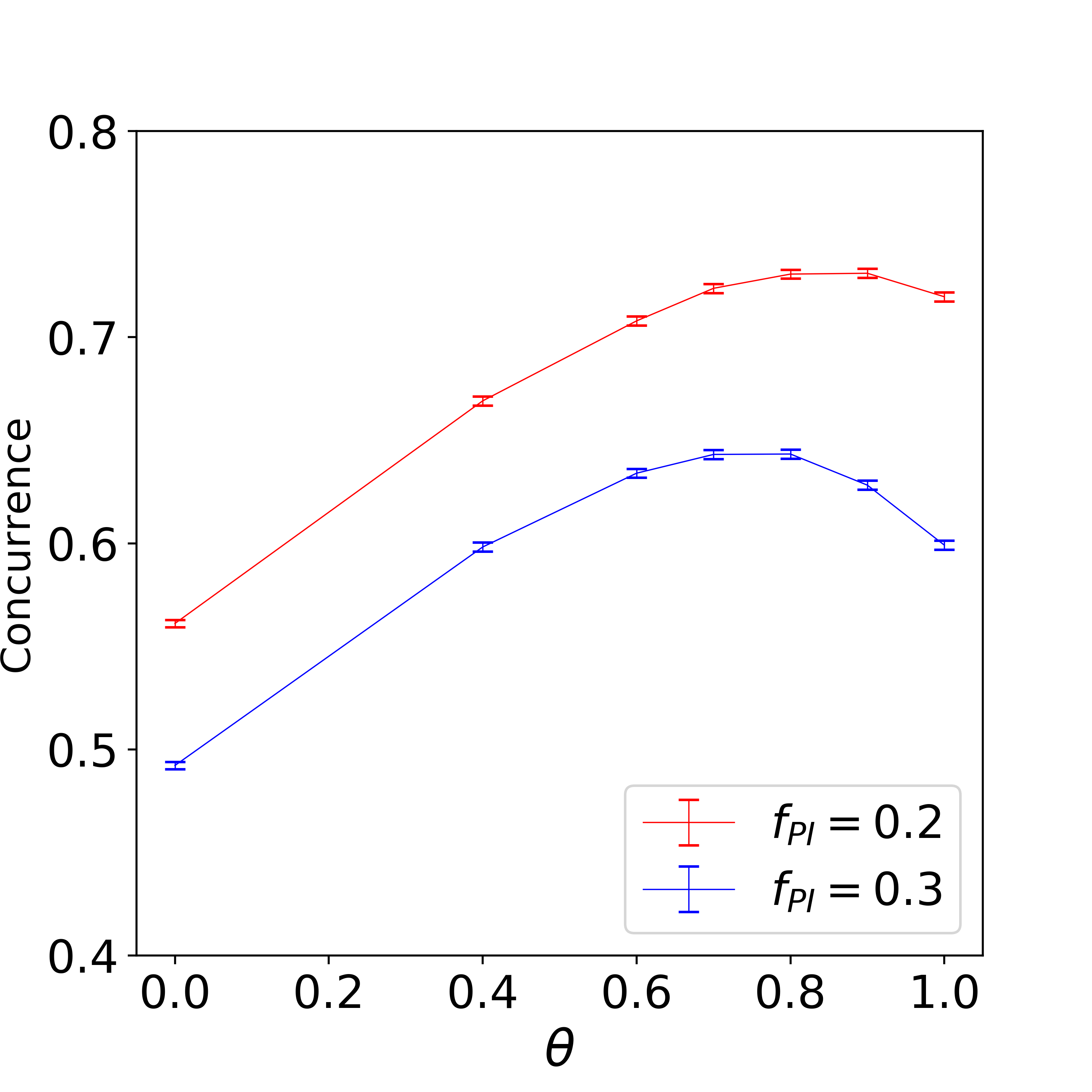}
	\label{fig:tq_theta_conc}}\\
\subfigure[]{
	\includegraphics[width=0.3\columnwidth]{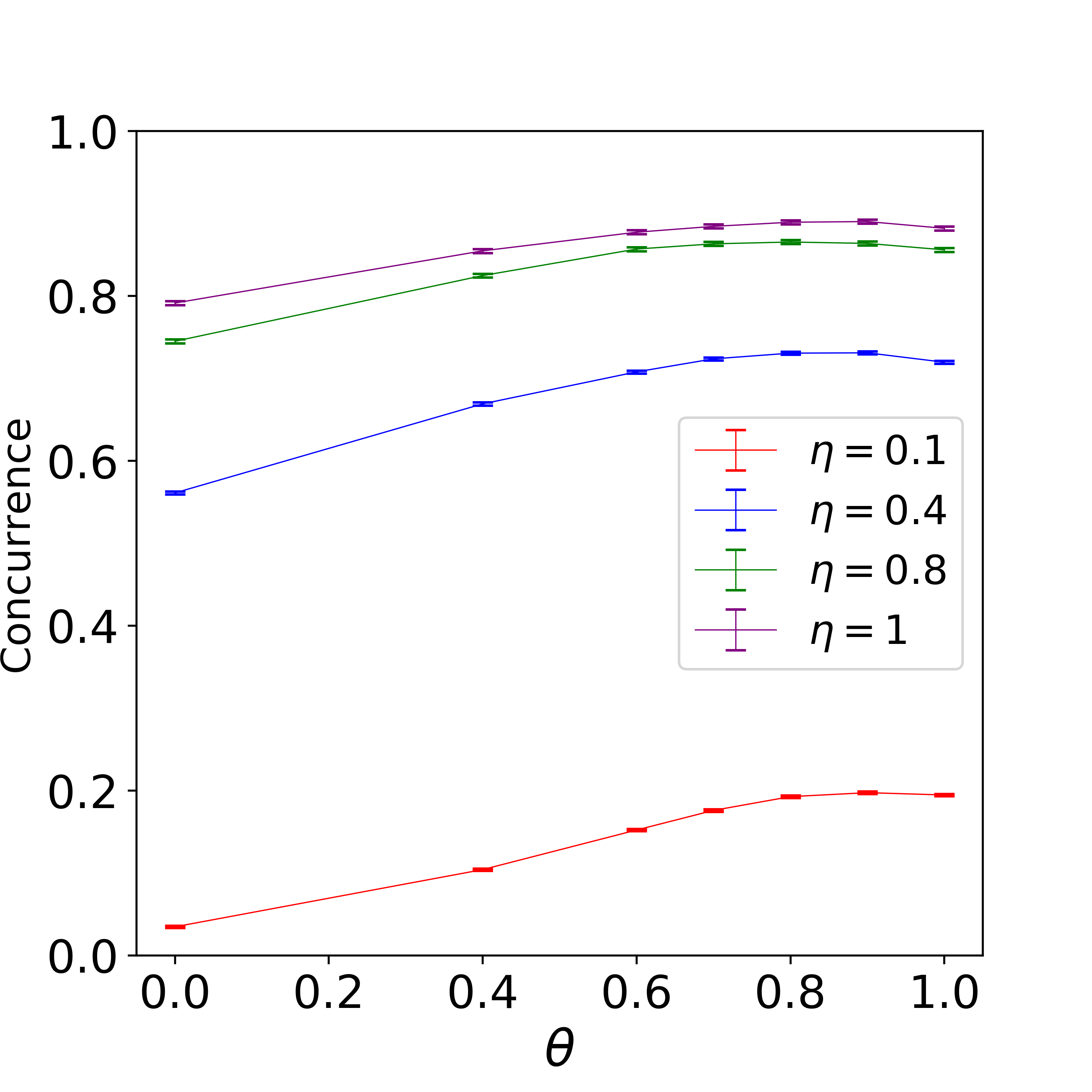}
	\label{fig:tq_theta_eta_conc}}
\subfigure[]{
	\includegraphics[width=0.3\columnwidth]{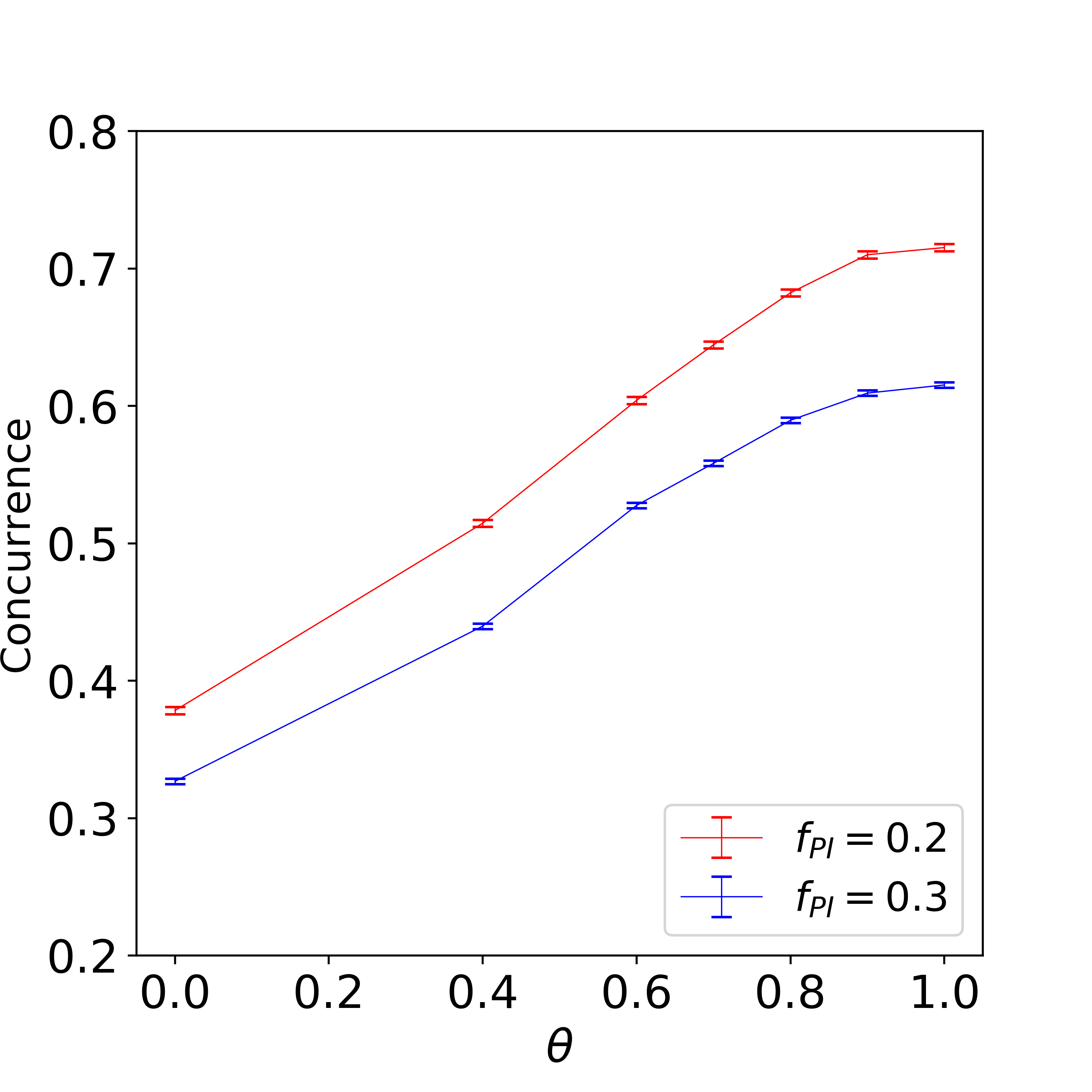}
	\label{fig:h0.5}}
\caption{Steady state average concurrence dependence on integration time and relative weight of P and I control in PI control, measured by the mixing ratio $\theta$, Eq.~\eqref{eq:mixing}.
For all calculations shown here, the initial state is taken to be the unentangled state $T_1$ and all results are averaged over 8000 trajectories.
(a) Steady state average concurrence vs. integration time $\tau_I$ for pure integral (I) control, for intrinsic Hamiltonian parameters $h_1=h_2=h=0.1$, measurement efficiency $\eta=0.4$ and two different integral feedback coefficient values $\alpha_i$. 
(b) Steady state average concurrence vs. mixing ratio $\theta$ for intrinsic Hamiltonian parameters $h_1=h_2=h=0.1$,  measurement efficiency $\eta=0.4$ and PI feedback with different values of total feedback strength $f_{PI}$. For $f_{PI}=0.3$ (blue line), the integration time for the integral component is  $\tau_I=1$, for $f_{PI}=0.2$, (red line) $\tau_I=3 $. 
(c) Steady state average concurrence vs. the mixing ratio $\theta$ for intrinsic Hamiltonian parameters $h_1=h_2=h=0.1$, and total feedback strength $f_{PI}=0.2$ with integration time $\tau_I=3$, under various values of measurement efficiency. 
(d) Steady state average concurrence vs. mixing ratio $\theta$ for  intrinsic Hamiltonian parameters $h_1=h_2=h=0.5$, and measurement efficiency $\eta=0.4$ under PI feedback. For $f_{PI}=0.3$ (blue line) the integration time parameter is $\tau_I=1$ and for $f_{PI}=0.2$ (red line), $\tau_I=3$. The key difference with panel (b) is that here $h=0.5$, and in this case I feedback is superior to any mixture of P and I feedback. }
\label{fig:tq_ss_conc}
\end{figure}

It is clear that the integration time $\tau_I$ is an important parameter for the integral control strategy. 
Optimization of this parameter involves a tradeoff between smoothing and time delay in the feedback action as $\tau_I$ increases. Specifically, we can expect that a longer integration time $\tau_I$ will improve the concurrence, due to the reduced fluctuations, but because the signal is being averaged over a longer time window, it will take longer for deviations away from the target value to affect the averaged value, resulting in a time delay in the feedback action. To illustrate the resulting trade-off between short and long integration time choices, Fig. \ref{fig:tq_conc_tau} plots the steady state average concurrence as a function of the filter integration time $\tau_I$ for I feedback. Note that the $\tau_I=0$ reference value refers to the proportional feedback strategy with no delay. The generic behavior shown here is found for any value of the feedback strength $\alpha_i$, i.e, for all $\alpha_i$ values we see that the concurrence shows a maximum value at a non-zero optimal filter integration time. This optimal value of $\tau_I$ decreases as the control parameter $\alpha_i $ increases (not shown).  
We also find that the system takes increasingly longer times to reach steady state as the feedback strength $\alpha_i$ goes to zero, or
as $\tau_I $ gets larger.  

Finally, we explore in more detail the possibility of full PI feedback, \ie  combining proportional and integral feedback for the problem of entangled state generation with inefficient measurements in this two qubit system. In Fig. \ref{fig:intPro} we already showed that there was a small benefit to combining both strategies for a particular set of coefficients. To study the performance of the combined strategy more systematically, we write 
\begin{align}
    \alpha_p=(1-\theta) f_{PI}, \,\,\, \alpha_i=\theta f_{PI}
\label{eq:mixing}
\end{align}
where $f_{PI}$ is the total feedback strength and $\theta\in[0,1]$ is a mixing ratio quantifying the combination of the two strategies. In Fig. \ref{fig:tq_theta_conc} we now plot the steady state average concurrence versus this strategy mixing ratio $\theta$ for PI feedback, while keeping the total feedback strength $f_{PI}$ constant. The plot shows the existence of an optimal mixing ratio $\theta_o$ located between $\sim 0.7$ and $\sim 0.9$, \ie  the optimal strategy is to have mostly integral control with some admixture of proportional control. The precise value of this optimal mixing ratio depends on the total feedback strength $f_{PI}$. However, as shown in Fig. \ref{fig:tq_theta_eta_conc}, $\theta_o$ is quite robust to variations in efficiency. Note that the maximum concurrence obtained by this PI feedback strategy for perfect efficiency, $\eta=1$, is less than that obtained using the globally optimal P feedback strategy with time-dependent proportionality constant $\alpha_p(t)$ ~\cite{martin_deterministic_2015, martin2017optimal}. 

These results show that the advantage of PI control relative to pure I or pure P control increases as the total feedback strength parameter $f_{PI}$ increases. This can be seen by comparing the difference in steady state average concurrence between P, I and PI with optimal $\theta_o$ for $f_{PI}=0.2$ (red line) and $f_{PI}=0.3$ (blue line) in Fig.~\ref{fig:tq_theta_conc}. Finally, we note that the optimal mixing ratio also depends on the system Hamiltonian, in particular, the value of $h$. In this case, for larger values of $h$, the optimal mixing parameter $\theta_o\rightarrow 1$ and the optimal feedback strategy becomes just I feedback. We show the concurrence versus $\theta$ curves for $h=0.5$ in Fig. \ref{fig:h0.5}, for comparison with Fig.~\ref{fig:tq_theta_conc}.

\section{Harmonic Oscillator State Stabilization}\label{sec:HO}
%%%%%%%%%%%%%%%%%%%%%%%%%%%%%%%%%%%%%%%%%%%%%%%%%%%%%%%%%%%%%%%%%%%%%%%%%%%%%%%%%%%%%%

State stabilization of a quantum harmonic oscillator is a canonical quantum feedback control problem that has been studied for several decades \cite{PhysRevA.60.2700, PhysRevB.68.235328, habib_quantum_2002, doherty2012quantum, hamerly_coherent_2013}. This problem has many practical applications, including the cooling and manipulation of trapped cold ions~\cite{jordan2019near} or atoms~\cite{meng2018near}, and cooling of nanoscale~\cite{qiu2020laser} or even macroscopic~\cite{rossi2018measurement,abbott2009observation} mechanical systems. Purely proportional feedback control schemes have been developed for this problem \cite{PhysRevA.60.2700, PhysRevB.68.235328, habib_quantum_2002, doherty2012quantum}.  In the following, we investigate whether adding integral control adds any benefit in terms of control accuracy.

The system is a quantum harmonic oscillator with mass $m$ and angular frequency $\omega$.
We apply a continuous measurement of the oscillator position $x$ with strength $k$ (\ie $c=\sqrt{k}x$ in the notation of the Sec. \ref{sec:formalism}) and efficiency $\eta$.  The SME describing the system under measurement is \cite{doherty2012quantum}
\begin{equation}
 \mathrm{d}\rho=-i[H_0,\rho]\mathrm{d}t+2\gamma(N+1)\mathcal{D}[a]\rho\mathrm{d}t+2\gamma N\mathcal{D}[a^\dagger]\rho \mathrm{d}t
 +k\mathcal{D}[x]\rho\mathrm{d}t +\sqrt{\eta k}\mathcal{H}[x]\rho\mathrm{d}W, 
\end{equation}
 where  $H_0=p^2/(2m)+m\omega^2 x^2/2$, $p$ is the oscillator momentum operator and $a$ is the annihilation operator. The terms proportional to $\gamma$ describe damping and excitation due to coupling to a bosonic thermal bath with mean occupation $N$. The associated measurement signal is
\begin{equation}
	j(t)=2\langle x\rangle(t) +\xi(t)/\sqrt{k\eta}.
\end{equation}
We shall consider two types of feedback for this system. First, we consider linear feedback in both $x$ and $p$, in which case we have two feedback operators:
\begin{equation}
	F_{1}=x, \quad\quad\quad
	F_{2}=p.
\end{equation}
We will attach (time-dependent) proportional coefficients ($\alpha_{p1}, \alpha_{p2}$) and integral coefficients ($\alpha_{i1}, \alpha_{i2}$) to each of these feedback operators. The total feedback operator is then
\begin{align}
 [\alpha_{p1}(t)e(t-\tau_P)+\alpha_{i1}(t)\mathcal{J}(t)]x+[\alpha_{p2}(t)e(t-\tau_P)+\alpha_{i2}(t)\mathcal{J}(t)]p.  \nn
\end{align}

Applying $F_1$ is usually considerably easier than $F_2$, since the former corresponds to applying a force on the oscillator. Therefore, we will also consider the setting where only $F_1$ is available, in which case we have only the coefficients $\alpha_{p1}, \alpha_{i1}$. 
Given the simplicity of the harmonic system, it is possible to set up analytic candidate control laws that are specified in terms of choices for the coefficients $\alpha_{p1},\alpha_{p2},\alpha_{i1},\alpha_{i2}$, and to then assess whether they are consistent with P, I or PI feedback. The SMEs of harmonic oscillator for both cases ($\tau_P>0$ and $\tau_P=0$) are given by Eq. ~\ref{eq:harmonic_sme_1} and Eq. ~\ref{eq:harmonic_sme_2} in Appendix~\ref{sec:SMEHO}.

In the simplest setting where the system starts in a Gaussian state, the state remains Gaussian when evolved according to the above measurement and feedback dynamics since all operators acting on the density matrix are linear or quadratic in $x,p$ \cite{PhysRevA.60.2700, doherty2012quantum}. A Gaussian state is completely determined by its first moments ($\Braket{x}$,$\Braket{p}$) and second moments ($V_x\equiv\Braket{(x-\Braket{x})^2}$, $V_p\equiv\Braket{(p-\Braket{p})^2}$, $C_{xp}\equiv \frac{1}{2}\Braket{xp+px}-\Braket{x}\Braket{p}$). The evolution of the second moments under the above measurement, thermal damping, and feedback is independent of the feedback, and evolve deterministically, independent of the measurement noise, $\xi(t)$~\cite{PhysRevA.60.2700}. The equations of motion for the second moments are given by Eq. \ref{eq:second_order_gamma} in Appendix~\ref{sec:firstsecondmoment}. We will assume in the following that these equations are solved in advance and therefore that $V_x(t), V_p(t)$ and $C_{xp}(t)$ are known functions of time. In all of the examples treated in this section, we shall take the initial state to be a coherent state with $V_x(0)=V_p(0)=0.5$ and $C_{xp}(0)=0$.  

Under the measurement and feedback dynamics described in Eqs. \eqref{eq:harmonic_sme_1} and \eqref{eq:harmonic_sme_2}, the evolution of the first moments are given by
$\text{tr}[x\drm \rho(t)]$ and $\text{tr}[p\drm \rho(t)] $: 
\begin{subequations}
\label{eq:lab_dyn}
\begin{align}
%\begin{aligned}
	\mathrm{d}\langle x\rangle(t) &= \frac{1}{m}\langle p \rangle(t)\mathrm{d}t -\gamma\langle x\rangle(t)\drm t + \alpha_{i2}\mathcal{J}(t)\mathrm{d}t
	+ \alpha_{p2}\left(2\langle x\rangle (t-\tau_P)-g(t-\tau_P)\right) \mathrm{d}t \nonumber\\
	& +\sqrt{\eta k}(2V_x(t)\mathrm{d}W(t) + \frac{\alpha_{p2}}{k\eta}\mathrm{d}W(t-\tau_P)),\\
	 \mathrm{d}\langle p\rangle(t) &= -m\omega^2\langle x\rangle(t) \mathrm{d}t-\gamma\langle p\rangle(t)\drm t -\alpha_{i1}\mathcal{J}(t)\mathrm{d}t -\alpha_{p1}\left(2\langle x\rangle (t-\tau_P)-g(t-\tau_P)\right)\mathrm{d}t \nonumber\\
	 & +\sqrt{\eta k}(2C_{xp}(t)\mathrm{d}W(t)-\frac{\alpha_{p1}}{k\eta} \mathrm{d}W(t-\tau_P)),
%\end{aligned}
\end{align}
\end{subequations}
where $\tau_P\geq 0$. In the limit of zero time delay, the equations of motion for the first moments are the same as above, with $\tau_P=0$ (this reduction for the evolution of the first moments of the quadratures is a special case since as noted above, taking  $\tau_P=0$ in Eq. \eqref{eq:sme_1} does not yield Eq. \eqref{eq:sme_2}).

Our overall control goal is state stabilization, where the aim is to center the state at an arbitrary stationary (time-independent) value of the two quadrature means in the rotating frame of the oscillator, notated $(X_g, P_g)$. In the laboratory frame this control goal is specified by the mean quadrature values ($x_g(t)$, $p_g(t)$), which are related to ($X_g$, $P_g$) by the transformation 
\begin{subequations}
\label{eq:ho_tran}
\begin{align}
	x_g(t) &=X_g\cos(\omega t)+P_g\sin(\omega t)/(m\omega), \\
	p_g(t) &=-m\omega X_g \sin(\omega t) +P_g\cos(\omega t).
\end{align}
\end{subequations} 
We note that the oscillator cooling problem \cite{PhysRevA.60.2700} can be viewed as a special case of this state stabilization with the control goal ($X_g=0$, $P_g=0$). 

The evolution of the first order moments in the rotating frame is given by
\begin{subequations}
\label{eq:evo_XP}
\begin{align}
    \drm \langle X\rangle(t)=&\left(\drm\langle x\rangle(t)-\frac{1}{m}\langle p\rangle(t)\drm t\right)\cos(\omega t)
    -\left(m\omega^2 \langle x\rangle(t)+\drm \langle p\rangle(t)\right)\sin(\omega t)/m\omega\\
    \drm \langle P\rangle(t)=&m\omega\left(\drm \langle x\rangle(t)-\frac{1}{m}\langle p\rangle(t)\drm t\right)\sin(\omega t)
    +\left(m\omega^2\langle x\rangle(t)+\drm \langle p\rangle (t)\right)\cos(\omega t).
\end{align}    
\end{subequations}
For later convenience we define the deviations from the target mean values in the rotating frame by $\tilde{X}(t) = \expect{X}(t) - X_g$ and $\tilde{P}(t) = \expect{P}(t) - P_g$ and put these deviations together in a vector $Z(t) \equiv [\tilde{X}(t), \tilde{P}(t)]^{\mathsf{T}}$.

We must choose an error signal, $e(t)$, that is based on this control goal and the measurement signal that we have access to. According to the description above, there are two components to the target state in this problem, one for each quadrature of the oscillator, \ie  $X_g$ and $P_g$. However, since our measurements are made in the laboratory frame and we measure only the $x$-quadrature, from now on we shall specify the goal function to be $g(t) = 2x_g(t)$, so that the error signal is then $e(t) = j(t) - 2x_g(t)$. 

Finally, we note that in this work we shall restrict ourselves to the regime of weak measurement and damping $k,\gamma \ll m\omega^2$, 
where the measurement extracts some information about the system at each timestep but does not completely distort the harmonic evolution. Similarly, the system is under-damped by the thermal bath. In this limit, it is valid to still define the characteristic period of the oscillator as $T = 2\pi/\omega$. 

In the following subsections we consider first the case of $x$ measurement with feedback controls in both $x$ and $p$ (section \ref{sec:oscillator_xp}) and then the case of $x$ measurement with feedback control only in $x$ (section \ref{sec:oscillator_true_pro}).

%%%%%%%%%%%%%%%%%%%%%%%%%%%%%%%%%%%%%
\subsection{$x$ and $p$ Control}
\label{sec:oscillator_xp}
We now analyze the case of $x$ measurement with feedback controls in both $x$ and $p$.
%%%%%%%%%%%%%%%%%%%%%%%%%%%%%%%%%%%%%
\subsubsection{Proportional feedback}
\label{sec:oscillator_xp_Pfb}
We first consider proportional feedback only, \ie $\alpha_{i1}=\alpha_{i2}=0$ in Eq. \eqref{eq:lab_dyn}. We shall show that the quadrature expectations of any state can be driven to the target values ($X_g$, $P_g$)
by setting $\alpha_{p1}(t)=2k\eta C_{xp}(t)$, $\alpha_{p2}(t)=-2k\eta V_x(t)$ and $\tau_{P}=0$.
However, in order to compensate for the thermal damping, we also need to add a term $\gamma (x_g(t) p + p_g(t) x )$ to the Hamiltonian $H_0$ (note that this is not a feedback term, since it is not dependent on the measurement record). The evolution of the first moments of the oscillator with these settings is given in Eqs. \eqref{eq:lab_xp2} in Appendix \ref{app:xp_p_eqns}, and when these equations are transformed into the rotating frame, the deviations $\tilde{X}$ and $\tilde{P}$ evolve as:
\begin{subequations}
\label{eq:xp_p_devs}
\begin{align}
	\mathrm{d} \tilde{X} &= -\gamma\tilde{X}\drm t -4k\eta \Big( V_x(t)\cos(\omega t)-C_{xp}(t)\sin(\omega t)/m\omega\Big)\Big[\tilde{X}\cos(\omega t)+\tilde{P}\sin(\omega t)/m\omega\Big] \mathrm{d}t, \\
	\mathrm{d}\tilde{P}&= -\gamma\tilde{P}\drm t - 4k\eta \Big(m\omega V_x(t)\sin(\omega t)+C_{xp}(t)\cos(\omega t)\Big)\Big[\tilde{X}\cos(\omega t)+\tilde{P}\sin(\omega t)/m\omega\Big]\mathrm{d}t.
\end{align}
\end{subequations}
We now see that our choice of proportional feedback coefficients $\alpha_{p1}(t)$ and $\alpha_{p2}(t)$ has allowed the feedback to completely cancel all measurement noise contributions (captured by the $\mathrm{d}W$ terms), resulting in deterministic equations for the evolution of the mean values $\langle x\rangle(t)$ and $\langle p\rangle(t)$. The fact that such cancellation is possible was already noted in the early studies of feedback cooling of quantum oscillators \cite{PhysRevA.60.2700}. In addition, as we shall prove explicitly below, these coefficients make use of the thermal and measurement induced dissipation to steer the system to the target quadrature mean values. 

\begin{figure}[t!]
\subfigure[~Proportional feedback.]{
\label{fig:oscillator_Pxp}
\includegraphics[width=0.4\columnwidth]{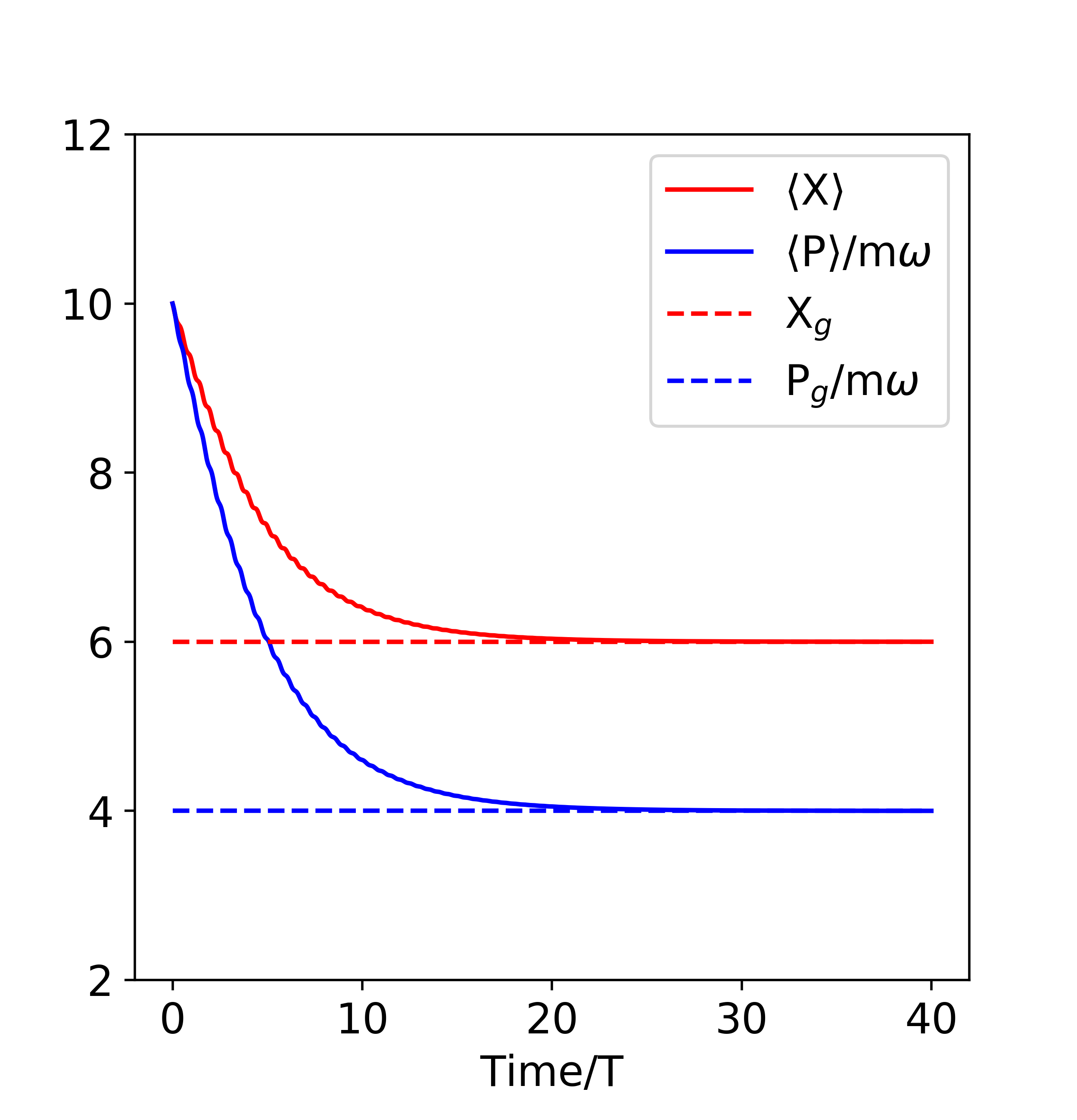}}
\subfigure[~Integral feedback.]{
\label{fig:oscillator_Ixp}
\includegraphics[width=0.4\columnwidth]{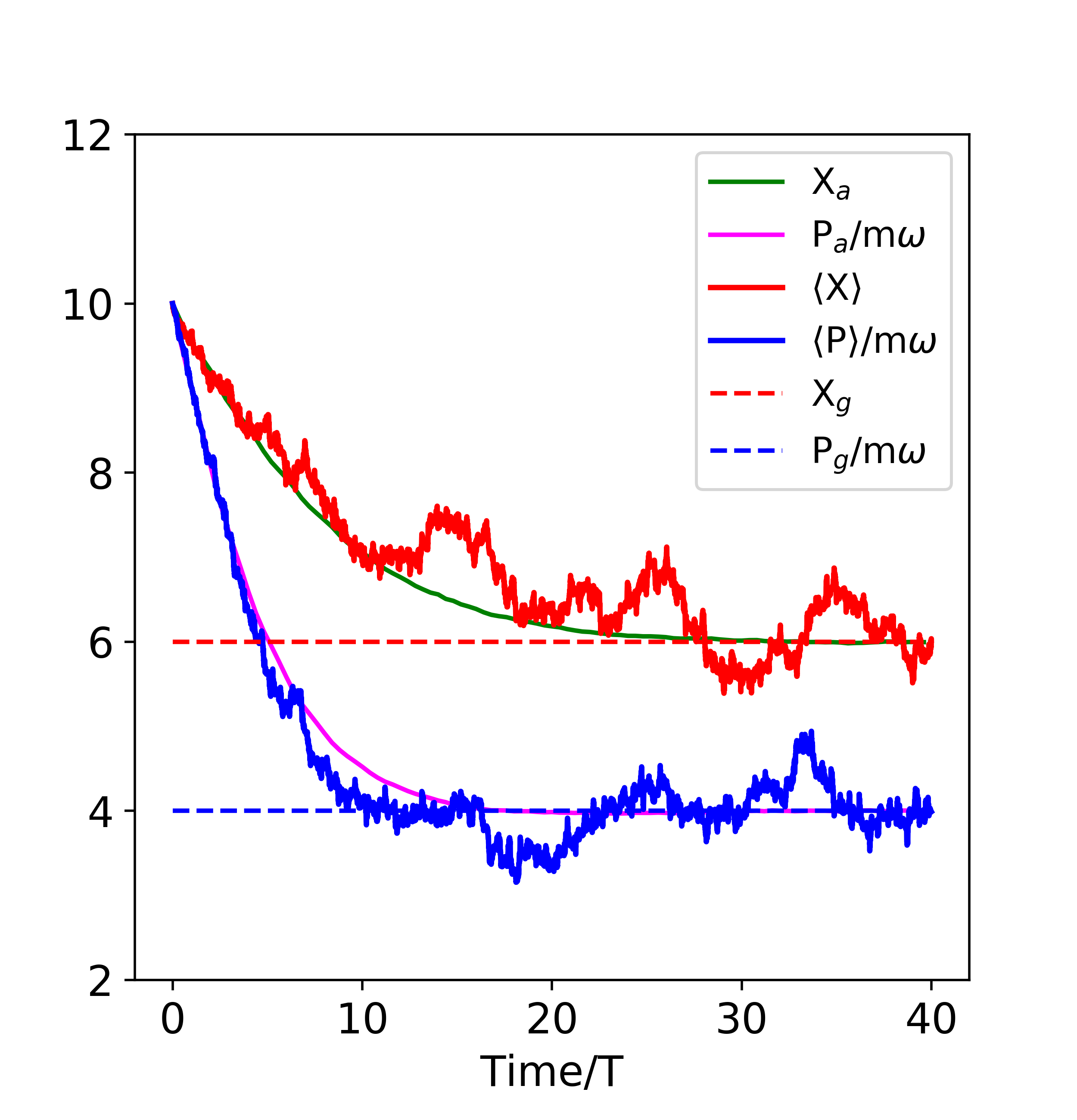}}
\caption{Evolution of expectation values of the quadratures of an oscillator in the rotating frame, subject to continuous measurement, $x$ and $p$ feedback control and thermal damping. The parameters of the system are as follows: $m =\omega=N=1$, $\gamma=k=m\omega^2/(50)$, $\eta=0.4$. The initial state is set to $\langle X \rangle =10$, $\langle P\rangle=10m\omega$ and the target values are set to $X_g=6$, $P_g=4m\omega$ (marked by dotted lines in both subfigures). For these simulations we used $dt=T/250=0.0251$.  (a) Proportional feedback.  The equations for $\langle X \rangle, \langle P/{m\omega} \rangle$ are deterministic (Eq.~\eqref{eq:xp_p_devs}) and converge exponentially to the target values.
(b) Integral feedback. The characteristic time $\tau_I$ for the exponential filter is set to $0.15T$. The red and blue solid lines show the evolution of the expectations $\langle X \rangle, \langle P/{m\omega} \rangle,$ for one trajectory. This evolution is now subject to measurement noise and is not deterministic (Eq. \eqref{eq:xp_int_devs}). The green and purple lines show the behavior of the ensemble average over 1000 trajectories, $X_a(t) = \mathbb{E}\langle X \rangle(t)$ and $P_a/{m\omega} = \mathbb{E}\langle P/m\omega\rangle (t)$. The maximum standard deviation of the trajectories $\langle X \rangle(t)$ and $\langle P \rangle(t)$ increases with $\tau_I$, saturating at 0.7610 at long times. 
}
\label{fig:oscillator_PIxp}
\end{figure}

Fig. \ref{fig:oscillator_Pxp} shows the evolution of the mean values of the quadratures in the rotating frame under this control law for an arbitrary initial state (specified in the caption).  The evolution behavior suggests
that this proportional control law yields exponential convergence to the goal quadrature values. 
To understand why this particular control law works and to prove the exponential nature of the convergence to the target state, we begin by noting that the coefficients in the system of differential equations in Eqs. \eqref{eq:xp_p_devs} display fast oscillations through the $\cos(\omega t)$ and $\sin(\omega t)$ terms, while the changes in the other time-dependent terms, $V_x(t), V_p(t)$ and $C_{xp}(t)$ are small over the timescale of these oscillations. Therefore we may approximate this evolution by another system with new coefficients defined by time-averaging the coefficients in Eqs. \eqref{eq:xp_p_devs} over one oscillator period $T$, and treating all time-varying quantities other than $\cos(\omega t)$ and $\sin(\omega t)$ as constants. For example, $V_x(t)\cos^2(\omega t) \approx \frac{V_x(t)}{2}$ since $\frac{1}{T}\int_0^T \drm t \cos^2(\omega t) = \frac{1}{2}$, and $C_{xp}(t) \sin(\omega t) \cos(\omega t) \approx 0$ since $\frac{1}{T}\int_0^T \drm t \cos(\omega t) \sin(\omega t) = 0$. We refer to this approximation as \emph{period-averaging}, but note that it is equivalent to the rotating wave approximation, since it amounts to dropping fast rotating terms in the evolution operator in the rotating frame. In Appendix \ref{sec:per_avg} we show that this is a very good approximation in the regime $k,\gamma \ll m\omega$. The period-averaged dynamics for the above system, written in matrix form is $\dot{Z}(t) \approx A(t) Z(t)$ (recall that $Z(t) = [\tilde{X}(t), \tilde{P}(t)]^{\mathsf{T}}$), with
\begin{align}
	A(t) = \begin{bmatrix}
  -\gamma -2k\eta V_x(t) & \frac{2k\eta}{(m\omega)^2} C_{xp}(t)\\
  - 2k\eta C_{xp}(t) & -\gamma - 2k\eta V_x(t)
\end{bmatrix}.
\label{eq:xp_p_devs_approx}
\end{align}
The deviation from the target mean values at time $t$ is given by $Z(t) = \exp\left(\int_0^t A(\tau) d\tau \right) Z(0)$. The matrix $A(t)$ has eigenvalues $-\gamma - 2k\eta V_x(t) \pm i\frac{2k\eta C_{xp}(t)}{m\omega}$, for which the real parts are negative for all $t$. Hence, this is a stable system that converges exponentially towards the $\tilde{Z}=0$ fixed point. We may view the $\tilde{Z}(t)$ as a vector Lyapunov function guaranteeing the stability of the final state ~\cite{bellman1962vector}. This shows that for this choice of proportional feedback parameters one can completely cancel the measurement noise and obtain a deterministic system that exponentially stabilizes an arbitrary initial state. 

The P feedback strategy developed above requires $\tau_P=0$, a condition that is experimentally challenging to achieve due to the finite bandwidth of any feedback control loop. Therefore, we have also tested the performance of the feedback law when $\tau_P>0$, on order to investigate the robustness of this strategy.  
The effect of finite time delay on individual trajectories and on the average state evolution is shown in Appendix \ref{sec:timedelays} for several values of $\tau_P$. We find that 
the ensemble average over trajectories for both quadratures, $\mathbb{E}\langle X(t) \rangle$ and $\mathbb{E}\langle P(t) \rangle$, still converge to 
the target values, 
although over a longer timescales than for the ideal $\tau_P=0$ setting (here $\mathbb{E}[\cdot]$ denotes an expectation over trajectories (measurement outcomes)). However, the individual trajectories no longer converge for finite $\tau_p$ values, and fluctuate around 
the target values.
 A detailed analysis of this behavior is given in Appendix  \ref{sec:timedelays}. This general behavior of ensemble averages converging to the target while individual trajectories show final state fluctuations about the average, resembles the stabilization performance under the I feedback strategy that we discuss in the next subsection. 

\subsubsection{Integral feedback}
\label{sec:xp_int_fb}
Now we examine the dynamics obtained by setting $\alpha_{p1}=\alpha_{p2}=0$ in Eq. \eqref{eq:lab_dyn}, which corresponds to applying only integral control. 
The measurement current $j(t)$ provides a noisy estimate of the oscillator position, so it is necessary to filter this in order to obtain a smoothed estimate of the error signal $e(t)$. We use the following exponential filter with memory $\tau_I$:
\begin{equation}
	\mathcal{J}(t)=\frac{1}{\tau_I}\int_{t-\tau_I}^{t}(j(s)-2x_g(s))\exp\Big(-\frac{(t-s)}{\tau_I}\Big)\mathrm{d}s.
\label{eq:filter_Ixp}
\end{equation}
Our choices for the coefficients $\alpha_{i1}$ and $\alpha_{i2}$ in the presence of such an integral filter are motivated by the same factor as in the P feedback case above, namely to cancel as much of the the measurement noise as possible. While it is not possible to do this exactly with I feedback, we show below that the choice $\alpha_{i1}(t) = 2k\eta C_{xp}(t)$ and $\alpha_{i2}(t) = -2k\eta V_x(t)$ does provide exponential convergence of the quadratures to their target values on average. As in the proportional feedback case, we also add a term $\gamma (x_g(t) p + p_g(t) x )$ to the Hamiltonian $H_0$ to compensate for thermal damping.

The evolution of $\drm\langle x\rangle$ and $\drm\langle p\rangle$ with these feedback settings is given in Appendix \ref{app:xp_i_eqns}. Converting these to the rotating frame and writing equations of motion for the deviations $\tilde{X}$ and $\tilde{P}$ yields:
\begin{equation}
  \begin{aligned}  
  \drm \tilde{X}(t) &=-\gamma\tilde{X}(t)\drm t -2\sqrt{k\eta }(\sqrt{k\eta}\mathcal{J}(t)\drm t-\drm W(t))(V_x(t)\cos(\omega t)-C_{xp}(t)\sin(\omega t)/m\omega),\\
  \drm \tilde{P}(t)&=-\gamma\tilde{P}(t)\drm t-2\sqrt{k\eta }(\sqrt{k\eta}\mathcal{J}(t)\drm t-\drm W(t))(m\omega V_x(t)\sin(\omega t)+C_{xp}(t)\cos(\omega t)).
  \label{eq:xp_int_devs}
  \end{aligned}
\end{equation}
A typical evolution (trajectory), started from the same initial state as for the P feedback above, is shown in Fig. \ref{fig:oscillator_Ixp}. We now see random fluctuations in the evolution of the quadrature expectations because the measurement noise has not been exactly cancelled by the I feedback.  Indeed this is now not possible, since the measurement noise term $\mathrm{d}W(t)$ is arbitrarily varying while the integral feedback term is not. Consequently, single trajectories will fluctuate around the target values, preventing perfect state stabilization of individual evolutions. 
However, the average values of the quadratures (marked by the solid lines labeled $X_a$ and $P_a/{m\omega}$ in Fig. \ref{fig:oscillator_Ixp}) do converge exponentially to the goal values. 
In Fig. \ref{fig:oscillator_Ixp} and in subsequent figures where we show stochastic trajectories, we will state the ``maximum standard deviation" at steady state for these trajectories. The standard deviations of $\langle X \rangle (t)$ and $\langle P \rangle (t)$ (calculated over multiple trajectories) are the same but time-dependent and oscillatory at long times. However, this standard deviation is within a narrow range and thus we quote the maximum value over a time window in the steady state region (which is defined as when $\mathbb{E}\langle X \rangle (t)$ and $\mathbb{E}\langle P \rangle (t)$ reach constant values).

To analyze this behavior and prove the exponential convergence of the average over trajectories, we again write Eqs. \eqref{eq:xp_int_devs} in matrix form as $dZ(t) = A Z(t)\drm t + b(t) \drm t + c(t) \drm W(t)$, with 
\begin{align}
	A &= \begin{bmatrix} -\gamma & 0 \\ 0 & -\gamma \end{bmatrix}, \quad b(t) = \begin{bmatrix} -2k\eta\mathcal{J}(t)(V_x(t)\cos(\omega t)-C_{xp}(t)\sin(\omega t)/m\omega) \\ -2k\eta\mathcal{J}(t)(m\omega V_x(t)\sin(\omega t)+C_{xp}(t)\cos(\omega t))\end{bmatrix} \nn \\
	c(t) &= \begin{bmatrix} 2\sqrt{k\eta}(V_x(t)\cos(\omega t)-C_{xp}(t)\sin(\omega t)/m\omega) \\ 2\sqrt{k\eta}(m\omega V_x(t)\sin(\omega t)+C_{xp}(t)\cos(\omega t)) \end{bmatrix}. \nn
\end{align}
The solution to this system can be formally written as 
\begin{align}
Z(t) = e^{-\gamma t}Z(0) + \int_0^t \drm \tau e^{-\gamma(t-\tau)} b(\tau) + \int_0^t \drm W(\tau) e^{-\gamma(t-\tau)} c(\tau)
\end{align}
Note that as before, the second order moments evolve slower than $\cos(\omega t), \sin(\omega t)$. Furthermore, since $\mathcal{J}(t)$ is a smoothed measurement current, it also evolves slowly on the timescale of an oscillator period, $T$. Thus, we may neglect the second term since the integral over the rapidly oscillating sinusoidal terms will average to zero for $t\gg T$. We cannot make the same argument for the third term, since $\drm W(t)$ does not have finite variation over any interval. This third term is in fact what causes fluctuations of individual quadrature trajectories around their setpoint values in Fig. \ref{fig:oscillator_Ixp}. However, note that since $c(\tau)e^{-\gamma (t-\tau)}$ is a non-anticipating function (alternatively, an adapted process that depends only on current and prior times, and independent of the Wiener process), we may conclude that the third term vanishes when averaged over many trajectories, \ie $\mathbb{E}\{\int_0^t\drm W(\tau) e^{-\gamma(t-\tau)}c(\tau)\}=0$ \cite{gardiner_handbook_2004}. This leaves only the first, exponentially decaying term, for the average quadrature values and is therefore the reason for the exponential convergence of the ensemble average to the target values. This analysis also shows that the rate of convergence is slower for I feedback than for P feedback, for which there is an additional contribution of $-2k\eta V_x(t)$ to the convergence rate, see Eq. \eqref{eq:xp_p_devs_approx}.

This first analysis of control of state stabilization for the harmonic oscillator has shown that when access to both $x$ and $p$ control is given, the performance of purely proportional feedback with zero time delay is not improved by adding integral feedback. Indeed, both P and I feedback strategies converge exponentially to the target state when an ensemble average over I feedback trajectories is taken. This shows that state estimation~\cite{PhysRevA.60.2700} is not necessary to drive a harmonic oscillator to an arbitrary quantum state in the presence of thermal noise.  However, when comparing the P and I strategies, it is evident that the P feedback is advantageous for two reasons.  The first is that with zero time delay there exists a proportional feedback law that can perfectly cancel the measurement noise perturbations to the system for each individual trajectory, whereas this can only be approximately canceled under an integral feedback strategy for an individual trajectory, resulting in fluctuations about the target mean quadrature values for any given trajectory.  
The second is a faster convergence for P feedback. 
Given the superior performance of P feedback over I feedback in this setting, we conclude that is not advantageous to consider a more general PI feedback protocol when P feedback with zero time delay is possible. 

For time delays greater than the ideal $\tau_P=0$ the stabilization performance of P feedback strategy degrades, with individual trajectories fluctuating around the target quadrature expectation values and these fluctuations having greater variance as the time delay is increased, although the ensemble average still converges to the target state (Appendix \ref{sec:timedelays}). For time delay values $\gtrapprox 0.2T$ the I feedback strategy  becomes preferable due to the larger deviations from the target values for the P feedback strategy..

%%%%%%%%%%%%%%%%%%%%%%%%%%%%%%%%%%%%%%%%%%%%%%%%%%%%%%%%%%%%

\subsection{$x$ Control only}
\label{sec:oscillator_true_pro}
Our second analysis of control of state stabilization for the harmonic oscillator considers the case of $x$ measurement with only a single control, namely feedback control in $x$.
Under $x$ control only, we set $\alpha_{p2}=\alpha_{i2}=0$, and therefore have a single feedback operator, $F_1=x$. 

\subsubsection{Proportional control}
\label{sec:oscillator_x_Pfb}
As before, we first consider proportional control alone, \ie $\alpha_{i1}$ is also set to zero. Our feedback operator is $x$, and thus the feedback applies a force. Ideally we want this force to be proportional to $-(\expect{p}(t)-p_g(t))$ in order to cancel the measurement noise. However, since we are measuring only the position, we do not have direct access to the momentum observable. This is manifest in the dynamical equations in Eq. \eqref{eq:lab_dyn} by the fact that the only deterministic term involving $\alpha_{p1}$ is the term $-\alpha_{p1}(2\langle x\rangle (t-\tau_P) - g(t-\tau_P))\mathrm{d}t$ in the equation for $\mathrm{d}\langle p\rangle(t)$. This term does not appear to be useful for controlling the oscillator momentum, because it contains information about $\expect{x}$ rather than $\expect{p}$. 
Indeed, we find that the trajectories for evolution of the mean values do not show convergent behavior when implementing proportional $x$ feedback with $\tau_P=0$. 
Noting that for a harmonic oscillator the average position and momentum have a $T/4$ relative delay (see also \cite{doherty2012quantum}),
in the weak measurement and damping limit ($k, \gamma\ll m\omega^2$) we can take a delayed signal term $\langle x\rangle (t-T/4)$ to be a good approximation to the 
the scaled oscillator momentum $-\langle p\rangle (t)/(m\omega)$.
This allows formulation of a 
good control law based on delayed proportional feedback with $\tau_P=T/4$. One can then follow the same line of reasoning  outlined above in Section \ref{sec:oscillator_xp} to tune the strength and offset of the feedback coefficient in order to achieve noise cancellation. 
Specifically, 
we set $\alpha_{p1}= -2k\eta V_x m\omega$ with $\tau_{P}=T/4$. We similarly add a term $\gamma p_g(t)x$ to $H_0$ in order to compensate for thermal damping. 
Note that full compensation of the effects of thermal damping requires adding a term $\gamma (x_g(t) p + p_g(t) x )$, however, consistent with the assumption in this subsection that there is no direct control over the oscillator momentum, we add only the term $\gamma p_g(t)x$.
The resulting dynamical equations for the mean quadratures are given in Appendix \ref{app:x_p_eqns}. When these are transformed into the rotating frame the evolution of the deviations become:
\begin{subequations}
\label{eq:dyn_xcon_pro_XP}
 \begin{align}
 \drm\tilde{X}(t) \approx& ~ 4k\eta V_x(t)\left[-m\omega\tilde{X}(t)\sin(\omega t)+\tilde{P}(t)\cos(\omega t)\right]\sin(\omega t)/m\omega \drm t -\gamma\tilde{X}(t) \drm t \nn
 \\& +[-\gamma X_g\cos^2(\omega t)-\gamma P_g\sin(\omega t)\cos(\omega t)/m\omega]\drm t \nn
 \\
 & +2\sqrt{k\eta }V_x(t)\drm W(t)\cos(\omega t)-2\sqrt{k\eta}\left(C_{xp}(t)\drm W(t)-m\omega V_x(t)\drm W(t-\frac{T}{4})\right)\sin(\omega t)/m\omega\\
 \approx& ~ \left[-2k\eta V_x(t) \tilde{X}(t) -\gamma\tilde{X}(t)  -\frac{\gamma}{2} X_g\right] \drm t \nn
 \\
 & + \left(2\sqrt{k\eta }V_x(t)\cos(\omega t)-2\sqrt{k\eta}C_{xp}(t)\sin(\omega t)/m\omega\right)\drm W(t)+2\sqrt{k\eta}V_x(t)\sin(\omega t)\drm W(t-\frac{T}{4})\\
 \drm\tilde{P}(t) \approx&~ 4k\eta V_x(t)\left[-\tilde{P}(t)\cos(\omega t)+m\omega \tilde{X}(t)\sin(\omega t)
 \right]\cos(\omega t)\drm t - \gamma\tilde{P}(t) \drm t\nn\\
 & + [-\gamma P_g\sin^2(\omega t)-\gamma m\omega X_g\sin(\omega t)\cos(\omega t)]\drm t \nn
 \\
 & +2m\omega \sqrt{k\eta}V_x(t)\drm W(t)\sin(\omega t)+2\sqrt{k
 \eta}\left(C_{xp}(t)\drm W(t) -m\omega V_x(t)\drm W(t-\frac{T}{4})\right)\cos(\omega t) \\
 \approx& ~ \left[-2k\eta V_x(t) \tilde{P}(t) -\gamma\tilde{P}(t)  -\frac{\gamma}{2} P_g \right]\drm t \nn
 \\
 &+\left(2m\omega \sqrt{k\eta}V_x(t)\sin(\omega t) +2\sqrt{k
 \eta}C_{xp}(t)\cos(\omega t) \right)\drm W(t)-2\sqrt{k
 \eta}m\omega V_x(t) \cos(\omega t)\drm W(t-\frac{T}{4}),
\end{align}   
\end{subequations}
where in the second line of each equation we have applied the period-averaging approximation to the deterministic terms, and regrouped the stochastic terms.

The inability to actuate the oscillator momentum in this situation introduces two negative features into these equations relative to Eqs. \eqref{eq:xp_p_devs}, for which both $x$ and $p$ control are available. The first is that we cannot perfectly cancel the measurement noise, resulting in the presence of stochastic terms. The second is that we cannot simply compensate for the thermal damping of oscillator momentum by adding a term $\gamma x_g p$ to $H_0$. This leads to the $-\frac{\gamma}{2}X_g \drm t$ and $-\frac{\gamma}{2}P_g \drm t$ terms in the period-averaged equations. The first point is not a serious hindrance to stabilization, because in the weak measurement limit the effect of the noise is small and leads primarily to fluctuations around the target values. However, the second point is more serious, since the inability to suppress thermal damping means that the system will be driven to a state that is different from the target state. In Appendix \ref{sec:evo_xcon_only} we show how the target state values $X_g$ and $P_g$ can simply be scaled to compensate for this incorrect steady state.

With this compensation trick solving the thermal damping issue for this constrained control setting, we can obtain very good stabilization behavior of individual trajectories to the desired target values, with relatively small fluctuations about these, as shown in Fig. \ref{fig:oscillator_truePx} in Appendix.~\ref{sec:evo_xcon_only}. 
This figure shows a typical trajectory under this proportional control law, incorporating the above scaling of the target $X_g$ and $P_g$ values. It is important to note that we are simulating the dynamics here without any of the approximations used in the above analysis; \ie using the time-delayed feedback current and without invoking the period-averaging approximation.
    Fig. \ref{fig:oscillator_truePx} shows that the time-delayed signal does indeed provide a good estimate of the oscillator momentum, evidently resulting in some but not complete suppression of measurement noise, as well as exponential convergence of the quadrature means to their goal values.
   Thus despite the reduced number of control degrees of freedom, one can nevertheless still achieve exponential convergence of the quantum expectations to their target values using P feedback, with zero bias from the target values and relatively small standard deviation (see Fig. \ref{fig:oscillator_truePx}.)
 
As in the case of $x$ and $p$ actuation, this P feedback strategy requires a precise value for the feedback loop time delay $\tau_P$. Here the desired value of $\tau_P$ is non-zero, and is thus experimentally less demanding to realize than the ideal P feedback strategy with x and p actuation for which $\tau_P =0$ (Sec. \ref{sec:oscillator_xp_Pfb}). However, it might still be challenging to engineer a feedback loop with a precise value of delay $\tau_P=T/4$. To assess the robustness of the strategy with respect to uncertainties in $\tau_P$, we also analyzed the stabilization performance of this P feedback strategy for larger time delays, \ie $\tau_P = T/4 + \epsilon$. Results for several values of $\epsilon$ are shown in Appendix \ref{sec:timedelays}, where it is seen that in this case the stabilization performance degrades for all $\epsilon>0$. The fluctuations of individual trajectories of quadrature expectation increase with $\epsilon$, and there is also a bias in the long-time values of these expectations; \ie the ensemble average values $\mathbb{E}\langle X(t) \rangle$ and $\mathbb{E}\langle P(t) \rangle$ do not converge to the target values. 
This error in convergence is appreciable even for offsets as small as $\epsilon = 0.05T$ and increases with $\epsilon$.

\subsubsection{Integral control}
\label{sec:oscillator_x_Ifb}
We now study the case of integral feedback when only one feedback operator is available, again choosing $F_1=x$. On setting $\alpha_{p2}=\alpha_{i2}=\alpha_{p1}=0$ in Eq. \eqref{eq:lab_dyn}, it is apparent that the only control handle into the system now comes from the $-\alpha_{i1}\mathcal{J}(t)$ term.
As we learned above, the key to stabilizing the system with $F_1=x$ alone is to construct an estimator of the oscillator momentum. For P feedback we used a time delay to achieve this. Here we will construct an estimator with the integral filter. 

Following Doherty \emph{et al.} \cite{doherty2012quantum}, we first modulate the measurement signal to form estimates of the oscillator quadrature deviations in the rotating frame:
\begin{subequations}
\label{eq:I_xp}
\begin{align}
	\mathcal{J}_X(t)&= \frac{1}{\tau_I}\int_{t-\tau_I}^{t}(j(s)-2x_g(s))\cos(\omega s)\mathrm{d}s \approx \tilde{X}(t),\\
	\mathcal{J}_P(t)&= \frac{m\omega}{\tau_I}\int_{t-\tau_I}^{t}(j(s)-2x_g(s))\sin(\omega s)\mathrm{d}s \approx \tilde{P}(t),
\end{align}
\end{subequations} 
Using Eqs. \eqref{eq:ho_tran} these integrals of the measurement record can be combined to yield an estimator of the error between $\langle p\rangle(t)$ and $p_g(t)$: 
\begin{equation}
\label{Eq:J_p}
    \mathcal{J}(t)=-m\omega \mathcal{J}_X(t)\sin(\omega t)+\mathcal{J}_P(t)\cos(\omega t).
\end{equation}
We choose $\alpha_{i1}(t)=4k\eta V_x(t)$ 
to achieve measurement noise cancellation and convergence to the target state. The resulting dynamic evolution of the quadrature means are given in Appendix \ref{app:x_i_eqns}, and the corresponding evolution of the deviations in the rotating frame is given by:
\begin{subequations}
    \label{eq:dyn_xdev_int}
    \begin{align}
    \drm\tilde{X}(t)&=\left[-\gamma\tilde{X}(t) - \gamma X_g\cos^2(\omega t) -\frac{\gamma P_g}{m\omega}\sin(\omega t)\cos(\omega t) + \frac{4k \eta V_x(t)}{m \omega}\mathcal{J}(t)\sin(\omega t)\right]\drm t \nn \\
    &\quad\quad  +2\sqrt{k \eta }(V_x(t)\cos(\omega t)-C_{xp}(t)\sin(\omega t)/m\omega)\drm W(t) \nn \\
    &\approx \left[- 2k \eta V_x(t)\tilde{X}(t) -\gamma\tilde{X}(t) - \frac{\gamma}{2} X_g  \right]\drm t +2\sqrt{k \eta }(V_x(t)\cos(\omega t)-C_{xp}(t)\sin(\omega t)/m\omega)\drm W(t) \\
    \drm \tilde{P}(t)&=\left[ -\gamma \tilde{P}(t) - \gamma P_g\sin^2(\omega t)  -\gamma m\omega X_g\sin(\omega t)\cos(\omega t) - 4k \eta V_x(t)\mathcal{J}(t)\cos(\omega t)\right]\drm t \nn \\
    &\quad\quad +2\sqrt{k \eta }(m\omega V_x(t)\sin(\omega t)+C_{xp}(t)\cos(\omega t))\drm W(t) \nn \\
    &\approx \left[ - 2k \eta V_x(t) \tilde{P}(t)-\gamma \tilde{P}(t) - \frac{\gamma}{2} P_g \right]\drm t +2\sqrt{k \eta }(m\omega V_x(t)\sin(\omega t)+C_{xp}(t)\cos(\omega t))\drm W(t),
    \end{align}
\end{subequations}
where in the second line of each equation we have used the period-averaging approximation and the approximations $\mathcal{J}_X(t) \approx \tilde{X}(t)$ and $\mathcal{J}_P(t) \approx \tilde{P}(t)$.

These equations have the same form as Eqs. \eqref{eq:dyn_xcon_pro_XP}, including exactly the same deterministic terms. Therefore, as shown in Appendix \ref{sec:evo_xcon_only} we know that the ensemble average steady state for this evolution will not be the target values $(X_g, P_g)$. However, as in that case, we can compensate for these scale factors by adjusting the target values. Once this compensation is made, the system converges exponentially towards the target values with fluctuations.  This is evidenced in the simulations shown in Fig. \ref{fig:oscillator_trueIx} which show similar relatively small fluctuations as for P feedback, with standard deviation $0.1676 \pm 0.002$ about the target values at long times.

\begin{figure}[t]
	\centering
		\includegraphics[width=0.4\columnwidth]{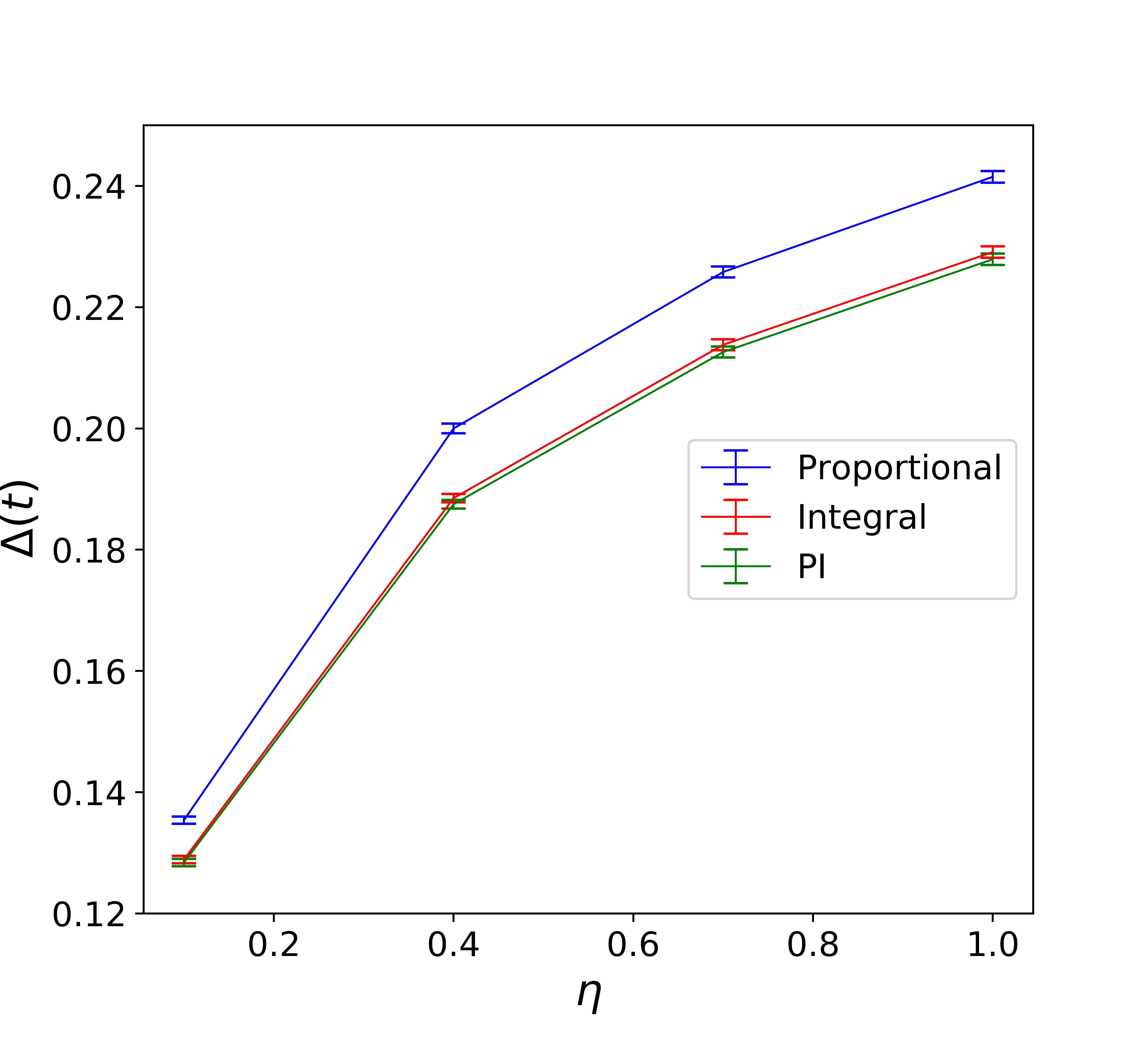}
	\caption{The long time control error, $\Delta(t\rightarrow \infty)$, as a function of the measurement efficiency, $\eta$, for the case of feedback with $F_1=x$ actuation only. The P feedback with time delay strategy is shown in blue, the I strategy is shown in red, and the PI strategy with $\theta=0.8$ (see main text for explanation of this mixing ratio $\theta$) is shown in green. The parameters of the harmonic oscillator are $m=\omega=N=1$, $k=\gamma=m\omega^2/50$, and the target values are set to $X_g=6$, $P_g/{m\omega}=4$. For P feedback, the time delay is $\tau_P=T/4$. For I feedback, the integration time parameter is $\tau_I=T/2$. The error 
	is calculated in all cases by averaging over $20,000$ trajectories.}
	\label{fig:oscillator_e_eta}
\end{figure}

Both the P and I feedback trajectories shown in Fig. \ref{fig:oscillator_truePIx} show stochastic noise. Since the feedback in the integral strategy is conditioned on a tempered version of the noise instead of on the instantaneous noise, we can expect that this smoothing of the noise should give the integral strategy a relative advantage over the purely proportional strategy here.  
While the noise does appear smaller in the I trajectory (compare Fig. \ref{fig:oscillator_trueIx} with \ref{fig:oscillator_truePx}), 
it is difficult ascertain the effect of this on the overall performance of the control strategy by examining single trajectories.  
To enable a quantitative comparison between the performance of the two control strategies in this situation, we therefore define the following average error metric that quantifies the deviation from the control goals when averaged over all measurement trajectories:
\begin{equation}
	\Delta(t)=\sqrt{\frac{1}{2}\mathbb{E}[m\omega\tilde{X}(t)^2+\frac{\tilde{P}(t)^2}{m\omega}]}. 
\end{equation}
We estimate this error by simulating a large ensemble of trajectories with P or I feedback control. In Fig. \ref{fig:oscillator_e_eta} we plot the long-time value of this average error, \ie when it reaches a constant value, as a function of the measurement efficiency, $\eta$. This plot shows that I feedback consistently gives a smaller error and thus performs better than P feedback over essentially the full range of measurement efficiency $\eta$.

In summary, when we only have access to the $F_1=x$ control operator, we do not have sufficient control degrees of freedom to follow the strategy of both cancelling the noise and engineering convergence to the target values, as was possible for P feedback in Sec. \ref{sec:oscillator_xp}. However, we have seen that by forming momentum estimators (via use of time delay in the P feedback case, and via integral approximations of the quadratures in the I feedback case), we can still achieve effective control, with exponential convergence as before. We find that with this approach, both P and I feedback achieve similar control accuracy, with I feedback performing slightly better on average and the difference increasing with greater measurement efficiency $\eta$.
Moreover, both of these P and I feedback strategies show the same rate of convergence to the target quadrature mean values, as is evident from the fact that (within the period-averaging approximation) Eqs. \eqref{eq:dyn_xcon_pro_XP} and \eqref{eq:dyn_xdev_int} have the same deterministic terms. However, neither of these strategies guarantee convergence of individual trajectories. Also we note that the P feedback strategy is very sensitive to the exact value of the time delay, for which the ideal value is $\tau_P=T/4$. Deviations from this ideal value result in inadequate stabilization performance, with failure to reach the target state even on average.

Given the similar performance of P and I feedback in this scenario and the lack of robustness of P feedback to variations in the time delay, 
one might not expect a significant benefit to combining the two to construct a PI feedback strategy. To assess this, we write  $\tilde{\alpha}_{p1}(t)=(1-\theta) \alpha_{p1}(t)$ and $\tilde{\alpha}_{i1}(t)=\theta \alpha_{i1}(t)$ where $\alpha_{p1}(t)$ and $\alpha_{i1}(t)$ are the values determined above, and $\theta\in[0,1]$ is a mixing ratio quantifying the combination of the two strategies. In Fig. \ref{fig:oscillator_e_eta} we plot the long time control error for $\theta=0.8$ (the long time control error is minimum, and almost the same, for any value of $\theta$ in the interval $[0.8,1]$.) and note that indeed, there is little statistically significant benefit to combining P and I feedback in this scenario.

%%%%%%%%%%%%%%%%%%%%%%%%%%%%%%%%%%%%%
\section{Discussion and Conclusions}
\label{sec:discussion}
%%%%%%%%%%%%%%%%%%%%%%%%%%%%%%%%%%%%%

We have presented and implemented a formalism for modeling proportional and integral (PI) feedback control in quantum systems for which, as in the case of classical PI feedback control, we allow the feedback to be tuned from a purely proportional feedback strategy (P feedback, including the possibility of delay) to a purely integral feedback (I feedback), with a combined strategy at any point in between (PI feedback).
In this approach both proportional and integral feedback components are defined in terms of the measurement outcomes only, \ie no dependence on knowledge of the quantum state is assumed.  Consequently we did not seek globally optimal protocols, rather the best performance within the options of P, I, and PI feedback, given the ability to feed back quantum operations based only on the measurement record.  For a given implementation we then first compared the performance of separate P feedback and I feedback control strategies, with and without the presence of time delay in the former, and then carried out a PI feedback strategy, following an assessment of whether or not this might be beneficial.

We implemented this quantum PI feedback approach in this work for two canonical quantum control problems, namely entanglement generation of remote qubits by non-local measurements with local feedback operations, and stabilization of a harmonic oscillator to arbitrary target values of its quadrature expectations when subject to thermal noise.
 
Our first case study was the generation of entanglement by measurement of collective operators of two non-interacting qubits, combined with local feedback operations, for arbitrary measurement efficiency $\eta \leq 1$ and time-independent proportionality constant $\alpha_p$. Unlike the situation for $\eta =1$ and more general time-dependent P feedback \cite{martin_deterministic_2015}, our more restricted -- but experimentally relevant -- case is unable to completely cancel measurement noise, regardless of the value of $\eta$. Here we found that an I feedback strategy can improve on P feedback and achieve superior performance, essentially because an I strategy is able to formulate a smoothed estimate of the error signal by means of the integral filter. This situation is reminiscent of PI feedback control in classical systems \cite{aastrom1995pid} and this case provides strong motivation for the formulation of a general PI feedback law that combines the P and I feedback strategies. We numerically determined an optimal mixing ratio between P and I feedback for this problem of remote entanglement generation, showed that this optimal value can depend on the overall feedback strength and system Hamiltonian, and demonstrated that PI feedback can be beneficial over both the I and P feedback strategies in some cases. 

In the case of the harmonic oscillator, as in previous work on cooling of a harmonic oscillator~\cite{PhysRevA.60.2700}, we studied two settings of feedback control based on measurement of the position degree of freedom $x$, which is generally easier to measure than the momentum $p$.  In the first setting, it is possible to actuate both $x$ and $p$ degrees of freedom of the oscillator, while in the second regime it is possible to only actuate $x$, \ie to apply a force.  The first setting allows formulation of a P feedback strategy that can perfectly cancel the measurement backaction noise entering the system, resulting in a deterministic evolution of the average state~\cite{PhysRevA.60.2700}.  In this setting, adding a Hamiltonian drive to compensate for thermal damping results in a P feedback strategy that allows any state to be exponentially driven to the target quadrature expectation values, without any measurement induced fluctuations.  In contrast, while an I feedback strategy that is exponentially convergent can also be formulated, integral feedback terms are regular and cannot completely cancel the measurement noise. This results in a somewhat slower rate of stabilization and considerable fluctuations in the quadrature expectations for individual trajectories, implying that a PI feedback strategy is not as effective as a P feedback strategy with zero time delay. 
However the ensemble average does converge exponentially to the target quadratures, indicating zero bias 
of the ensemble in the long-time quadrature expectations. 

In the second harmonic oscillator setting, with control only over the $x$ degree of freedom, complete cancellation of the measurement noise can no longer be made, even in a P feedback strategy. However, by using a time delay in P feedback and integral filters in I feedback to obtain estimates of the time-dependent oscillator momentum, we found that it is nevertheless still possible to formulate good feedback control laws that achieve exponential convergence of quadrature expectation values on average, with relatively small measurement noise induced fluctuations of individual trajectories around their target values. 
In this case, we consistently found a small advantage of I feedback over P feedback for all efficiencies $\eta$, with the former also showing smaller fluctuations around the goal. This was seen to stem from the fact that I feedback can derive a smoother estimate of the oscillator momentum through use of a integral filter, and thus allows us to engineer a system with more controlled and smaller fluctuations around the target quadrature mean values. 

Thus for the harmonic oscillator state stabilization, we find the best performance with a pure P strategy when both $x$ and $p$ controls are available, and the best performance with a pure I feedback strategy when only $x$ control is available.  We found little significant advantage in formulating a general mixed PI feedback strategy for the harmonic oscillator state stabilization.
Although we make no claims about the optimality of any of these feedback control strategies for the harmonic oscillator, a significant feature of our analysis is the proof that all of them lead to exponential convergence of the expectation values of the oscillator quadratures to their goal values. We emphasize that this convergence analysis has been restricted to the parameter regime where a period-averaging (\ie rotating wave) approximation is valid. It is possible that this landscape of PI feedback performance could change outside this regime, which is a potential topic for further study.

We also examined the robustness of the P feedback strategies to imperfect time delay, investigating the effects of larger values of $\tau_P$ than specified by the ideal control law.  
We found that the harmonic oscillator state stabilization example when both $x$ and $p$ actuation are available is the most robust to finite time delays, with the quadrature expectations at long time having zero bias from their target values (\ie $\mathbb{E}\langle X \rangle (t\rightarrow \infty)=X_g$ and $\mathbb{E}\langle P \rangle (t\rightarrow \infty)=P_g$), but with fluctuations from the target values that increase with time delay.
Meanwhile, both the harmonic oscillator with only $x$ actuation and the two qubit remote entanglement example are very sensitive to deviations of $\tau_P$ away from the ideal specified value, with performance degrading rapidly as the deviation increases. For the latter cases, the I feedback strategy will therefore be preferred when the perfect time delay condition can not be met.

These case studies reveal a key difference between the benefits of PI feedback in the quantum and classical domains.  In the quantum case,  there is an unavoidable correlation between the noise experienced by the system and the noise in the measurement signal.
This is not always the case in classical systems, where the ``process noise" that the system experiences is often independent of the measurement noise. This difference means that P feedback strategies can play a unique and potentially more powerful role in the quantum domain than they typically do in the classical domain. In particular, in some circumstances, depending on the feedback actuation degrees of freedom, a P feedback strategy can perfectly cancel the noise that the system experiences, while an I feedback strategy can only approximately cancel this noise. Of course, one can get the same behavior in special classical systems, \eg linear systems with zero process noise. In cases where this perfect cancellation is not possible, whether this is due to time delay or other constraints on the feedback action, we saw that I feedback can outperform P feedback, because it provides a smoothed version of the measurement/process noise.  This beneficial value of I feedback is similar to that seen in classical PI feedback control. 

Several possibilities for extending this work are immediately evident. Firstly, formulating \emph{optimal} forms of PI feedback in the quantum domain would be beneficial, even for paradigmatic systems that are analytically tractable like the harmonic oscillator example treated here. The results in the current work indicate that such optimality studies would be particularly useful for feedback control in situations with inefficient measurements (see e.g., \cite{jiang_optimality_2019}).
Secondly, the development of heuristic methods for tuning the optimal proportions of P and I feedback for any system, analogous to those that exist for classical PI feedback control \cite{aastrom1995pid} is an interesting direction. Here, it would be of interest to determine the optimal strategies under constraints of finite measurement and feedback bandwidth, in contrast to the infinite bandwidth controls implicitly assumed in this work, but still without state estimation. Exploration of robust methods to address the implementation of differential control terms to allow implementation of quantum PID control would also be valuable.  Finally, our demonstration of the beneficial effects of integral control strategies for generation of entangled states of qubits under inefficient measurements within the range of current capabilities \cite{PhysRevLett.112.170501}, indicate good prospects for experimental demonstration of quantum PI feedback in the near future.

\begin{acknowledgements}
H.C. acknowledges the China Scholarship Council (grant 201706230189) for support of an exchange studentship at the University of California, Berkeley.

H.L., F.M., and K.B.W. were supported in part by the DARPA QUEST program. 

L.M. was supported by the National Science Foundation Graduate Fellowship Grant No. 1106400 and the Berkeley Fellowship for Graduate Study.  

M.S. was supported by the U.S. Department of Energy, Office of Science, Office of Advanced Scientific Computing Research, under the Quantum Computing Application Teams.

Sandia National Laboratories is a multimission laboratory managed and operated by National Technology \& Engineering Solutions of Sandia, LLC, a wholly owned subsidiary of Honeywell International Inc., for the U.S. Department of Energy's National Nuclear Security Administration under contract {DE-NA0003525}. This paper describes objective technical results and analysis. Any subjective views or opinions that might be expressed in the paper do not necessarily represent the views of the U.S. Department of Energy or the United States Government.
\end{acknowledgements}

%%%%%%%%%%%%%%%%%%%%%%%%%%%%%%%%%%%%%%%%%%%%%%%%%%%%%%%%%%%%%%%%%%%%%%%%%%%%%%%%%%%%%%%%
\bibliography{reference}

\clearpage
\appendix

%%%%%%%%%%%%%%%%%%%%%%%%%%%%%%%%%%%%%

\section{Two-Qubit Entanglement Generation}
\label{sec:app_TwoQubit}

The stochastic master equation that describes the evolution of the two-qubit system for $\tau_P=0$ is 
\begin{equation}
\begin{split}
 \text{d}\rho(t)&=\Big\{-i[H, \rho]+k\mathcal{D}[L_z]\rho - i\alpha_{p}[L_x, L_z\rho+\rho L_z]+\frac{\alpha_{p}^2}{k\eta }\mathcal{D}[L_x]\rho -i\alpha_{i} \mathcal{J}(t)[L_x, \rho] \Big\} \drm t\\
&+\drm W\mathcal{H}[\sqrt{\eta k}L_z-\frac{i\alpha_{p}}{\sqrt{\eta k}}L_x]\rho , 
\end{split}
\label{eq:two_qubits}
\end{equation}
and for $\tau_P>0$
\begin{equation}
\begin{split}
\text{d}\rho(t) &=\Big\{-i[H, \rho]+k\mathcal{D}[L_z]\rho - i\alpha_{p}j(t-\tau_P)[L_x, \rho] 
+\frac{\alpha_{p}^2}{k\eta}\mathcal{D}[L_x]\rho-i\alpha_{i}  \mathcal{J}(t)[L_x, \rho] \Big\} \drm t
\\ &+\sqrt{\eta k}\text{d}W\mathcal{H}[L_z]\rho,
\label{eq:two_qubits_tau}
\end{split}
\end{equation}
where $\alpha_p$ and $\alpha_i$ are the proportional and integral feedback coefficients, and as mentioned 
in the main text above, we have set the goal $g(t)=\bra{T_0}L_z\ket{T_0}=0$.

In this Appendix we write in full the non-linear stochastic equation of motion of the triplet state populations and coherences for the two-qubit example treated in the main text. We keep things general and do not assume $h_1=h_2$ in this Appendix. 

The measurement current is 
\begin{equation}
    I(t)=2\langle L_z\rangle(t)+\xi(t)/\sqrt{k\eta }.
\end{equation}

We denote the populations of the triplet and singlet two-qubit states as $T_{\pm}=\text{tr}(\rho \vert T_\pm\rangle\langle T_\pm\vert)$,  $T_0=\text{tr}(\rho \vert T_0\rangle\langle T_0\vert)$, and $T_S=\text{tr}(\rho\vert S\rangle\langle S\vert)$.
If the initial state is in the triplet subspace, the subsequent evolution will stay within this subspace under the action of the half-parity measurement and local feedback operations $\propto \sigma_{x1} + \sigma_{x2}$.

The evolution of the triplet state populations is given by

\begin{small}
\begin{equation}
\begin{aligned}
\text{d}T_{-1}=&\left[-\sqrt{2}\alpha_{p} \text{Im}T_{0,-1}+\frac{\alpha_{p}^2}{2k\eta}(T_0-T_{-1}-\text{Re}T_{-1,1})+\sqrt{2}\alpha_i \mathcal{J}(t)\text{Im}T_{0,-1}\right]\text{d}t\\
&-\left[2\sqrt{\eta k}T_{-1}(1+\langle{L_z\rangle(t)})-\frac{\sqrt{2}\alpha_p}{\sqrt{\eta k}}\text{Im}T_{0,-1}\right]\text{d}W(t)
\end{aligned}
\end{equation}
\end{small}

\begin{small}
\begin{equation}
\begin{aligned}
\text{d}T_{0}=&\left[2(h_1-h_2)\text{Im}T_{0,S}+\sqrt{2}\alpha_p(\text{Im}T_{0,-1}-\text{Im}T_{0,1})+\frac{\alpha_p^2}{2k\eta}(T_{-1}+T_1-2T_0+2\text{Re}T_{-1,1})\right.\\
&\left.-\sqrt{2}\alpha_i \mathcal{J}(t)(\text{Im}T_{0,1}+\text{Im}T_{0,-1})\right]\text{d}t-\left[2\sqrt{\eta k}T_0\langle L_z\rangle(t)+\frac{\sqrt{2}\alpha_p}{\sqrt{\eta k}}(\text{Im}T_{0,1}+\text{Im}T_{0,-1})\right]\text{d}W(t)
\end{aligned}
\end{equation}
\end{small}

\begin{small}
\begin{equation}
\begin{aligned}
\text{d}T_{1}=&\left[\sqrt{2}\alpha_p\text{Im}T_{0,1}+\frac{\alpha_p^2}{2k\eta}(T_0-T_1-\text{Re}T_{1,-1})+\sqrt{2}\alpha_i \mathcal{J}(t)\text{Im}T_{0,1}\right]\text{d}t\\
&+\left[2\sqrt{\eta k}T_{1}(1-\langle{L_z\rangle(t)})+\frac{\sqrt{2}\alpha_p}{\sqrt{\eta k}}\text{Im}T_{0,1}\right]\text{d}W(t)
\end{aligned}
\end{equation}
\end{small}

The corresponding coherence terms within the triplet subspace are
\begin{small}
\begin{equation}
\begin{split}
\text{d}T_{1,-1}=&\left\{2[i(h_1+h_2)-k]T_{1,-1}+\frac{i\alpha_p}{\sqrt{2}}(T_{0,-1}+T_{1,0})
+\frac{\alpha_p^2}{2k\eta}[T_0-T_{1,-1}-\frac{1}{2}(T_1+T_{-1})]\right.\\
&\left.-\frac{i\alpha_i}{\sqrt{2}} \mathcal{J}(t)(T_{0,-1}-T_{1,0})\right\}\text{d}t+\left[\frac{i\alpha_p}{\sqrt{2\eta k}}(T_{1,0}+T_{0,-1})-2\sqrt{\eta k}\langle L_z\rangle(t) T_{1,-1}\right]\text{d}W(t)
\end{split}
\end{equation}
\end{small}

\begin{small}
\begin{equation}
\begin{aligned}
\text{d}T_{0,1}=&\left\{-[i(h_1+h_2)+\frac{1}{2}k]T_{0,1}+i(h_1-h_2)T_{S,1}-i\sqrt{2}\alpha_p T_1+\frac{\alpha_p^2}{2k\eta}(T_{-1,0}+T_{1,0}-\frac{1}{2}T_{0,-1}-\frac{3}{2}T_{0,1})\right.\\
&\left.-\frac{i\alpha_i}{\sqrt{2}} \mathcal{J}(t)(T_1-T_0
+T_{1,-1})\right\}\text{d}t+\left[\frac{i\alpha_p}{\sqrt{2\eta k}}(T_0-T_1-T_{-1,1})+\sqrt{\eta k}(1-2\langle L_z\rangle(t))T_{0,1}\right]\text{d}W(t)
\end{aligned}
\end{equation}
\end{small}

\begin{small}
\begin{equation}
\begin{aligned}
\text{d}T_{0,-1}=&\left\{[i(h_1+h_2)-\frac{1}{2}k]T_{0,-1}+(h_1-h_2)T_{S,-1} +i\sqrt{2}\alpha_p T_{-1} +\frac{\alpha_p^2}{2k\eta}(T_{1,0}+T_{-1,0}-\frac{3}{2}T_{0,-1}-\frac{1}{2}T_{0,1})\right.\\
&\left.-\frac{i\alpha_i}{\sqrt{2}} \mathcal{J}(t)(T_{1,-1}+T_{-1}-T_0)\right\}\text{d}t
+[\frac{i\alpha_p}{\sqrt{2\eta k}} (T_0-T_{-1}-T_{1,-1})
-\sqrt{\eta k}(1+2\langle L_z\rangle(t))T_{0,-1}]\text{d}W(t)
\end{aligned}
\end{equation}
\end{small}
$T_{s,i}$ are coherences with the singlet state, which come into play if $h_1\neq h_2$.

This system of nonlinear SDEs cannot be solved for directly, except in the special case of P feedback with zero time delay, \ie $\alpha_p>0$, $\alpha_i=0$, $\tau_P=0$. In this case, the SDEs are Markovian and we can directly get the evolution equations for the ensemble average by simply discarding the stochastic terms \cite{wisemanSqueezing1994}. In this case, we can solve for the steady state of $\mathbb{E}T_0(t)$ to get
\begin{equation}
\label{eq:SST0}
    \mathbb{E}T_{0}(t\rightarrow \infty)=\frac{4\eta(h_1+h_2)^2+k^2\eta+8\eta^2k^2+\alpha_p^2}{12(h_1+h_2)^2+3k^2\eta+8\eta^2k^2+3\alpha_p^2}
\end{equation}
This expression shows that the steady state average population in the desired state increases with decreasing $\alpha_p$. However, from simulations we also see that the system takes longer to converge to the steady state as $\alpha_p$ decreases. Finally, we note that $\mathbb{E}[T_{0}(t\rightarrow \infty)]<1$ always.

\section{SME of harmonic oscillator}
\label{sec:SMEHO}
For $\tau_P>0$, the evolution of the system with PI feedback control is obtained from Eq.~\eqref{eq:sme_1} as
\begin{align}
	\label{eq:harmonic_sme_1}
	\drm \rho(t)=& \Big\{-i[H_0,\rho(t)] + k\mathcal{D}[x]\rho(t) +2\gamma(N+1)\mathcal{D}[a]\rho+2\gamma N\mathcal{D}[a^\dagger]\rho+\frac{\alpha_{p1}^2}{k\eta}\mathcal{D}[x]\rho(t)+\frac{\alpha_{p2}^2}{k\eta}\mathcal{D}[p]\rho(t)\nn\\
	&-i \Big(\alpha_{i1} \mathcal{J}(t) + \alpha_{p1} e(t-\tau_P)\Big)[x,\rho]
	-i \Big(\alpha_{i2} \mathcal{J}(t) + \alpha_{p2} e(t-\tau_P)\Big)[p,\rho]
	\Big\}\drm t \nn \\
	 &+\sqrt{k\eta}\mathcal{H}[x]\rho(t)\drm W(t) .
\end{align} 
For $\tau_P=0$, the evolution of the system with PI feedback control is obtained from Eq.~\eqref{eq:sme_2} as
\begin{align}
	\drm \rho(t)=& \Big\{-i[H_0,\rho(t)] + k\mathcal{D}[x]\rho(t) +2\gamma(N+1)\mathcal{D}[a]\rho+2\gamma N\mathcal{D}[a^\dagger]\rho 
	+\frac{\alpha_{p1}^2}{k\eta}\mathcal{D}[x]\rho(t) 
	+ \frac{\alpha_{p2}^2}{k\eta}\mathcal{D}[p]\rho(t)\nn\\
	&-i \Big(\alpha_{i1} \mathcal{J}(t) - \alpha_{p1} g(t)\Big)[x,\rho]
	-i \Big(\alpha_{i2} \mathcal{J}(t) - \alpha_{p2} g(t)\Big)[p,\rho]
	 -i\sqrt{k}\alpha_{p1}[x,x\rho(t)+\rho(t)x]  
	 \nn \\ &-i\sqrt{k}\alpha_{p2}[p,x\rho(t)+\rho(t)x] \Big\}\drm t 
	 + \mathcal{H}[\sqrt{k\eta}x - i\frac{\alpha_{p1}x + \alpha_{p2}p}{\sqrt{k\eta}}]\rho(t) \drm W(t).
	\label{eq:harmonic_sme_2}
\end{align}
In both cases $g(t)$ is a goal that we define in the main text, with $e(t)$ the corresponding error signal. The proportional feedback component is the same as in Ref. \cite{PhysRevA.60.2700}, except that in this work we also consider a time delay $\tau_P>0$ in the feedback loop.

\section{Equations for first and second moments of harmonic oscillator under PI feedback}
\label{sec:firstsecondmoment}
\subsection{Second moments}

The equations of motion of the second moments of the oscillator can be derived by evaluation of ${\rm tr}[V_x d\rho(t)] = {\rm tr}[(x-\langle x \rangle)^2d\rho(t)]$, etc. These evolve as
\begin{subequations}
\label{eq:second_order_gamma}
    \begin{align}
    \dot{V}_x&=-2\gamma V_x+\gamma(2N+1)/m\omega+(2/m)C_{xp}-4k\eta V_x^2,\nn\\
    \dot{V}_p&=-2\gamma V_p+\gamma(2N+1)/m\omega-2m\omega^2C_{xp}-4k\eta C_{xp}^2+k,\nn\\
    \dot{C}_{xp}&=-4\gamma C_{xp}+V_p/m-m\omega^2V_x-4k\eta C_{xp}V_x.
    \end{align}
\end{subequations}

\begin{figure}[!h]
    \centering
    \includegraphics[width=0.3\columnwidth]{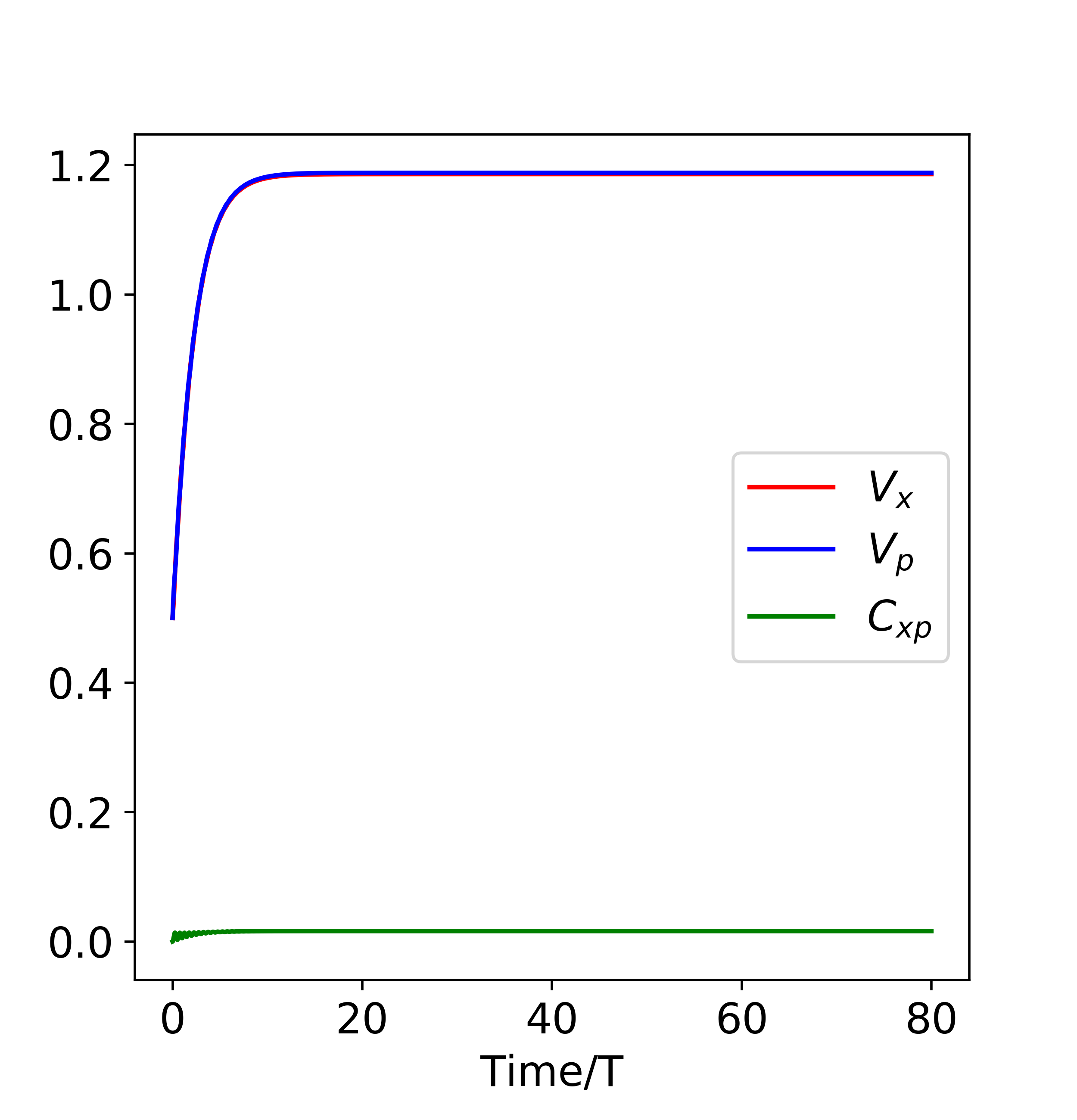}
    \caption{Example evolution of harmonic oscillator second order moments according to Eq. \eqref{eq:second_order_gamma} for $m=\omega=N=1, k=\gamma=1/50, \eta=0.4$.}
    \label{fig:secondmomments}
\end{figure}

Note that there is no dependence of these equations on the first moments, the feedback operator or on the measurement record. Fig. \ref{fig:secondmomments} shows representative evolution of these second moments for system parameters used in the main text ($m=\omega=N=1, k=\gamma=1/50, \eta=0.4$). 

\subsection{First moments: $x$ and $p$ control, Proportional feedback}
\label{app:xp_p_eqns}
The evolution of the first moments of the oscillator with $x$ and $p$ actuation and proportional feedback with the feedback coefficients specified in Sec. \ref{sec:oscillator_xp_Pfb} is given by
\begin{subequations}
\label{eq:lab_xp2}
\begin{align}
    \drm \langle x\rangle(t)=&\frac{1}{m}\langle p\rangle(t)\text{d}t-\gamma(\langle x\rangle(t)-x_g(t))\drm t
    -4k\eta V_x(t)(\langle x\rangle (t)-x_g(t))\drm t,\\
    \drm\langle p\rangle (t)=&-m\omega^2\langle x\rangle(t)\drm t-\gamma(\langle p\rangle(t) - p_g(t))\drm t
    -4k\eta C_{xp}(t)(\langle x\rangle(t)
    -x_g(t))\drm t.
\end{align}
\end{subequations}

\subsection{First moments: $x$ and $p$ control, Integral feedback}
\label{app:xp_i_eqns}

The evolution of the first moments of the oscillator with $x$ and $p$ actuation and integral feedback with the feedback coefficients specified in Sec. \ref{sec:xp_int_fb} is given by
\begin{subequations}
\label{eq:lab_xp_int}
\begin{align}
    \drm\langle x\rangle(t)=&\frac{1}{m}\langle p\rangle(t)\drm t-\gamma( \langle x\rangle(t)-x_g(t))\drm t+\alpha_{i2}\mathcal{J}(t)\drm t +2\sqrt{\eta k}V_x(t)\drm W(t) \\
    \drm\langle p\rangle(t)=&-m\omega^2\langle x\rangle(t)\drm t-\gamma(\langle p\rangle(t)-p_g(t))\drm t+\alpha_{i1}\mathcal{J}(t)\drm t +2\sqrt{\eta k}C_{xp}(t)\drm W(t)
\end{align}
\end{subequations}

\subsection{First moments: $x$ control, Proportional feedback}
\label{app:x_p_eqns}
The evolution of the first moments of the oscillator with $x$ actuation only and proportional feedback with the feedback coefficients specified in Sec. \ref{sec:oscillator_x_Pfb} is given by
\begin{subequations}
\label{eq:dyn_xcon_pro}
    \begin{align}
    \drm \langle x\rangle(t)=&\frac{1}{m}\expect{p}(t)\drm t  -\gamma\langle x\rangle(t)\drm t
    +2\sqrt{k\eta }V_x(t)\drm W(t)\\
    \drm\langle p\rangle(t)=&-m\omega^2\langle p\rangle(t)\drm t - \gamma(\langle p\rangle(t)-p_g(t))\drm t+4k\eta V_x(t) m\omega(\langle x\rangle(t-\frac{T}{4})-x_g(t-\frac{T}{4}))\drm t \nn \\
    &\quad \quad +2\sqrt{k\eta}\left(C_{xp}(t)\drm W(t)-m\omega V_x(t)\drm W(t-\frac{T}{4})\right)\\
    &\approx -m\omega^2\langle x\rangle(t)\drm t -\gamma(\langle p\rangle(t)-p_g(t))\drm t-4k\eta V_x(t) m\omega(\langle p\rangle(t)-p_g(t))\drm t \nn \\
    &\quad \quad +2\sqrt{k\eta}\left(C_{xp}(t)\drm W(t)-m\omega V_x(t)\drm W(t-\frac{T}{4})\right)
    \end{align}
\end{subequations}

\subsection{First moments: $x$ control, Integral feedback}
\label{app:x_i_eqns}
The evolution of the first moments of the oscillator with $x$ actuation only and integral feedback with the feedback coefficients specified in Sec. \ref{sec:oscillator_x_Ifb} is given by
\begin{equation}
 \begin{aligned}   
 \drm \langle x\rangle(t)=&\frac{1}{m}\langle p\rangle(t)\drm t -\gamma\langle x\rangle(t)\drm t+2\sqrt{k \eta}V_x(t)\drm W(t),\\
 \drm \langle p\rangle(t)=&-m\omega^2\langle x\rangle(t)\drm t -\gamma(\langle p\rangle(t)-p_g(t))\drm t -4k \eta V_x(t)\mathcal{J}(t)\drm t +2\sqrt{k \eta}C_{xp}(t)\drm W(t).
 \end{aligned}
 \label{eq:dyn_xcon_int}
\end{equation}

\section{Quality of the period-averaging approximation}
\label{sec:per_avg}
\begin{figure}[t!]
	\includegraphics[width=0.4\columnwidth]{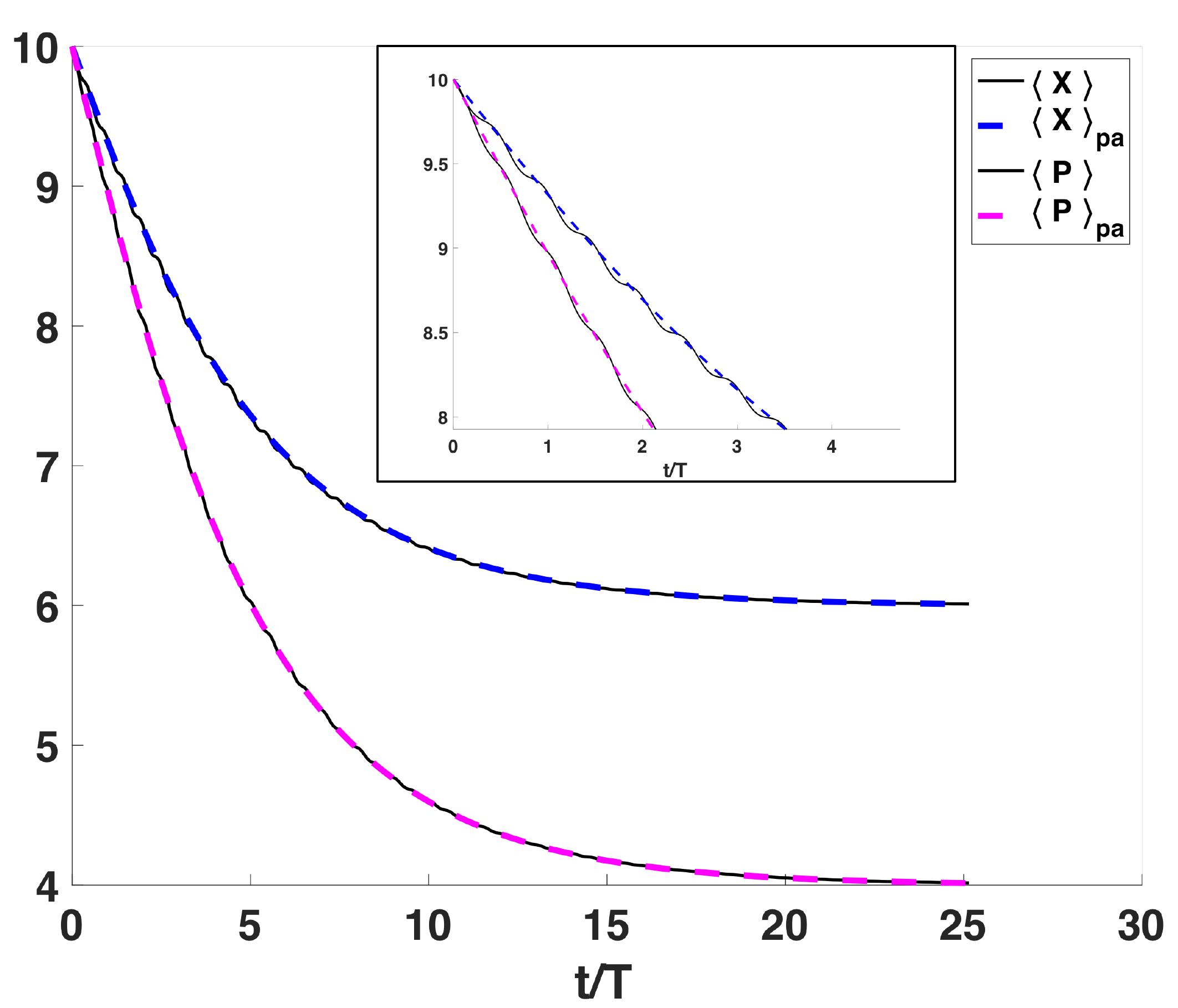}
	\caption{Evolution of $X$ and $P$ quadratures in the rotating frame of an oscillator subject to $x$ and $p$ proportional feedback controls, under the exact evolution (solid, black lines) and the period-averaged evolution (dashed, colored lines). The system parameters are $m=\omega=N=1, \kappa=\gamma=1/50, \eta=0.4$, initial conditions $X_0=P_0=10$, and target values $X_g=6, P_g=4$. The inset is a zoom into the early time scale when the deviation between exact and approximate evolution is greatest. \label{fig:per_avg}}
\end{figure}

Here we evaluate the quality of the period-averaging approximation used in section \ref{sec:HO} of the main text. Consider the deterministic evolution of the oscillator mean deviations $\tilde{X}(t)$ and $\tilde{P}(t)$ in Eqs. \eqref{eq:xp_p_devs}, and its period-averaged approximation in Eq. \eqref{eq:xp_p_devs_approx}. In Fig. \ref{fig:per_avg} we plot the evolution of the oscillator means in the rotating frame, $X(t)$ and $P(t)$, under the exact dynamical equation and its period-averaged approximation for oscillator parameters $m=\omega=N=1, \kappa=\gamma=1/50, \eta=0.4$, initial conditions $X_0=P_0=10$, and target values $X_g=6, P_g=4$. We see that there is very good agreement between the exact and approximate evolution.

\section{Steady-state compensation for harmonic oscillator with $x$ actuation only}
\label{sec:evo_xcon_only}

As stated in the main text, under $x$ actuation only, time-delayed proportional feedback results in the following evolution equations for the deviations in the rotating frame: 
\begin{subequations}
\label{eq:dyn_xcon_pro_XP_app}
 \begin{align}
 \drm\tilde{X}(t) \approx& ~ 4k\eta V_x(t)\left[-m\omega\tilde{X}(t)\sin(\omega t)+\tilde{P}(t)\cos(\omega t)\right]\sin(\omega t)/m\omega \drm t -\gamma\tilde{X}(t) \drm t \nn
 \\& +[-\gamma X_g\cos^2(\omega t)-\gamma P_g\sin(\omega t)\cos(\omega t)/m\omega]\drm t \nn
 \\
 & +2\sqrt{k\eta }V_x(t)\drm W(t)\cos(\omega t)-2\sqrt{k\eta}\left(C_{xp}(t)\drm W(t)-m\omega V_x(t)\drm W(t-\frac{T}{4})\right)\sin(\omega t)/m\omega\\
 \approx& ~ \left[-2k\eta V_x(t) \tilde{X}(t) -\gamma\tilde{X}(t)  -\frac{\gamma}{2} X_g\right] \drm t \nn
 \\
 & + \left(2\sqrt{k\eta }V_x(t)\cos(\omega t)-2\sqrt{k\eta}C_{xp}(t)\sin(\omega t)/m\omega\right)\drm W(t)+2\sqrt{k\eta}V_x(t)\sin(\omega t)\drm W(t-\frac{T}{4})\\
 \drm\tilde{P}(t) \approx&~ 4k\eta V_x(t)\left[-\tilde{P}(t)\cos(\omega t)+m\omega \tilde{X}(t)\sin(\omega t)
 \right]\cos(\omega t)\drm t - \gamma\tilde{P}(t) \drm t\nn\\
 & + [-\gamma P_g\sin^2(\omega t)-\gamma m\omega X_g\sin(\omega t)\cos(\omega t)]\drm t \nn
 \\
 & +2m\omega \sqrt{k\eta}V_x(t)\drm W(t)\sin(\omega t)+2\sqrt{k
 \eta}\left(C_{xp}(t)\drm W(t) -m\omega V_x(t)\drm W(t-\frac{T}{4})\right)\cos(\omega t) \\
 \approx& ~ \left[-2k\eta V_x(t) \tilde{P}(t) -\gamma\tilde{P}(t)  -\frac{\gamma}{2} P_g \right]\drm t \nn
 \\
 &+\left(2m\omega \sqrt{k\eta}V_x(t)\sin(\omega t) +2\sqrt{k
 \eta}C_{xp}(t)\cos(\omega t) \right)\drm W(t)-2\sqrt{k
 \eta}m\omega V_x(t) \cos(\omega t)\drm W(t-\frac{T}{4}),
\end{align}   
\end{subequations}
where in the second line of each equation we have applied the period-averaging approximation of Section~\ref{sec:per_avg} to the deterministic terms, and regrouped the stochastic terms.

We will show below that this system is driven to a steady state with ensemble average quadrature values (where the ensemble average is taken over many trajectories) given by $\mathbb{E}[\langle X\rangle(t\rightarrow \infty)] = \alpha X_g$ and $\mathbb{E}[\langle P\rangle(t\rightarrow \infty)] = \beta P_g$, with $\alpha<1$ and $\beta<1$. 

Inspection of Eq. \eqref{eq:dyn_xcon_pro_XP_app} shows that one can correct this incorrect ensemble average steady state of the evolution by scaling the target quadrature mean values $X_g$ and $P_g$ to compensate for $\alpha$, $\beta$, if these two coefficients can be determined. To do this, we write the solution of Eq. \eqref{eq:dyn_xcon_pro_XP_app} under the period-averaging approximation in matrix form as
\begin{align}
	Z(t) = e^{a(t)} Z(0) + \int_0^t \drm \tau e^{a(t-\tau)} b(\tau) + \int_0^t \drm W(\tau) e^{a(t-\tau)} c(\tau) + \int_{T/4}^t \drm W(\tau-T/4) e^{a(t-\tau)} d(\tau), 
	 \label{eq:x_prop_mateqn}
\end{align}
with 
\begin{align}
	a(t) &= -\gamma t - k\eta \int_0^t d\tau V_x(\tau),  &b(t) &= -\gamma [ X_g, P_g]^{\mathsf{T}}, \nn \\
	c(t) &= \begin{bmatrix}
 	2\sqrt{k\eta }V_x(t)\cos(\omega t)-2\sqrt{k\eta}C_{xp}(t)\sin(\omega t)/m\omega \\
 	2m\omega \sqrt{k\eta}V_x(t)\sin(\omega t) +2\sqrt{k
 \eta}C_{xp}(t)\cos(\omega t)
 \end{bmatrix},
 	&d(t) &= \begin{bmatrix}
 		2\sqrt{k\eta}V_x(t)\sin(\omega t) \\ 
 		-2\sqrt{k\eta}m\omega V_x(t) \cos(\omega t)
 \end{bmatrix} \nn
\end{align}
The first term is exponentially decaying to zero. The second term provides a deterministic offset from zero at long times, which is exactly what leads to the $\alpha$, $\beta$ scaling factors in the steady state. The third and fourth terms generate fluctuations on all trajectories. However, since $e^{a(t-\tau)}c(\tau)$ and $e^{a(t-\tau)}d(\tau)$ are non-anticipating functions (they are independent of the Wiener process), both of these terms will be zero when the expectation value over different measurement realizations are taken. Therefore, we can solve for the ensemble average steady state by dropping the stochastic terms and evaluating the $t\rightarrow \infty$ value of Eq. \eqref{eq:x_prop_mateqn} (or equivalently dropping the stochastic terms from Eqs \eqref{eq:dyn_xcon_pro_XP_app}(b) and \eqref{eq:dyn_xcon_pro_XP_app}(d) and solving for the steady state). Doing this yields $\alpha = \beta \approx (2k\eta V^{\rm ss}_x+\gamma/2)/(2k\eta V^{\rm ss}_x+\gamma)$, where $V^{\rm ss}_x$ is the steady state of this second moment. We note that this expression for $\alpha$ and $\beta$ is approximate, because we have solved for the steady state from Eq. (\ref{eq:x_prop_mateqn}) which was derived under the period-averaged evolution and we also assumed that $\expect{x}(t-T/4)\approx \expect{p}(t)/m\omega$ in formulating our control law. However, both of these approximations are very well justified in the $\gamma,\kappa \ll m\omega$ limit, so that the corresponding expressions provide excellent estimates of the average steady state for the trajectory. 
Knowing the values of $\alpha, \beta (=\alpha )$, we can then compensate for the thermal damping by setting $X_g=X_g^{\rm true}/\alpha$ and $P_g=P_g^{\rm true}/\alpha$, where $X^{\rm true}_g/P^{\rm true}_g$ are the true target values of the quadrature means. \footnote{We note that using such steady state compensation can be an alternative to the strategy of introducing  deterministic terms in the Hamiltonian to cancel thermal damping effects, \ie to introducing one or both of the terms $\gamma (x_g(t)p + p_g(t)x)$.}
Note that this implies a similar rescaling of the laboratory frame target values, \ie $x_g=x_g^{\rm true}/\alpha$, $p_g=p_g^{\rm true}/\alpha$.

\begin{figure}[t!]
\subfigure[~Proportional feedback]{
\includegraphics[width=0.4\columnwidth]{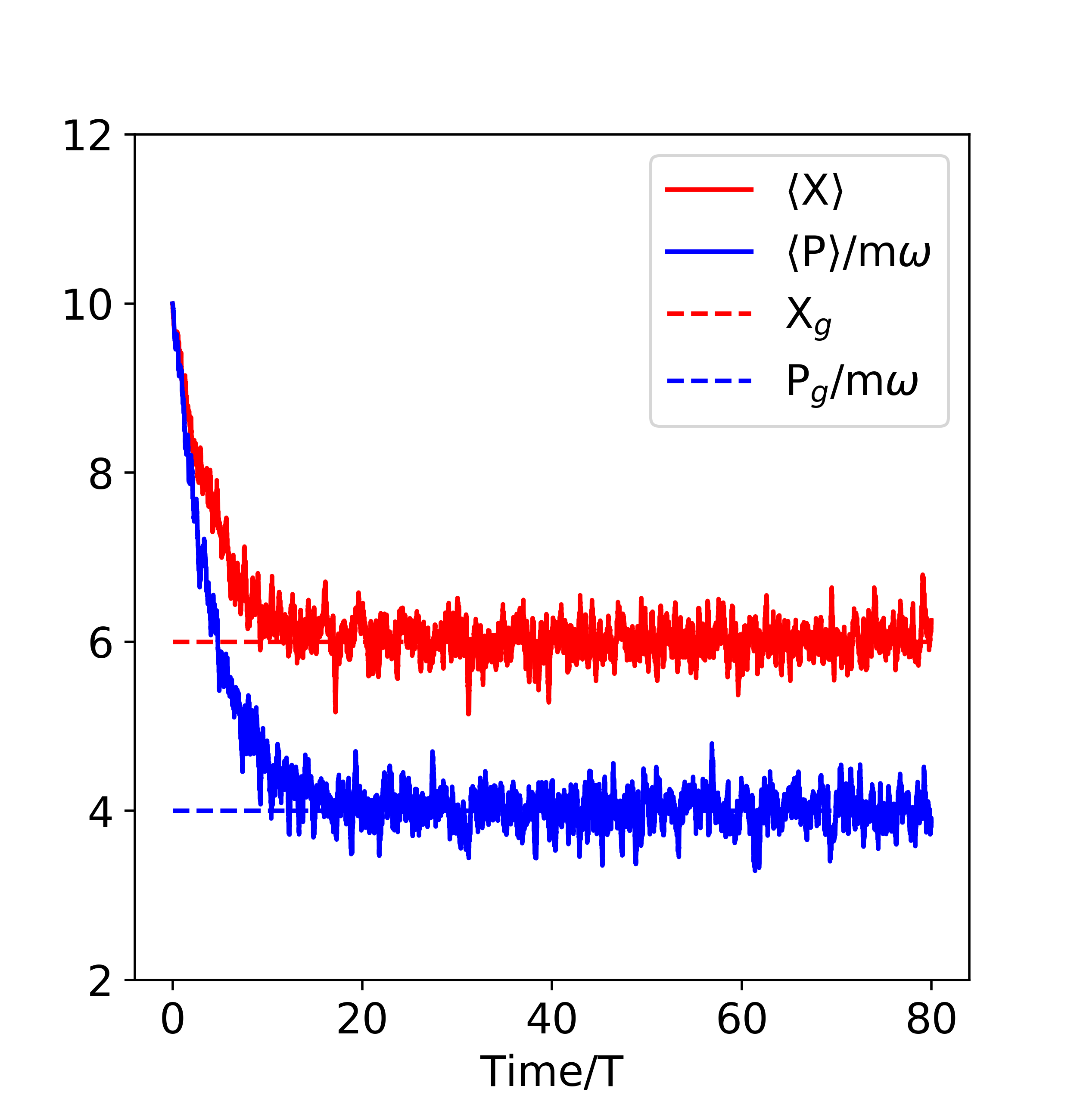}
\label{fig:oscillator_truePx}}
\subfigure[~Integral feedback]{
\includegraphics[width=0.4\columnwidth]{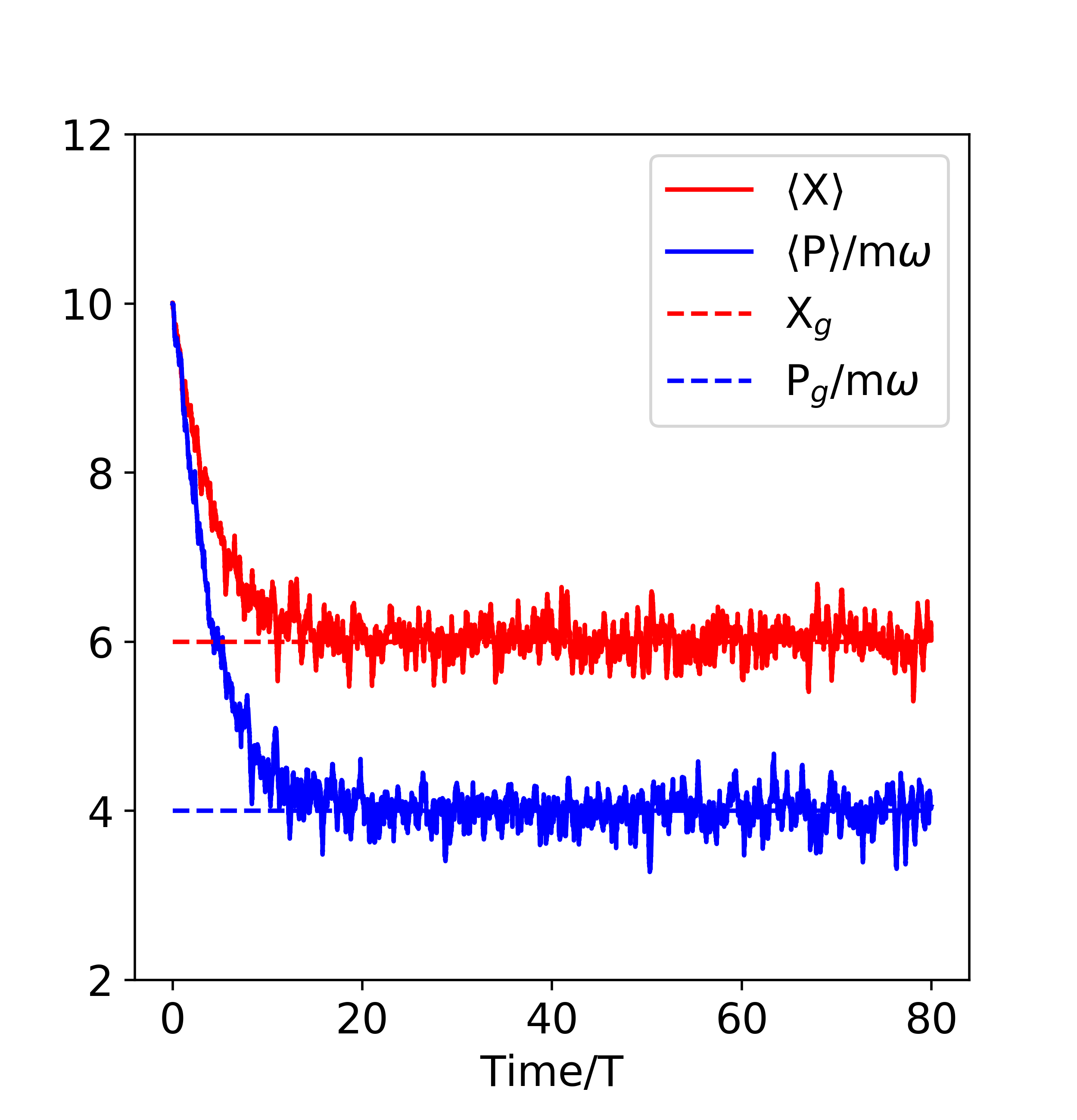}
\label{fig:oscillator_trueIx}}
\caption{Evolution of $X$ and $P$ quadratures in the rotating frame of an oscillator subject to an $x$ feedback Hamiltonian alone (representative individual trajectories). The parameters of the oscillator are as follows: $m= \omega=N=1$, $\eta=0.4$, $k=\gamma=m\omega^2/(50)$. The initial state is set to $\langle X\rangle=10$, $\langle P/{m\omega} \rangle =10$ and the target values are $X_g=6$, $P_g=4m\omega$ (marked by dotted lines in both panels). For these simulations we used $dt=T/500=0.0126$. (a) Proportional feedback control, simulated by Eq.~\eqref{eq:dyn_xcon_pro_XP} with time delay $\tau_P=T/4$. Maximum standard deviation of $\langle X\rangle$ and $\langle P/m\omega\rangle$ in steady state is $0.2420$. 
(b) Integral feedback control, simulated by Eq.~\eqref{eq:dyn_xdev_int} with $\tau_I^\prime=T/2$. Maximum standard deviation of $\langle X\rangle$ and $\langle P/m\omega\rangle$ in steady state is $0.2395$
The steady state compensation in terms of the $\alpha$, $\beta$ discussed in the text is incorporated into both of these simulations, with $\alpha=\beta \approx 0.7434$.
}
\label{fig:oscillator_truePIx}
\end{figure}

\section{Harmonic oscillator stabilization: effect of time delays on P feedback strategies}
\label{sec:timedelays}
For the harmonic oscillator state stabilization example presented in the main text, we derived effective P feedback strategies in the case of x and p actuation, and of x actuation only. In the former case, the P feedback strategy required zero time delay, $\tau_P=0$, while in the latter, formulating a momentum estimate required a time delay of $\tau_P=T/4$.

Given that any real feedback loop will have some time delay, and that sometimes it is difficult to make this delay small compared to the natural timescales of the system being controlled, we study the impact of larger-than-desired time delays on the P feedback strategies in this Appendix, to examine their robustness with respect to variations in $\tau_P$.  

\subsection{x and p Control}
\begin{figure}[t!]
\centering
\subfigure[~$\tau_P=0.05T$ ]{
\includegraphics[width=0.5\columnwidth]{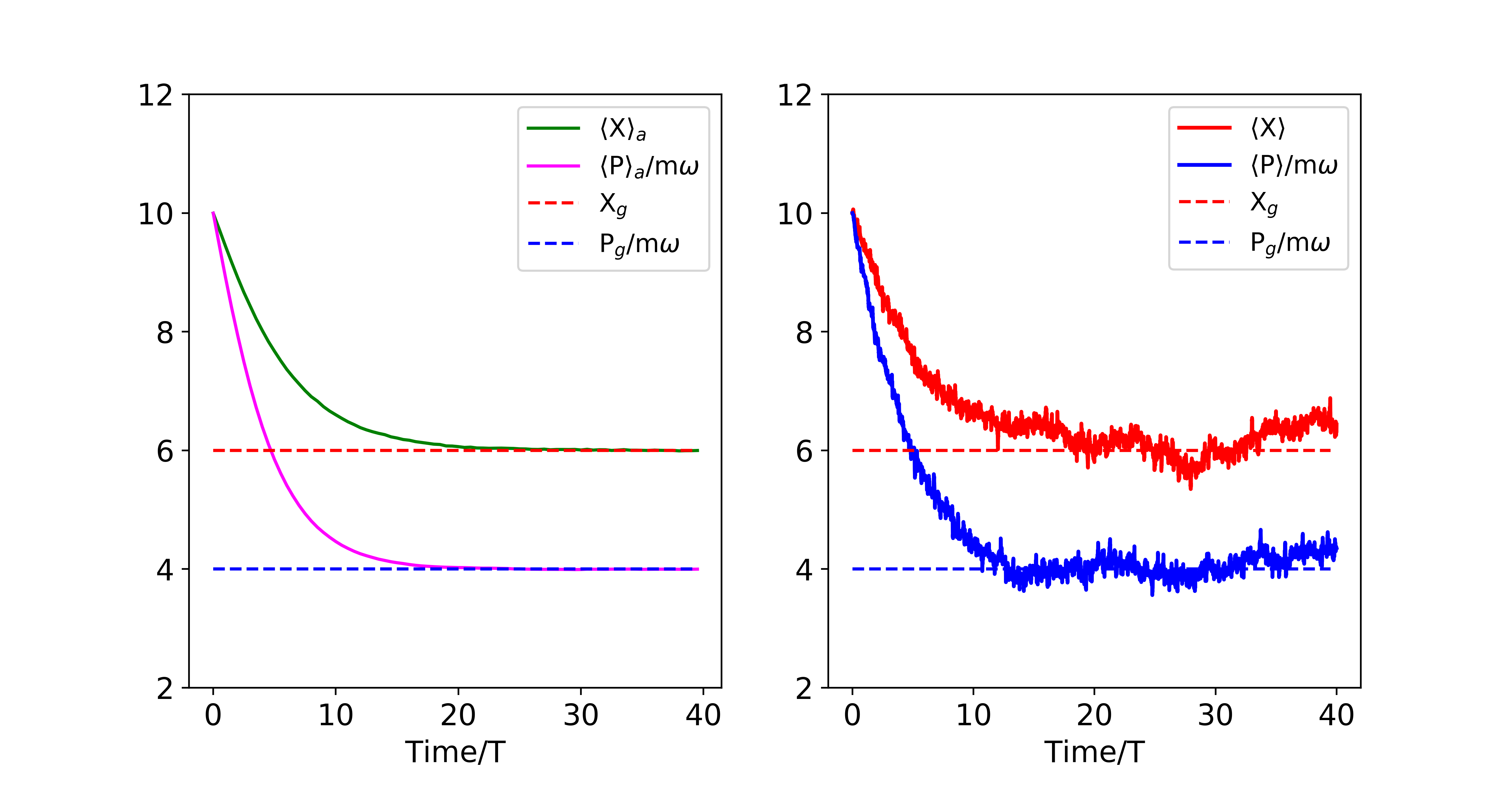}}
\subfigure[~$\tau_P=0.1T$]{
\includegraphics[width=0.5\columnwidth]{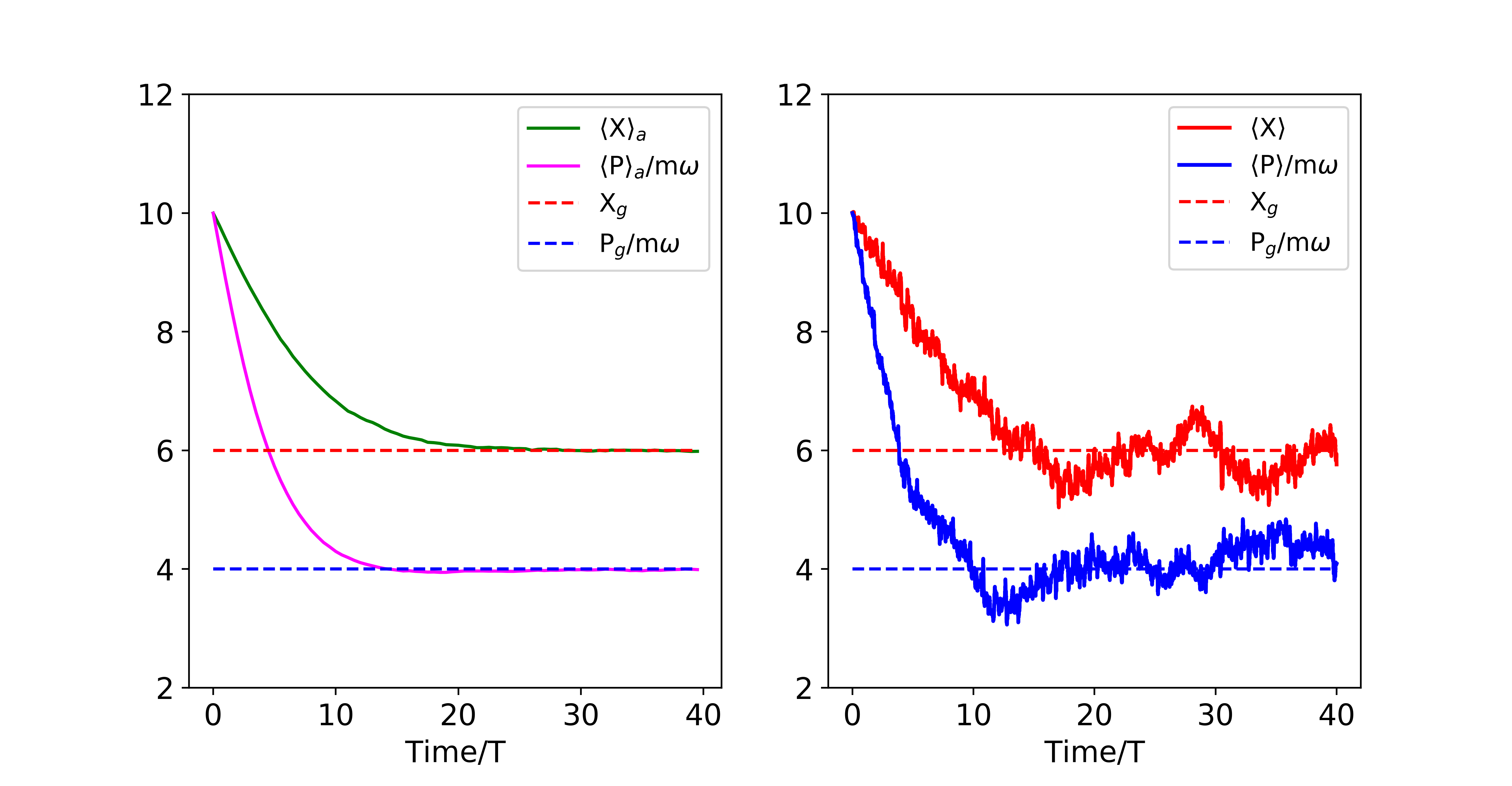}}
\subfigure[~$\tau_P=0.2T$]{
\includegraphics[width=0.5\columnwidth]{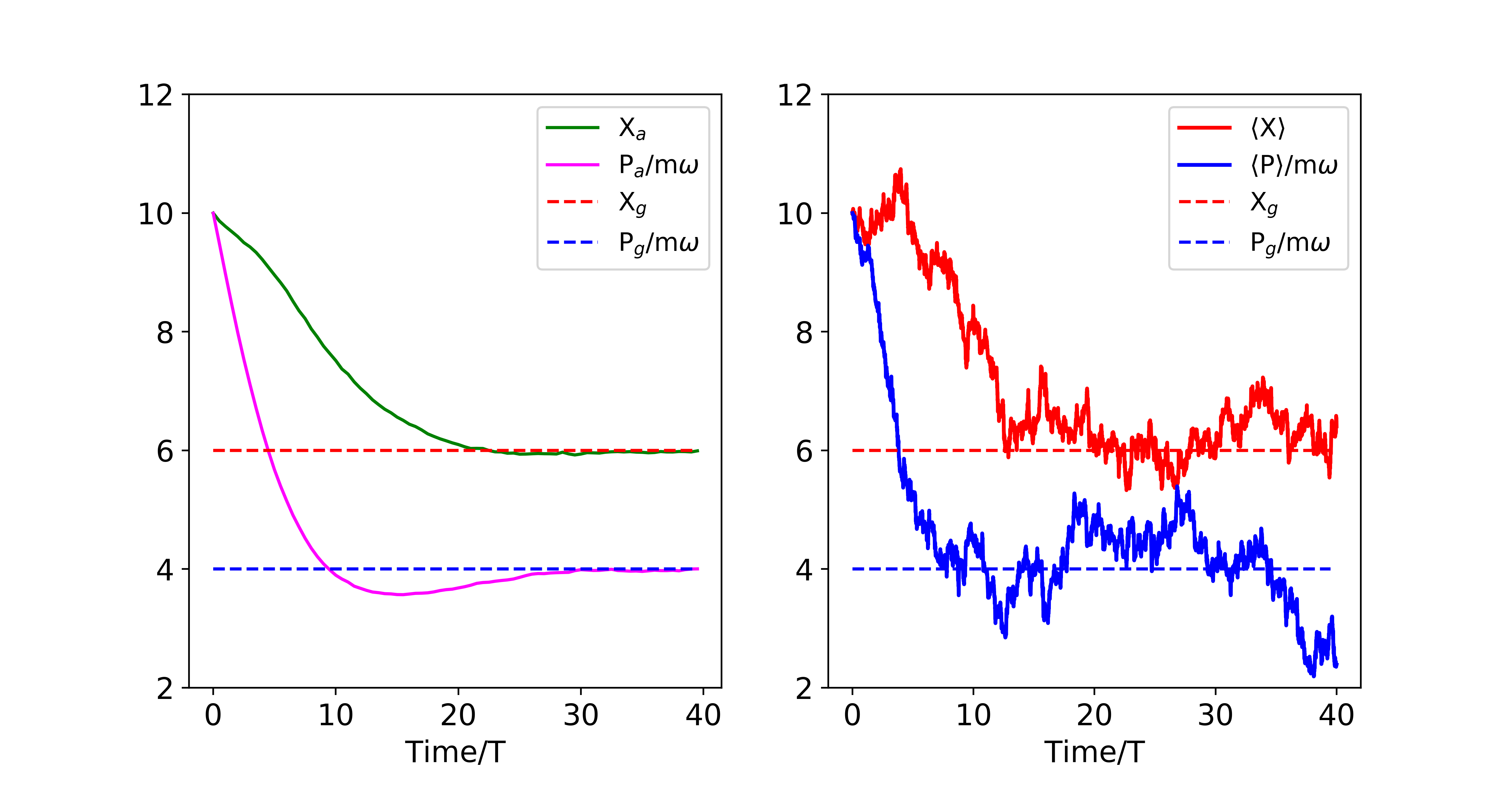}
}
\caption{The effect of time delays on the P feedback law when x and p actuation are available (Sec. \ref{sec:oscillator_xp_Pfb}). The subfigures show three values for the time delay. 
The left panel in each subfigure shows the ensemble average of the quadrature expectations, $\mathbb{E}\langle X(t) \rangle$ and $\mathbb{E}\langle P(t) \rangle$ over 1000 trajectories, and the right panel shows a representative trajectory.
The maximum standard deviation of the trajectories at long times are $0.245$, $0.397$, $0.837$, for $\tau_P=0.05T, 0.1T, 0.2T$, respectively. 
 \label{fig:xp_tau}}
\end{figure}

In the case where $x$ and $p$ actuation is available, Fig. \ref{fig:oscillator_Pxp} of the main text shows that the ideal P feedback strategy with $\tau_P=0$ achieves deterministic and exponential convergence of the quadrature expectations to their target values. In Fig. \ref{fig:xp_tau} we show the behavior of the quadrature expectations for finite delay times, $\tau_P>0$. The trajectories are very different from the case of $\tau_P=0$, showing increasing noise as $\tau_P$ increases. This is expected, since with a finite time delay we no longer exactly cancel the measurement-induced fluctuations. 
While the ensemble average of the trajectories still converges to the target state (left panels of Fig. \ref{fig:xp_tau}), albeit at a slower rate than for $\tau_P=0$, individual trajectories fluctuate around the target quadrature values. Thus with $\tau_P>0$ the long-time behavior of the quadrature expectations  
has zero bias from the target values (\ie $\mathbb{E}\langle X \rangle (t\rightarrow \infty)=X_g$ and $\mathbb{E}\langle P \rangle (t\rightarrow \infty)=P_g$), 
but non-zero variance.

The zero bias property of the quadrature expectations from their target values at long times can be proved rigorously. We return to the equations of motion for the quadratures in the presence of finite time delay, Eq. \eqref{eq:evo_XP}, transform to the rotating frame and then, consistent with averaging over an ensemble of trajectories, drop the stochastic terms to obtain coupled deterministic equations for the deviations $\tilde X$ and $\tilde P$. This is all done while retaining a finite value of $\tau_P$.  Following the notation of the main text, we then arrive at
\begin{equation}
    \dot{Z}(t)=-\gamma Z(t)+AZ(t-\tau_P),
    \label{eq:Z_time_delay}
\end{equation}
where $Z(t)=[\Tilde{X}(t)-X_g, \Tilde{P}(t)-P_g]^T$ and 
\begin{equation}
  A=-2k\eta\begin{bmatrix}
  V_s\cos(\omega\tau_P)-C_s\sin(\omega\tau_P)/m\omega & -V_s\sin(\omega\tau_P)/m\omega-C_s\cos(\omega\tau_p)/(m\omega)^2\\
  m\omega V_s\sin(\omega\tau_P)+C_s\cos(\omega\tau_P)& V_s\cos(\omega\tau_P)-C_s\sin(\omega\tau_P)/m\omega
  \end{bmatrix}.
\end{equation}
Note that we have replaced the second moments by their time-independent steady state values since we are going to be considering the long-time behavior of the system; $V_x(t) \rightarrow V_s, V_p(t) \rightarrow  V_s, C_{xp}(t)\rightarrow C_s$.
Consider the Laplace transform of $Z(t)$: $Z(s)=\int_0^\infty\text{d}t\text{e}^{-st}Z(t)$. The final value theorem says: 
\begin{equation}
 Z(t)\xrightarrow{t\to\infty}\lim_{s\to 0}sZ(s). 
\end{equation}

The Laplace transform of Eq. \ref{eq:Z_time_delay} is given by
\begin{equation}
\begin{aligned}
    &sZ(s)-Z(0)=-\gamma Z(s)+A\text{e}^{-\tau_P s}Z(s)\\
    \Rightarrow &sZ(s)-Z(0)=-\frac{\gamma}{s} sZ(s)+\frac{A\text{e}^{-\tau_Ps}}{s}sZ(s)\\
    \Rightarrow& sZ(s)\left(1+\frac{\gamma}{s}-\frac{Ae^{-\tau_P}s}{s}\right)=Z(0)\\
    \Rightarrow &sZ(s)=\left(I(1+\frac{\gamma}{s})-\frac{A\text{e}^{-\tau_Ps}}{s}\right)^{-1}Z(0) \equiv M^{-1}Z(0).
\end{aligned}
\end{equation}
Assuming for simplicity that $m=\omega=1$ (as in the main text), we have
\begin{equation}
    M
    =\begin{bmatrix}1+\frac{\gamma}{s}&0\\0&1+ \frac{\gamma}{s}
    \end{bmatrix}+\begin{bmatrix}2k\eta (V_s x-C_s y)\frac{\text{e}^{-\tau_Ps}}{s}&-2k\eta (V_s y+C_s x)\frac{\text{e}^{-\tau_Ps}}{s}\\2k\eta (V_s y+C_s x)\frac{\text{e}^{-\tau_Ps}}{s}&
    2k\eta (V_s x-C_s y)\frac{\text{e}^{-\tau_Ps}}{s}
    \end{bmatrix},
\end{equation}
where $x=\cos(\tau_P)$ and $y=\sin(\tau_P)$.
$M^{-1}$ can be explicitly computed and written as
\begin{equation}
\begin{aligned}
    M^{-1}=&\frac{s}{\left(s+\gamma+2k\eta(V_sx-C_sy)\text{e}^{-\tau_Ps}\right)^2 + \left(2k\eta(V_sy+C_sx)\text{e}^{-\tau_Ps}\right)^2}\\
    &\times\begin{bmatrix}s+\gamma+2k\eta(V_s x-C_sy)\text{e}^{-\tau_Ps}&-2k\eta (V_sy+C_sx)\text{e}^{-\tau_Ps}\\
    2k\eta (V_sy+C_sx)\text{e}^{-\tau_Ps}&s+\gamma+2k\eta(V_s x-C_sy)\text{e}^{-\tau_Ps} 
    \end{bmatrix}
    \\\equiv &~
    m_{inv} M_{inv}
\end{aligned}
\end{equation}
It then follows that
\begin{equation}
    Z(t\to\infty)=\lim_{s\to0}sZ(s)=\lim_{s\to0}M^{-1}Z(0)=0,
\end{equation}
since all matrix elements of $M_{inv}$ go to constant values as $s\to 0$, 
while $m_{inv}$ goes to zero $s\to 0$, so that  $\lim_{s\to0}M^{-1}=0$.

\begin{figure}[t!]
\centering
\subfigure[~$\epsilon=0.05T$ ]{
\includegraphics[width=0.5\columnwidth]{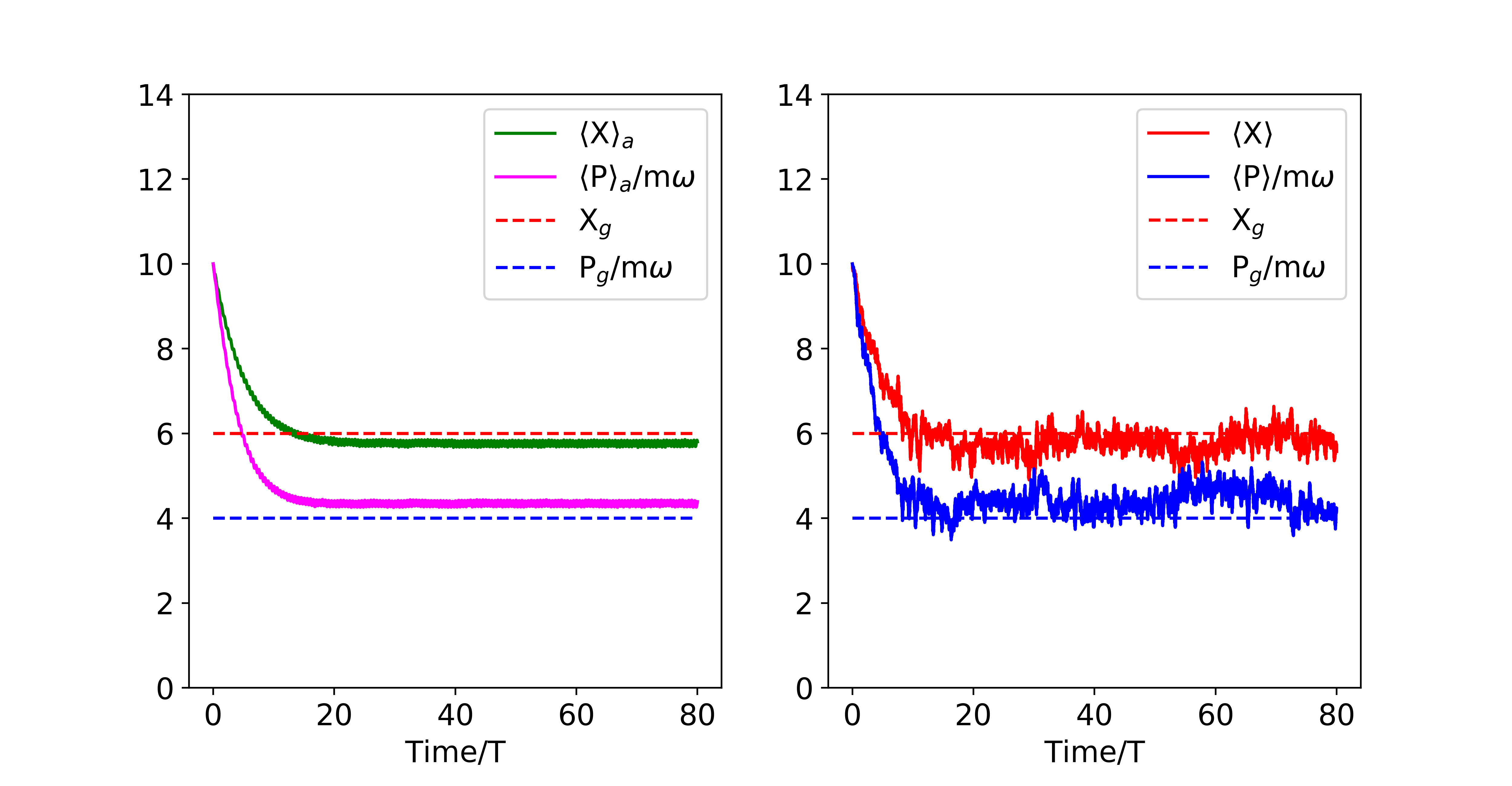}}
\subfigure[~$\epsilon=0.1T$]{
\includegraphics[width=0.5\columnwidth]{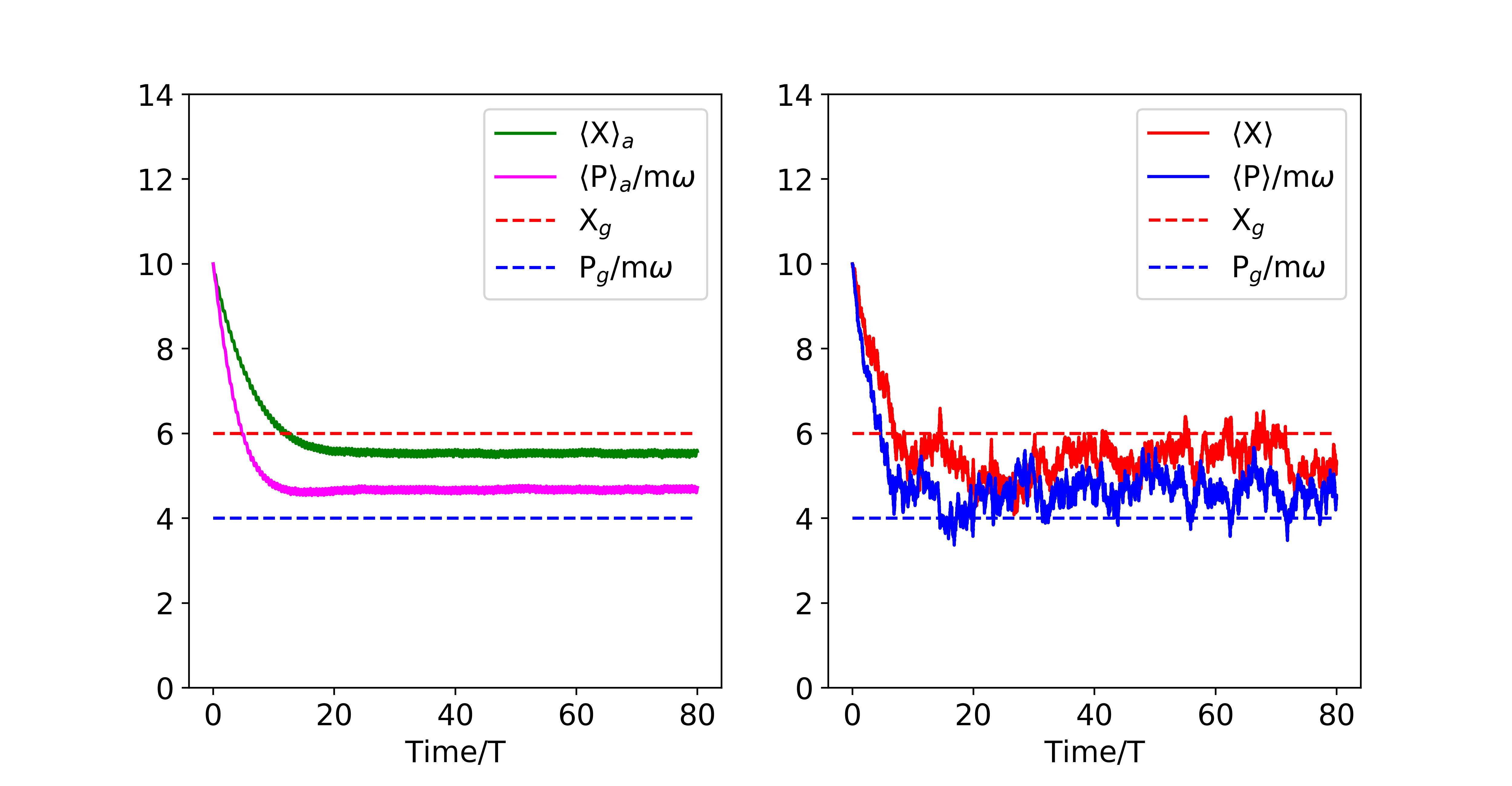}}
\subfigure[~$\epsilon=0.2T$]{
\includegraphics[width=0.5\columnwidth]{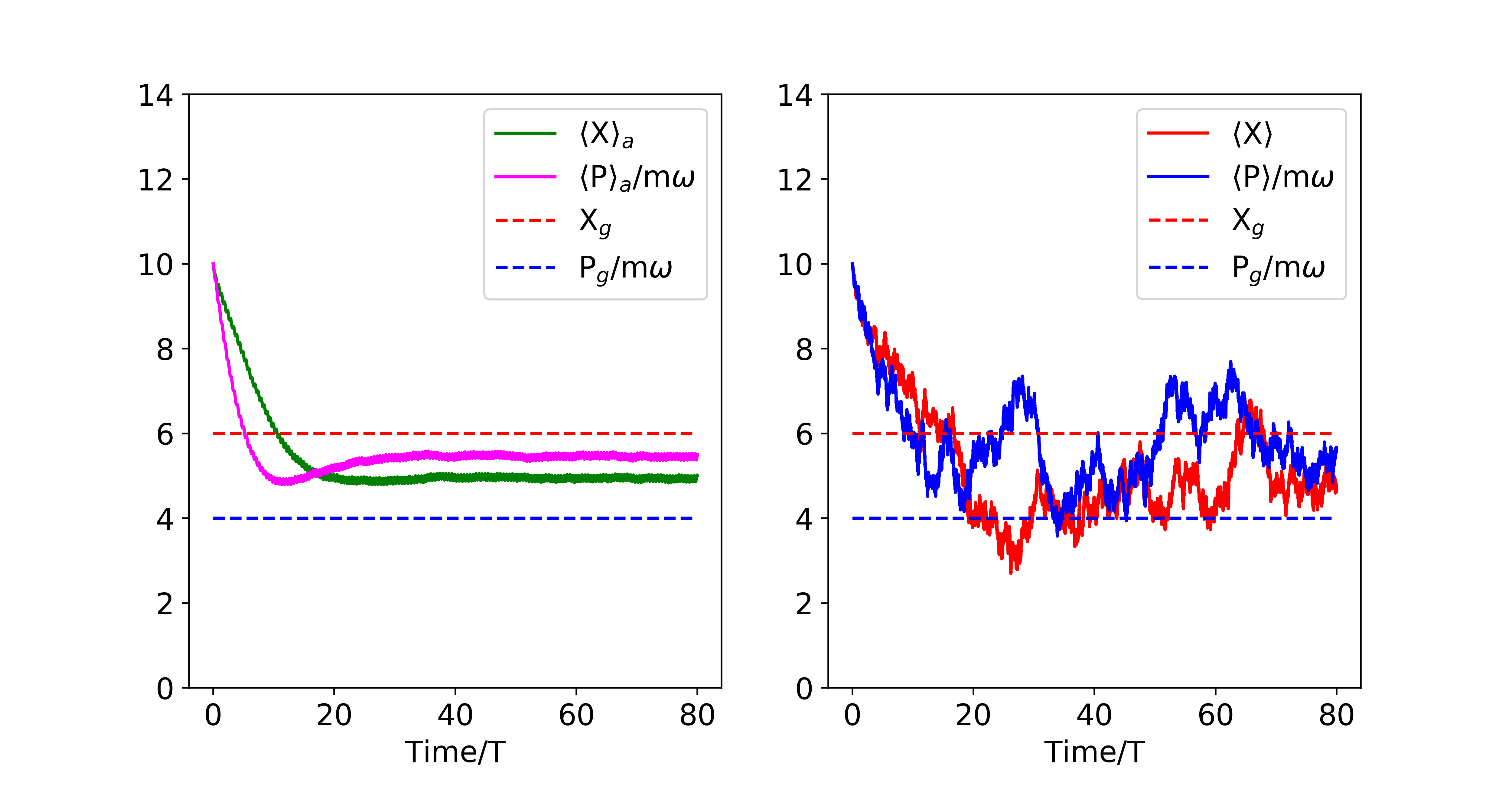}
}
	\caption{The effect of time delays on the P feedback law when only x actuation is available (Sec. \ref{sec:oscillator_x_Pfb}). 
The subfigures show three values for the time delay $\tau_P=T/4+\epsilon$. 
The left panel in each subfigure shows the ensemble average of the quadrature expectations, $\mathbb{E}\langle X(t) \rangle$ and $\mathbb{E}\langle P(t) \rangle$ over 1000 trajectories, and the right panel shows a representative trajectory. 
The maximum standard deviation of the trajectories at long times are $0.314$, $0.453$, $0.863$, for $\epsilon=0.05T, 0.1T, 0.2T$, respectively. 
}  \label{fig:x_tau}
\end{figure}

\subsection{x Control only}
In the case where only $x$ actuation is available, Fig. \ref{fig:oscillator_truePx} shows that the ideal P feedback strategy with $\tau_P=T/4$ achieves exponential convergence of the quadrature expectations to their target values, with a restricted amount of noise on the individual trajectories. In Fig. \ref{fig:x_tau} we now show the behavior of the quadrature expectations when the time delay is not exactly equal to $T/4$, i.e., for $\tau_P = T/4 + \epsilon$ with $\epsilon>0$. We see that in this situation the stabilization performance degrades for all values of $\epsilon$ -- the quadrature expectations deviate from their targets in expectations (show a bias) and the fluctuations in individual trajectories increase with $\epsilon$. 
Thus the performance of the time-delayed P feedback strategy with x control only is less robust to deviations from the ideal $\tau_P$ value than that of the P feedback strategy with both x and p control.

\end{document}